\documentclass[traditabstract]{aa}

\usepackage{graphicx}
\usepackage{amsmath}
\usepackage{color}
\usepackage{natbib}
\bibpunct{(}{)}{;}{a}{}{,}

\begin{document}

\title{Quantifying stellar radial migration in an N-body simulation: blurring, churning, and the outer regions of  galaxy discs}
\titlerunning{Blurring, churning, and the outer regions of  galaxy discs}

\author{A. Halle        \inst{1,3} 
        \and
       P. Di Matteo        \inst{1}
       \and
       M. Haywood    \inst{1}
       \and
       F. Combes \inst{2}
}

\offprints{A. Halle}

\institute{Observatoire de Paris, GEPI, 
        5 place Jules Janssen
        92190 Meudon, France
\and
Observatoire de Paris, LERMA, 61 Av. de
               l'Observatoire, 75014 Paris, France
   \and    UPMC-CNRS, UMR7095, Institut d’Astrophysique de Paris, 
      98 bis boulevard Arago,
   75014 Paris, France    \\
                \email{anaelle.halle@obspm.fr}
}

\date{}

\abstract{Radial stellar migration in galactic discs has received much attention in studies of galactic dynamics
and chemical evolution, but remains a dynamical phenomenon that needs to be fully quantified. In this work,
using a Tree-SPH simulation of an Sb-type disc galaxy, we quantify the effects of blurring (epicyclic excursions)
and churning (change of guiding radius).
We quantify migration (either blurring or churning) both in terms of flux (the number of migrators passing at a given
radius), and by estimating the population of migrators at a given radius at the end of the simulation compared to
non-migrators, but also by giving the distance over which the migration is effective at all radii.
We confirm that the corotation of the bar is the main source of migrators by churning in a bar-dominated galaxy, its intensity being
directly linked to the episode of a strong bar, in the first 1-3 Gyr of the simulation. We show that within
the outer Lindblad resonance (OLR), migration is strongly dominated by churning, while blurring gains progressively more importance
towards the outer disc and at later times.
Most importantly, we show that the OLR limits the exchange of angular momentum, separating the disc in two distinct parts
with minimal or null exchange,
 except in the transition zone, which is delimited by the position of the OLR at the epoch of the formation
of the bar, and at the final epoch. We discuss the consequences of these findings for our understanding
of the structure of the Milky Way disc. Because the Sun is situated slightly outside the OLR, we suggest
that the solar vicinity may have experienced very limited churning from the inner disc.}


\maketitle

\section{Introduction}

Radial stellar migration in the galactic discs has been attracting increasing attention, including some
theoretical \citep{sellwood02,minchev10, daniel14} or numerical works \citep{brunetti11, minchev11, minchev12, dimatteo13, roskar13, veraciro14}, and studies based on observations of the stars of the Milky Way \citep[e.g.][]{haywood08, schoenrich09, yu12}.

In galactic discs where non-axisymmetric potential perturbations such as bars or spiral arms occur, stars can gain or lose angular momentum \citep[e.g.]{lynden72}, leading to outward or inward migration, respectively. The result of this mechanism is that stars can be found at a galactocentric radius differing significantly from their birth radius. Another reason for apparent radial migration is simply the nature of orbits in axisymmetric or nearly axisymmetric potentials: stars oscillate radially around a guiding radius, the amplitude of the oscillations increasing with the radial velocity dispersion. \citet{sellwood02} studied the impact of transient spiral arms on the change in angular momentum of stars in N-body simulations. They found that stars nearly at corotation with a spiral pattern experience the largest changes in angular momentum, larger than what can be experienced by stars at the Lindblad resonances of the spiral pattern or in the rest of the disc. This prominent effect of corotation was confirmed, in particular by \citet{roskar12}, in N-body simulations exhibiting several spiral patterns amongst which the strongest one causes large changes in angular momentum for stars around its corotation radius. Other studies have shown the influence of the bars on radial migration \citep[e.g.]{dimatteo13, brunetti11, kubryk13}. In the case where both a bar and spiral arms are present, the bar-spiral resonance overlap studied by \citet{minchev10, minchev11, minchev12} can produce some non-linear coupling that can generate larger changes in angular momentum. In addition, while angular momentum can occur at corotation without any increase in radial kinetic energy, resonance overlap is expected to increase radial energy: stars that were originally on nearly circular orbits can thus have significantly more eccentric orbits once they have experienced a change in angular momentum that is due to a resonance overlap.

As it redistributes stars over the disc of a galaxy, radial migration  has been invoked as a possible explanation to a number of unanswered questions of galactic and extra-galactic stellar populations:
the upturn of the mean age in the outer parts of discs (\citet{roskar08} and e.g \citet{bakos08, zheng14}), the formation of the Galactic thick disc \citep{schoenrich09, loebman11}, the dispersion in metallicity at a given age observed on stars in the solar vicinity
\citep{haywood08, schoenrich09}.

Each of these problems requires a different level of migrations. For instance,  U-shaped age profiles would require
that stars from the inner disc of a galaxy are massively present in the outer regions. Similarly,
shifting the mean solar-vicinity metallicity towards higher values, as has been suggested to explain the flatness of the
age-metallicity relation at the solar vicinity \citep{loebman11}, would require massive migration of stars to the solar radius.
Explaining the range of metallicities in the solar vicinity is a more subtle problem because it may require only very
little migration as it depends on the local (6-10~kpc) metallicity gradient. If the gradient is steep, the presence of high-metallicity stars found at the solar radius may be explained by blurring alone, without any significant churning. That the gradient is steep is suggested by the data, which
now heavily indicate that the Sun is at the interface between an inner and an outer disc that each have different chemical properties, as advocated in \citet{haywood13}, and confirmed on more extended data by \citet{nidever14}. The recent results found for stars at the solar radius seem thus to dispute that strong migration by churning is necessary to explain the metallicity distribution. 

A common prediction of N-body models of galaxy  evolution is the significant migration of inner disc stars in the external parts of galaxy discs (see for example \citet{roskar08, minchev11, minchev14}).   In the case of the outer disc of the Milky Way, which contains stars with guiding radii greater than $\sim 10$~kpc -- some of them with pericentres small enough to reach the solar vicinity-- there is no evidence that this extreme migration has ever occurred at any time in the last 10~Gyr. At every age, stars of the outer disc are more metal poor than inner disc stars and more alpha-enhanced \citep{haywood13}, which indicates that the outer Galactic disc must have followed a different chemical evolutionary path \citep{snaith14}. Even without access to age information, large-scale spectroscopic surveys like APOGEE are confirming the substantial difference between the inner and the outer Milky Way disc, as is shown by the properties of their stars in the [$\alpha$/Fe]-[Fe/H]  plane or by the change of the metallicity distribution function with distance from the Galaxy centre \citep{anders14,nidever14}. It is striking in this context that regardless of the nature and formation of the outer disc of the Milky Way, stars from the inner thin disc have not been able to significantly pollute the outer parts of the Galactic disc,  when models of radial migration would predict this as a common scenario. This also
casts some doubts on the idea that migration due to the present bar/spiral is at the origin of the U-shaped age profiles observed in external galaxies, at least in bar-dominated systems where this inversion occurs outside of the outer Lindbland resonance (OLR) of the bar.

However, it is possible that the first pattern that occurred in the galaxies had a lower speed than the current speed and that the OLR was at larger radii, which would have extended the churning action at larger radii. During the mass assembly of galaxies, the concentration of mass increases, and consequently the pattern speed increases, implying a smaller OLR radius with time. The simulations of \citet{bournaud02}, which included gas accretion, showed such an increase in the bar pattern speed that
is due to a central mass accumulation that causes the increase of the azimuthal and radial frequencies, and subsequently an increase of the bar pattern speed.\\

In addition, the characteristics of the thick disc of the Milky Way hardly require
any significant migration, and they can be  well explained by formation from a disc
that is rich in gas and turbulent \citep{haywood13,lehnert14, haywood15}.
On the theoretical side, the uncertainties are equally important, and although phenomenological radial migration is
now implemented in several Galactic chemical evolution models \citep{schoenrich09, kubryk14a, kubryk14b}, there is still much uncertainty about the importance of the role of radial migration. For example, what
fraction of stars are subject to migration and on which distances?
Is the migration episodic or continuous? How are the different parts of the disc affected by migration? While tentative answers to these questions have been given in past works, from the simple model of recurring spiral perturbations in an isolated disc in \citet{sellwood02} to recent models in a cosmological context by \citet{minchev14}, the detailed quantitative effects of radial migration are still subject to debate.  \\

In this paper, we are interested in distinguishing between the effects of `blurring', or in other words, the radial migration that is due to epicyclic excursions around a fixed guiding radius, and `churning', which is the radial migration that is due to a change in this guiding radius (according to the terminology of \citet{schoenrich09}). In particular, we seek to quantify the fraction of the stars involved in the migration, the extent of the migration, the fraction of migrators at a given galactocentric radius, and the regions affected by the process. The study is performed using an N-body+SPH simulation of an Sb-type galaxy that was first presented in \citet{halle13}. A few results on radial migration for this simulations were presented in the Appendix of \citet{dimatteo14}. In the present paper, we first briefly list the parameters of the simulation in Sect.~2. In Sect.~3 we study the density resonances, and in Sect.~4 we  quantify the effects of blurring and churning, their strength, and their
extent. Section~5 discusses the behaviour of migrators near the outer Lindblad resonance, and in particular its role as a barrier for migrators. Finally, the kinematic characteristics of migrating stars are discussed in Sect.~6, and we conclude in Sect.~7.

\section{Numerical simulation}

\begin{table*}
\caption{Sb galaxy parameters}
\begin{flushleft}
\begin{tabular}{lcccccccccc}

\hline
Sb & M$_{h}$& M$_{d}$ & M$_{b}$& M$_{g}$   &  $r_h$ & $a_d$& $h_d$ & $r_b$ & $a_g$ & $h_g$ \\
& [M$_\odot$]  & [M$_\odot$] & [M$_\odot$] & [M$_\odot$]  & [kpc] & [kpc] & [kpc]  & [kpc]  & [kpc]  & [kpc]\\

\hline

  & 1.7 $10^{11}$ & 4.5 10$^{10}$ & 1.1 10$^{10}$ & 0.9 10$^{10}$ &   12  & 5  & 0.5 & 1 & 11.8 & 0.2 \\
    \hline

\end{tabular}
\end{flushleft}

\label{t-sim}
\end{table*}

We used one of the simulations presented in \citet{halle13}. We briefly summarize the main characteristics of the simulations. They were performed with the N-body SPH code Gadget-2 \citep{springelg2} and include stochastic star formation following a Schmidt law, kinetic feedback from core-collapse supernovae and some detailed cooling of the gas down to $100~\mathrm{K}$, with the possibility of including cooling by molecular hydrogen, whose local mass fraction is computed based on a semi-analytic recipe from \citet{kmtI,kmtII,kmtIII}.

The simulated Sb type galaxy includes
\begin{itemize}
\item a stellar ($s$) and a gas ($g$) discs that both have Miyamoto-Nagai density profiles:
\begin{multline}
\rho_{s,g}(R,z)=\frac{h_{s,g}^2 M_g}{4\pi} \\ 
\times \frac{a_{s,g} R^2+\left( a_{s,g}+3 \sqrt{z^2+h_{s,g}^2}\right) \left( a_{s,g}+\sqrt{z^2+h_{s,g}^2} \right)^2  }{ \left( R^2+ \left( a_{s,g}+ \sqrt{z^2+h_{s,g}^2}\right)^2  \right)^ {\frac{5}{2}} \left(z^2+h_{s,g}^2 \right)^{\frac{3}{2}} }
\end{multline}
\item a stellar bulge ($b$) and a dark matter halo ($h$) with Plummer density profiles:
\begin{equation}
\rho_{b,h}(r)=\frac{3M_{b,h}}{4\pi r_{b,h}}\left( 1+\frac{r^2}{{r_{b,h}}^2}\right)^{-\frac{5}{2}} 
.\end{equation}
\end{itemize}

The masses of the different components and density profiles parameters are shown in Table~\ref{t-sim}. There are initially 400~000 particles of each component with particle masses of m$_{\mathrm{DM}}$, m$_{\star}$ and m$_{g}$ for the dark matter, stars, and gas, listed in Table \ref{t-res}. The softening length $\epsilon$ used in the gravitational force computation and the smoothing length $h$ used for hydrodynamics are also listed in Table \ref{t-res}. Particle velocities were assigned in the following way (see \citet{halle13} for details): For gaseous and stellar disc particles, we computed circular velocities from gravitational accelerations and applied an analytic asymmetric drift correction to obtain a more realistic profile. Radial velocity dispersions were derived as $\sigma_r(r)=\dfrac{3.36GQ\Sigma(r)}{\kappa (r)}$, with $\kappa (r)$ and $\Sigma(r)$ the epicyclic frequency and the surface density at radius $r$, respectively, and adopting a Toomre parameter $Q$ equal to 1. The azimuthal velocity dispersions $\sigma_\theta(r)$ were computed from the radial velocity dispersions $\sigma_r(r)$  using the epicycle approximation, and the vertical velocity dispersion $\sigma_z(r)$ was derived by assuming isothermal equilibrium for the discs. For the spheroidal components, the velocity dispersion is isotropic and was derived from the second moment of the Jeans equation. It can be noted here that our initial conditions for the evolution of the disc are not representative of clumpy discs observed at high-redshifts \citep[e.g.][]{elmegreen07}. We focused on the effect of a bar on the evolution of an already formed thin disc. In clumpy discs with a chaotic evolution, resonant patterns causing significant churning are difficult to form because of the higher velocity dispersion, and the radial migration is more likely to be some blurring. The uncertainty on the morphology and kinematics of galactic discs before bar formation is large.
In addition, we did not take into account any thick-disc component that may have contributed significantly to the total mass of the disc at the epoch of bar formation in the Milky Way, for example \citep[e.g.][]{haywood13, dimatteo14,snaith14}. This kinematically hot component may have been less affected by radial migration \citep{loebman11,brunetti11}.

\begin{table}
\caption{Resolution of the simulation}
\begin{flushleft}
\begin{tabular}{ccccc}
\hline
$\epsilon$ & h & m$_{\mathrm{DM}}$ & m$_{\star}$ & m$_{g}$  \\
$[\mathrm{pc}]$ &  & [M$_\odot$]  &[M$_\odot$]  &[M$_\odot$] \\
\hline
100 & $\geq \dfrac{\epsilon}{10} $ & 3.7 $10^{5}$ & 1.4 10$^{5}$ & 2.5 10$^{4}$    \\
    \hline
\end{tabular}
\end{flushleft}
\label{t-res}
\end{table}

\begin{figure*}
\centering
\includegraphics[width=16cm]{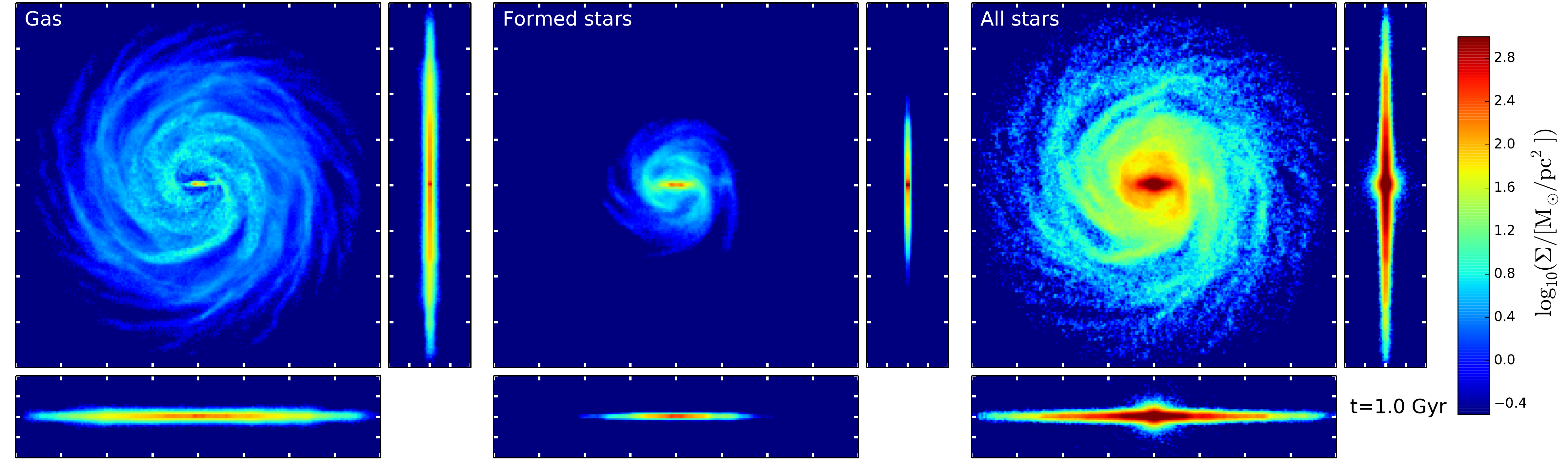} \\
\includegraphics[width=16cm]{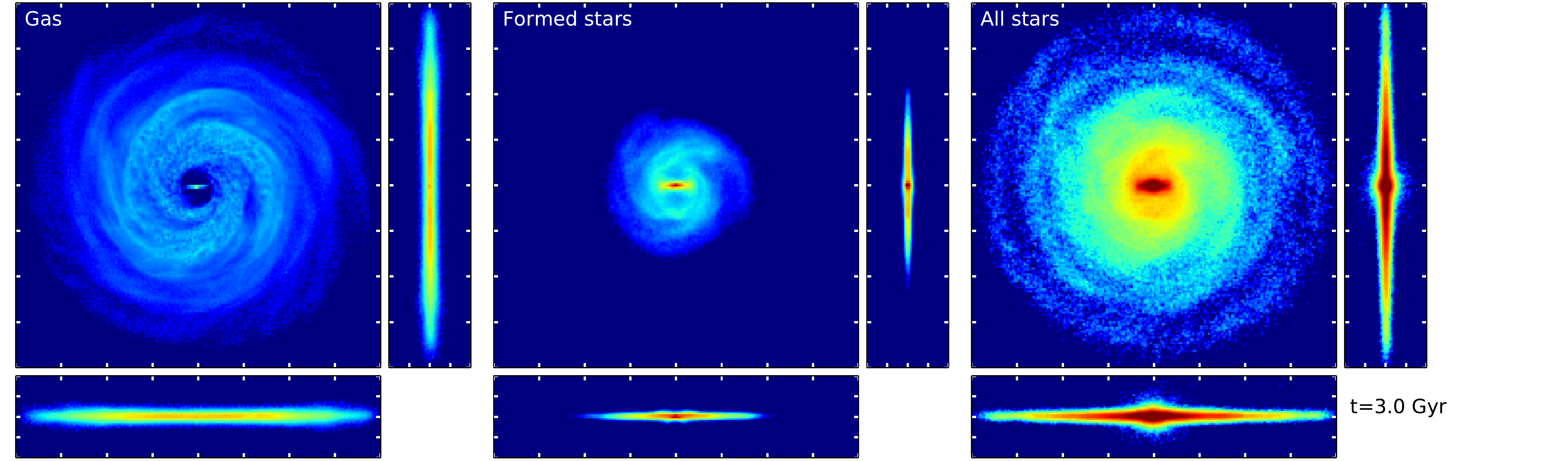} \\
\includegraphics[width=16cm]{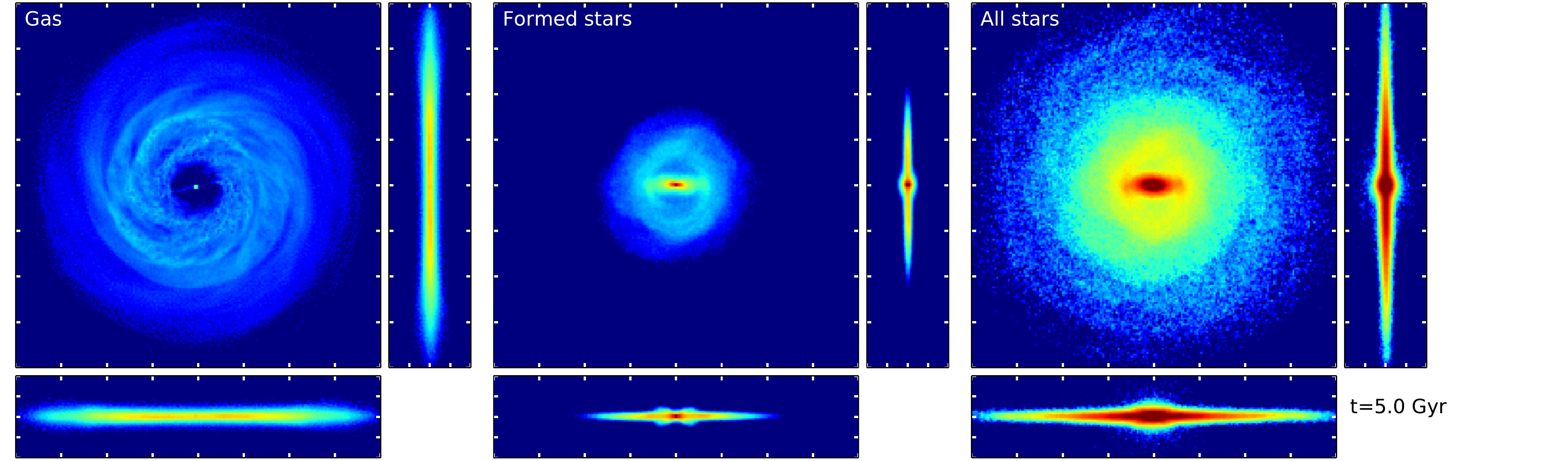} \\
\includegraphics[width=16cm]{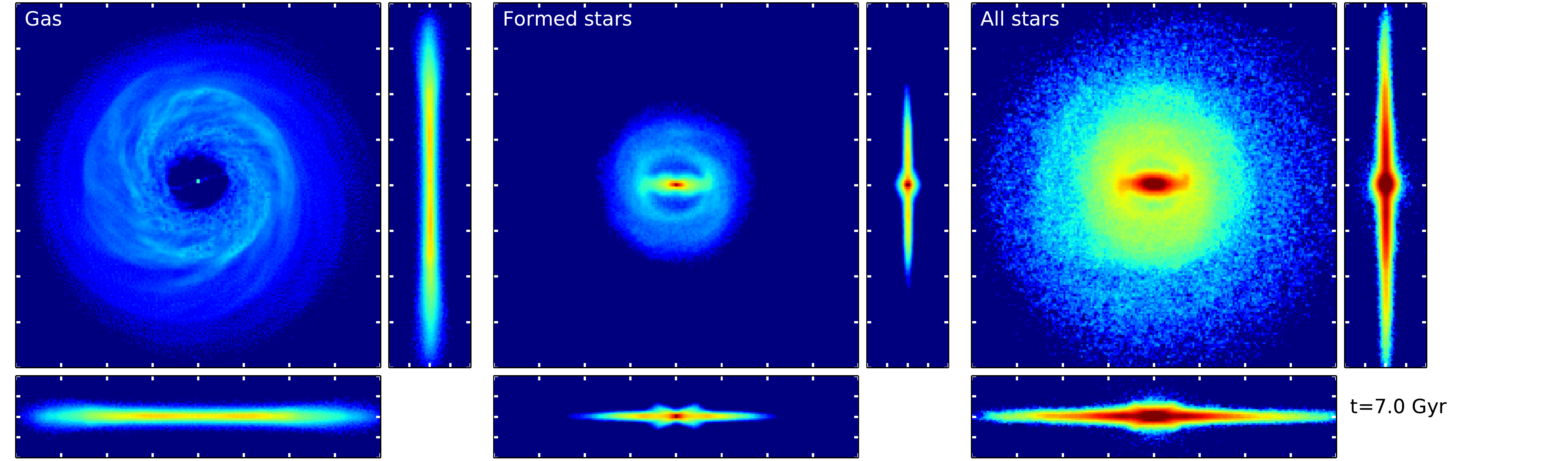} \\
\includegraphics[width=16cm]{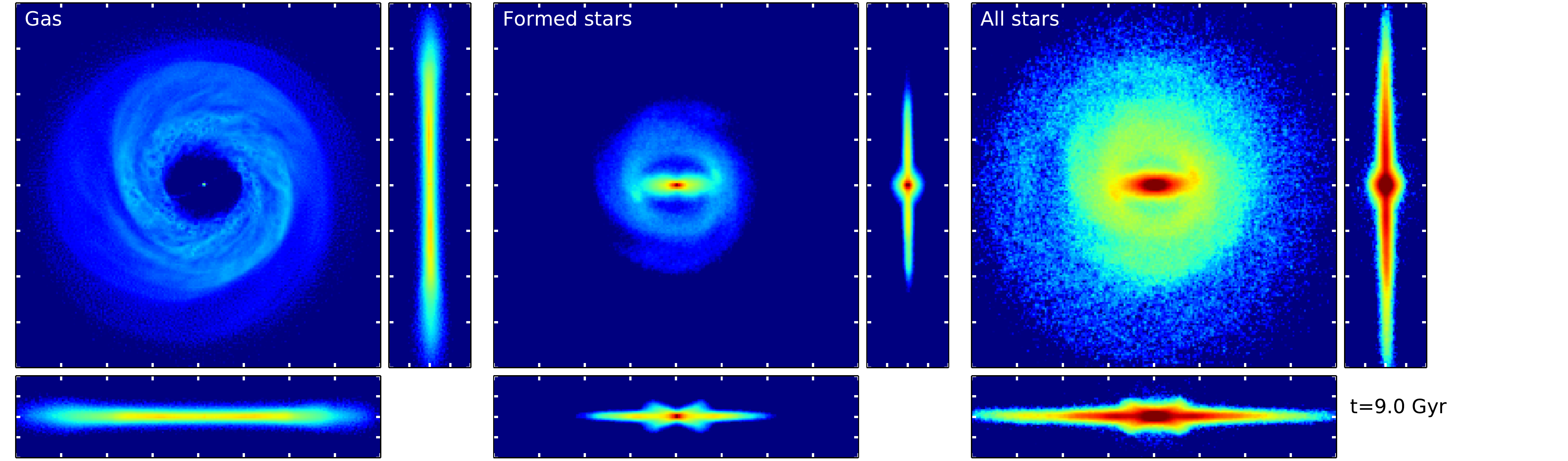} 
\caption{Surface density maps of the gas (left), stars formed during the simulation (middle), and all the stars (right). Face-on view boxes have a size of [80~kpc $\times$ 80~kpc] and edge-on views a size of [80~kpc $\times$ 20~kpc]. The colour scale, shown in the top right corner, is the same for all the views.}
\label{snap-fig}
\end{figure*}

\begin{figure}[h!]
\centering
\resizebox{\hsize}{!}{\includegraphics{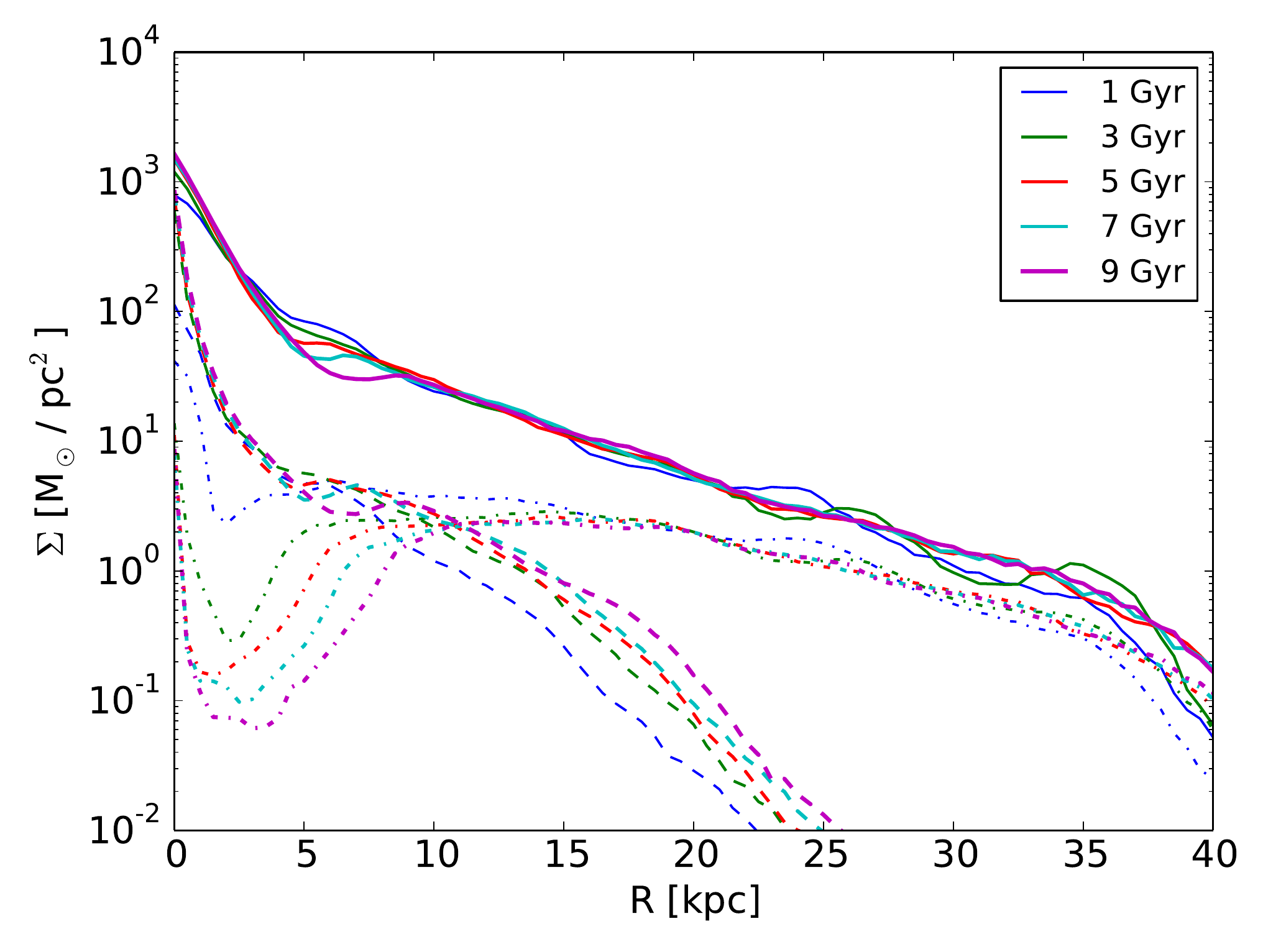} }
\caption{Time evolution of the azimuthally averaged surface density of the old stellar disc without the bulge (solid), the young stellar disc (dashed), and the gas disc (dot-dashed), as a function of the distance from the galaxy centre.}
\label{surfdens-fig}
\end{figure}

The simulated time-span we used is 9~Gyr, which is similar to the estimated time-span during which the Milky Way has been evolving with no major mergers \citep[e.g.]{haywood13, hammer07}. Figure~\ref{snap-fig} shows the surface density maps of different components of the disc (gas, all stars, and new stars formed during the simulation). A bar is formed during the first Gyr. In this simulation, the feedback efficiency is $10 \%$ and there is no molecular hydrogen cooling. With these parameters, the surface density of gas and stars is relatively smooth because this feedback efficiency and the absence of molecular hydrogen cooling prevent gas and stellar clumps from forming. We note that compared to many previous studies (for example \citet{minchev11, dimatteo13}), the stellar and gaseous discs are very extended (initial conditions were generated with an initial cut at $R=36~\mathrm{kpc}$ for both gas and stars). This choice also allows studying regions outside the OLR at all times, which in this model is located between 11 and 23 kpc from the centre (see next sections). The maximum extent of discs is difficult to determine observationally, be it at high or low redshifts, which means that considering an extended disc is not necessarily a singular case. The recent study by \citet{vandokkum14} showed, for example, that the disc of M101, previously strongly underestimated, extends to as much as 18 scale lengths in radius. Moreover, we here focused rather on the effect of radial migration with respect to the location of the main resonances than on the absolute sizes. Figure~\ref{surfdens-fig} shows the azimuthally averaged surface density of the simulated disc components as a function of the galactocentric radius. The old stellar disc dominates the surface density at all radii throughout the simulated time-span. Its surface density profile is very stable, with only mild departures from the initial conditions, which is due to the presence of the bar and spiral arms formed during the simulation.

\section{Resonances in the disc}

\begin{figure*}
\centering
\includegraphics[width=18cm]{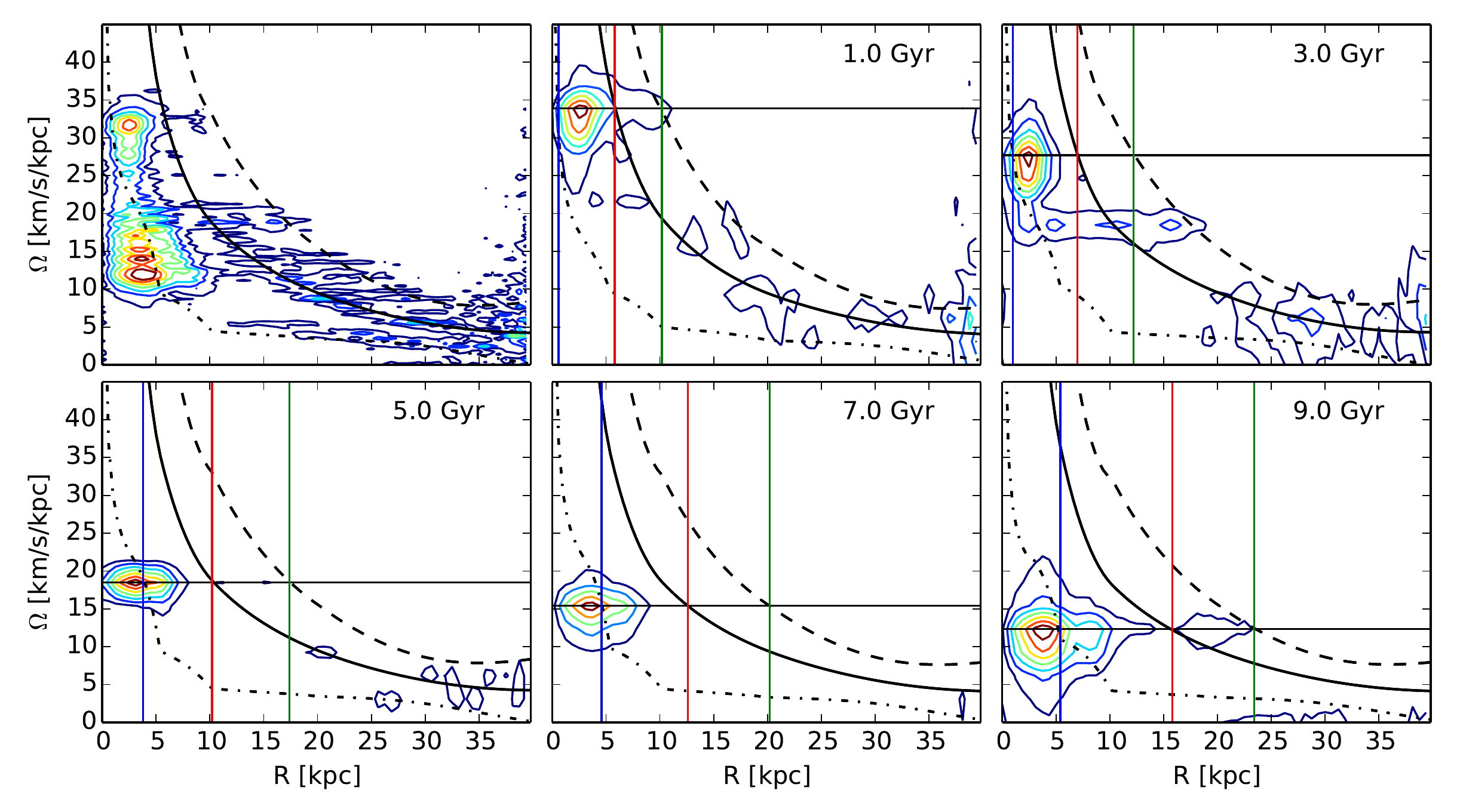}
\caption{Spectrograms of the $m=2$ Fourier mode. Top left plot: integration on 9~Gyr from 0.5~Gyr to 9.5~Gyr. Other plots: integrations on 1~Gyr centred on the times specified in the plots. The black curves are $\Omega(R)$ (solid), $\Omega(R) - \dfrac{\kappa(R)}{2}$ (dot-dashed), $\Omega(R) + \dfrac{\kappa(R)}{2}$ (dashed). The horizontal line represents the estimate of the bar pattern speed. The vertical lines represent the estimates of the bar ILR (blue), corotation (red), and OLR (green) radii.}
\label{fourres-fig}
\end{figure*}

The disc forms a central bar that persists during the whole simulated time-span. Transient spiral arms are also present. We determined the pattern speeds through a classic Fourier method:
\begin{itemize}
\item At times spaced by a constant interval of 10$~$Myr, we performed a spatial Fourier transform of the surface density $\Sigma(R,\theta)$ of the stars to obtain the dominant azimuthal modes in the different radial bins, 
\begin{equation}
S_m(R) \propto \int_{0}^{2 \pi} \Sigma(R, \theta) e^{i m \theta} \mathrm{d} \theta
.\end{equation}
\item We then performed a time Fourier analysis of the different modes in each radius bin to study their azimuthal speed,
\begin{equation}
T_m(R,\omega) \propto \int_{t_{\rm i}}^{t_{\rm f}}  S_m(R) e^{i \omega t} \mathrm{d} t
.\end{equation}
\end{itemize} 

\begin{figure}[!h]
\centering
\resizebox{\hsize}{!}{\includegraphics{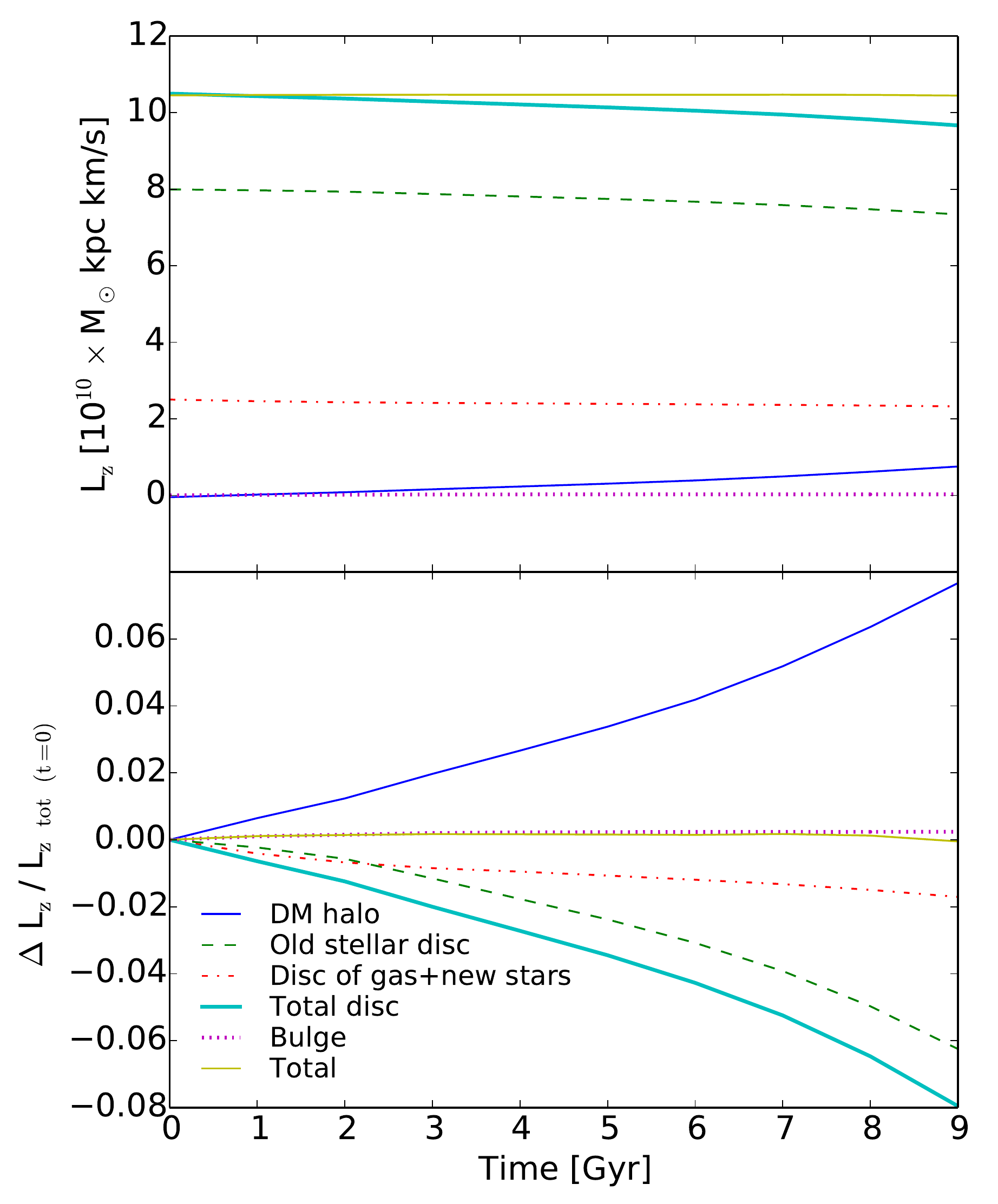} }
\caption{Top: Time evolution of the vertical component of the angular momentum $L_{\rm z}$ for different galaxy components. Bottom: Time variation of the fraction of the total angular momentum contained in each galaxy component $\dfrac{L_{\rm z}-L_{\rm z \; t=0}}{L_{ \rm z \; tot \; t=0}}$.}
\label{Lztrans-fig}
\end{figure}

In the spatial Fourier transforms, the $m=2$ mode usually dominates, which corresponds to $\pi$-periodic patterns: the central bar or two-armed spirals. The contours of the obtained power in the $R$-$\Omega$ plane for this mode are shown in Fig.~\ref{fourres-fig} for a time Fourier-integration performed on the total time intervals of 9~Gyr (first panel) and of 1~Gyr centred on 1, 3, 5, 7, and 9~Gyr. $\Omega=\dfrac{\omega}{m}$, with $m=2$ here, is the pattern speed. The integration on 9~Gyr shows that the contours at low radii, which correspond to the bar, fill a $\Omega$ range from $\simeq 10$ to $\simeq 30$~km/s/kpc. The time interval of 1~Gyr on which the time Fourier transform was then performed around specific times was chosen so as to optimise the determination of the pattern speeds: for a given time resolution, a too short integration time prevents correctly determining low pattern speeds and yields a poor resolution in frequency, while a too large integration time leads to a determination of overly averaged pattern speeds if they change significantly during this time. Figure~\ref{fourres-fig} shows that the bar slows down with time. The bar transfers angular momentum to the rest of the disc and to the stellar bulge and dark matter halo, as already discussed, among others, by \citet{debattista98, athanassoula02, martinez06, saha12, saha13, dimatteo14}. The angular momentum transfer between the different components of the galaxy is detailed in Fig.~\ref{Lztrans-fig}. At the beginning of the simulation, the angular momentum is contained in the disc because the DM halo and stellar bulge have no initial rotation. In 9~Gyr, the disc loses 8$\%$ of its angular momentum, which is transferred mainly to the DM halo, while a slight fraction is given to the bulge. 

The contours of Fig.~\ref{fourres-fig} show the main peaks of power in the $\Omega$-$R$ plane. The bar (power peak at low radii) is clearly the strongest $\pi$-periodic surface density perturbation with a rigid-body rotation in the integration periods of 1~Gyr, but spiral arms (peaks at larger radii) are visible as well. For example, the panel corresponding to $t=3$~Gyr indicates spiral patterns rotating at $\simeq 18~\mathrm{km/s/kpc}$ and $\simeq 7~\mathrm{km/s/kpc}$. These resonant modes with pattern speeds such that some of the resonance radii coincide might be sustained by non-linear mode coupling \citep[e.g.]{tagger87}. Some higher $m$-periodic features are also present ($m=3,4$...) with various pattern speeds, but they are weaker than the bar in the period from 1 to 9 Gyr.

\begin{figure}[h!]
\centering
\resizebox{\hsize}{!}{\includegraphics{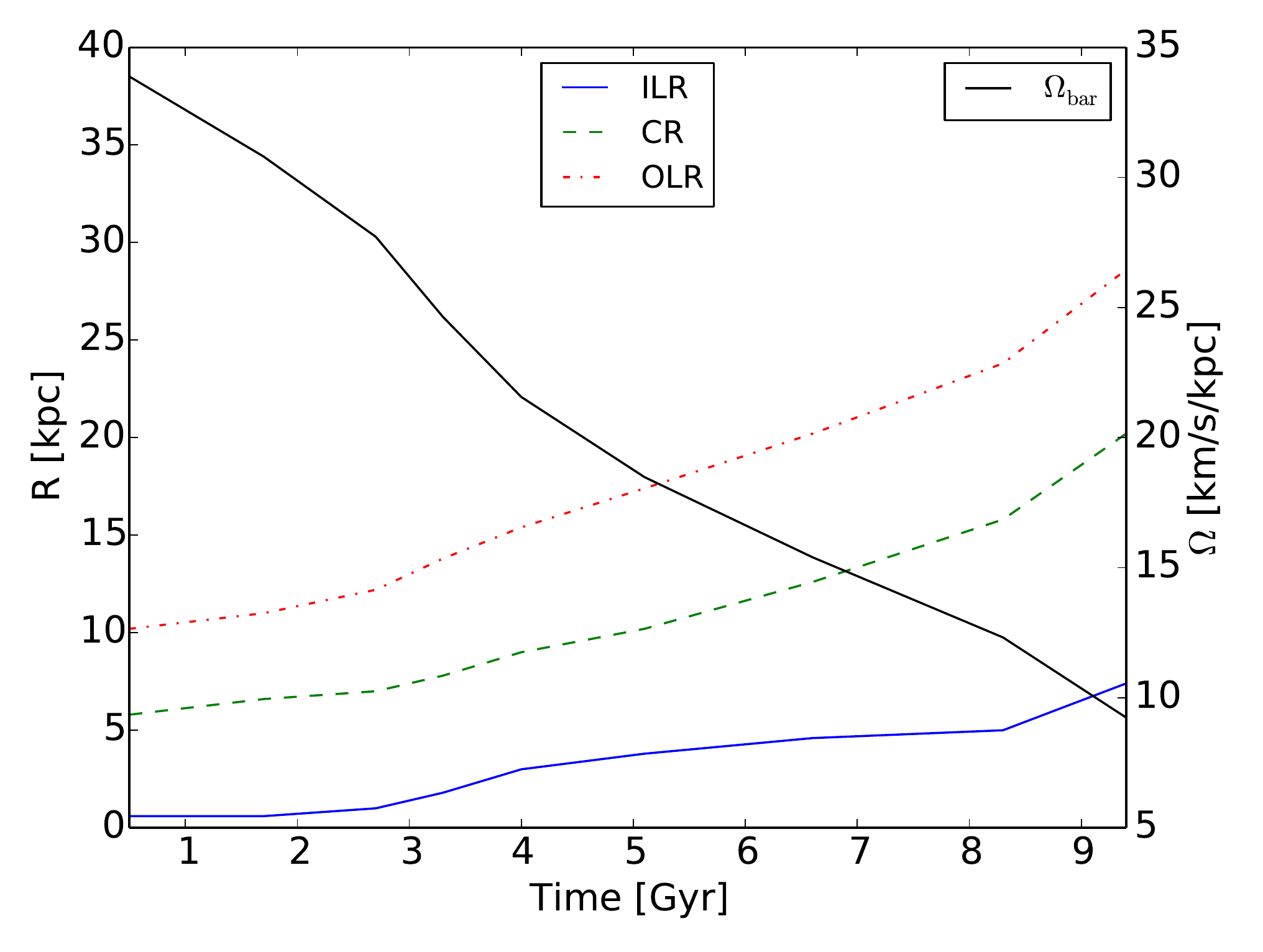} }
\caption{Time evolution of the ILR, CR, and OLR radii of the bar (left y-axis) and of the bar speed $\Omega$ (right y-axis).}
\label{radresev-fig}
\end{figure}

\begin{figure}[!h]
\centering
\resizebox{\hsize}{!}{\includegraphics{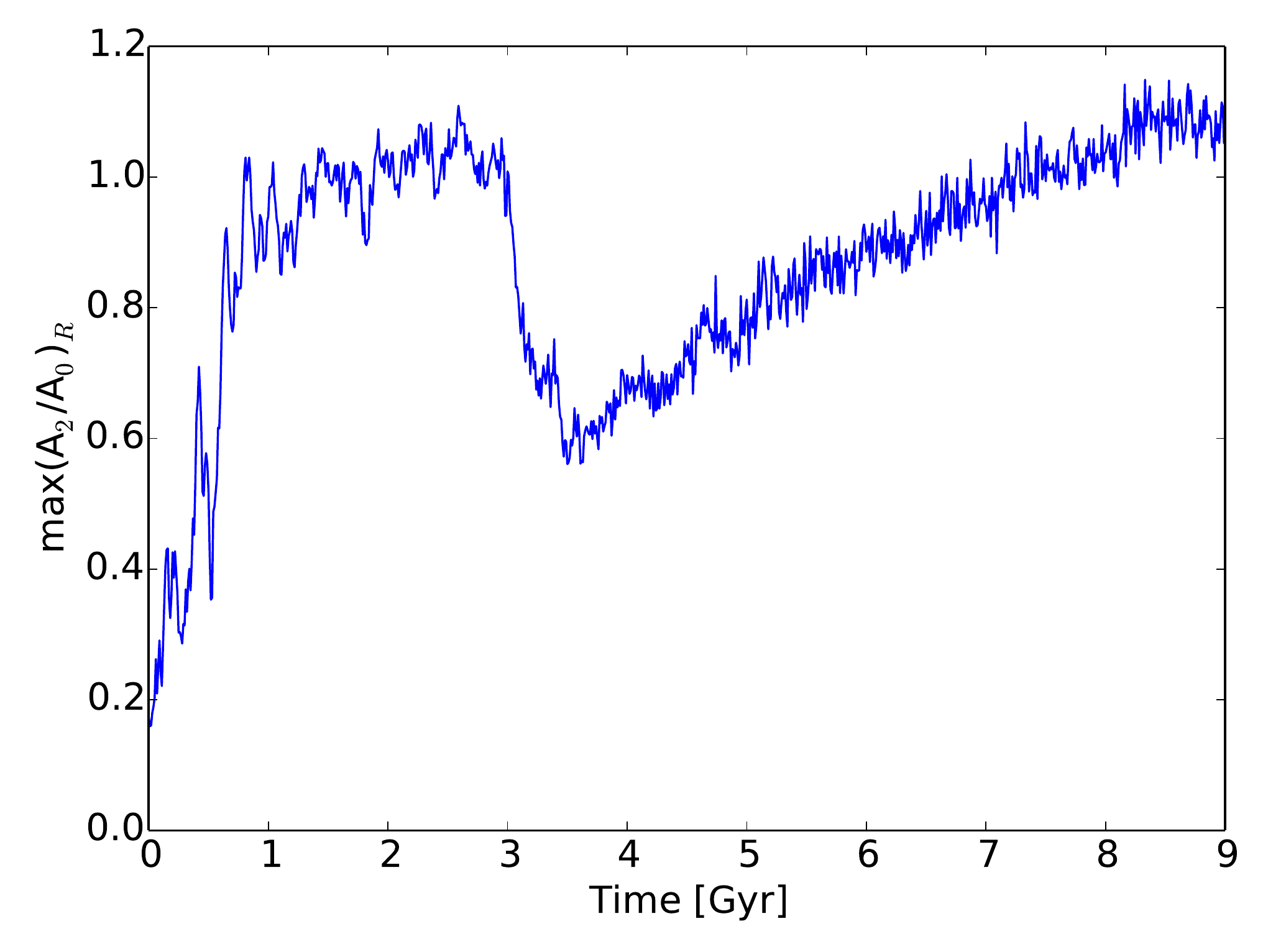} }
\caption{Time evolution of an estimate of the bar strength from the $m=2$ coefficient of the Fourier decomposition of the stellar surface density. The bar strengthens between 0 and 1~Gyr before weakening when a buckling instability occurs around 3~Gyr, and it then strengthens again.}
\label{barst-fig}
\end{figure}

By determining the angular speeds corresponding to the main contours, we obtained the radii at which stars on circular orbits are in resonance with the main patterns. We computed the angular speed $\Omega$ from the potential by $\Omega^2= \dfrac{1}{R}\dfrac{\mathrm{d} \phi}{\mathrm{d}R}_{z=0}$ and the epicyclic frequency $\kappa$ by $\kappa^2=\left(\dfrac{\mathrm{d^2} \phi}{\mathrm{d}R^2}+\dfrac{3}{R}\dfrac{\mathrm{d} \phi }{ \mathrm{d} R} \right)_{z=0}$. The corotation resonance (CR) occurs for $\Omega=\Omega_P$, the inner Lindblad resonance (ILR) for $\Omega-\Omega_P=\dfrac{\kappa}{2} $ , and the OLR for $\Omega-\Omega_P=-\dfrac{\kappa}{2}$. From a number of time Fourier transforms on 1~Gyr around times separated by 100~Myr, we estimated the values of the ILR, CR, and OLR radii of the bar as a function of time (see Fig.~\ref{radresev-fig}). The bar slow-down causes the bar corotation radius to increase and allows the bar to become more elongated. A measure of the bar strength using the same spatial Fourier decomposition of coefficients $A_m$ at each time is shown in Fig.~\ref{barst-fig}. The bar strength increases until $t=3~\mathrm{Gyr,}$ when it drops before increasing again as a result of the angular momentum exchange with the DM halo \citep{athanassoula02}. The drop in strength corresponds to the start of the buckling of the bar. This vertical instability, leading to a thick X-shaped bar, can be seen in the edge-on views of the new stars and the whole stellar component of Fig.~\ref{snap-fig}.

The various resonances are expected to locally affect the angular momentum of the stellar disc and thus to generate some radial migration. We study next the radial migration that occurs between 1 and 9~Gyr of evolution (after the initial building of the bar).

\section{Blurring and churning}

In this section, we quantify different types of radial stellar migration. We examine 
\begin{itemize}
\item migration in terms of galactocentric radius (hereafter simply radius), that is, the difference of radius between two different times. Blurring and churning can both cause a change in the galactocentric radius of a star, thus this definition captures both types of migrators and can be considered as the overall migration experienced by disc stars. The comparison between radii at two different times during disc evolution is widely used in the literature, see for example  \citet{roskar08,brunetti11, loebman11, bird12, dimatteo13, kubryk13, minchev12, minchev14}.
\item migration in terms of guiding radius, or churning, that
is, the difference of the guiding radius between two different times. This follows the nomenclature of \citet{schoenrich09}. Stars that increase the amplitude of their radial oscillations over time but do not change their guiding radius, that is, stars that only experience blurring,  are not considered as migrators according to this definition.

\end{itemize}
We recall that  our model is not intended to reproduce either the characteristics of the  Milky Way  disc -- as an example, the pattern speed of our simulated bar is lower than the pattern speed measured for the Galaxy (Gerhard 2011) and as a consequence, the main resonances are located at much larger distances from the centre than those measured for the Milky Way -- or of any other specific galaxy. As a result, the exact values of  spatial or kinematic variations experienced by stars in the model are
not directly applicable to any galaxy.  However, the effects described below can be considered as representative of those that any typical bar-dominated galaxy probably experiences. 

\subsection{Determination of guiding radii and amplitude of radial oscillations}

A stellar radius oscillates around a guiding radius. In the case of low eccentricities, it is possible to determine this guiding radius by finding the radius $R_{\rm c}$ at which a star has the same vertical component of the angular momentum and a circular trajectory: $L_{\rm z}=R_{\rm c} v_{\rm circ} (R_{\rm c}) $, where $v_{\rm circ}$ is the circular velocity obtained from the potential. The amplitude of the radial oscillations around the guiding radius is expected to scale as $\dfrac{\sigma_{\rm R}}{\kappa}$, where $\sigma_{\rm R}$ is the radial velocity dispersion, and $\sigma_{\rm R}$ and the epicyclic frequency $\kappa$ are both functions of the radius $R$. 

The output snapshots of our simulation are separated by $10~\mathrm{Myr}$, which is small enough to allow computing the guiding radius at any time $t$ by a simple method:
\begin{itemize}
\item determination of the relative minima and maxima of the oscillatory evolution of the radius.
\item use of a linear fit between the relative minima on the one hand and the relative maxima on the other to obtain a local minimum radius $R_{\rm min}(t)$ and a local maximum radius $R_{\rm max}(t)$ at time $t$.
\item definition of the guiding radius at $t$ by the average of these two radii: $\langle R \rangle (t)=\dfrac{R_{\rm min}(t)+R_{\rm max}(t)}{2}$.
\end{itemize}

\begin{figure}[!h]
\centering
\resizebox{\hsize}{!}{\includegraphics{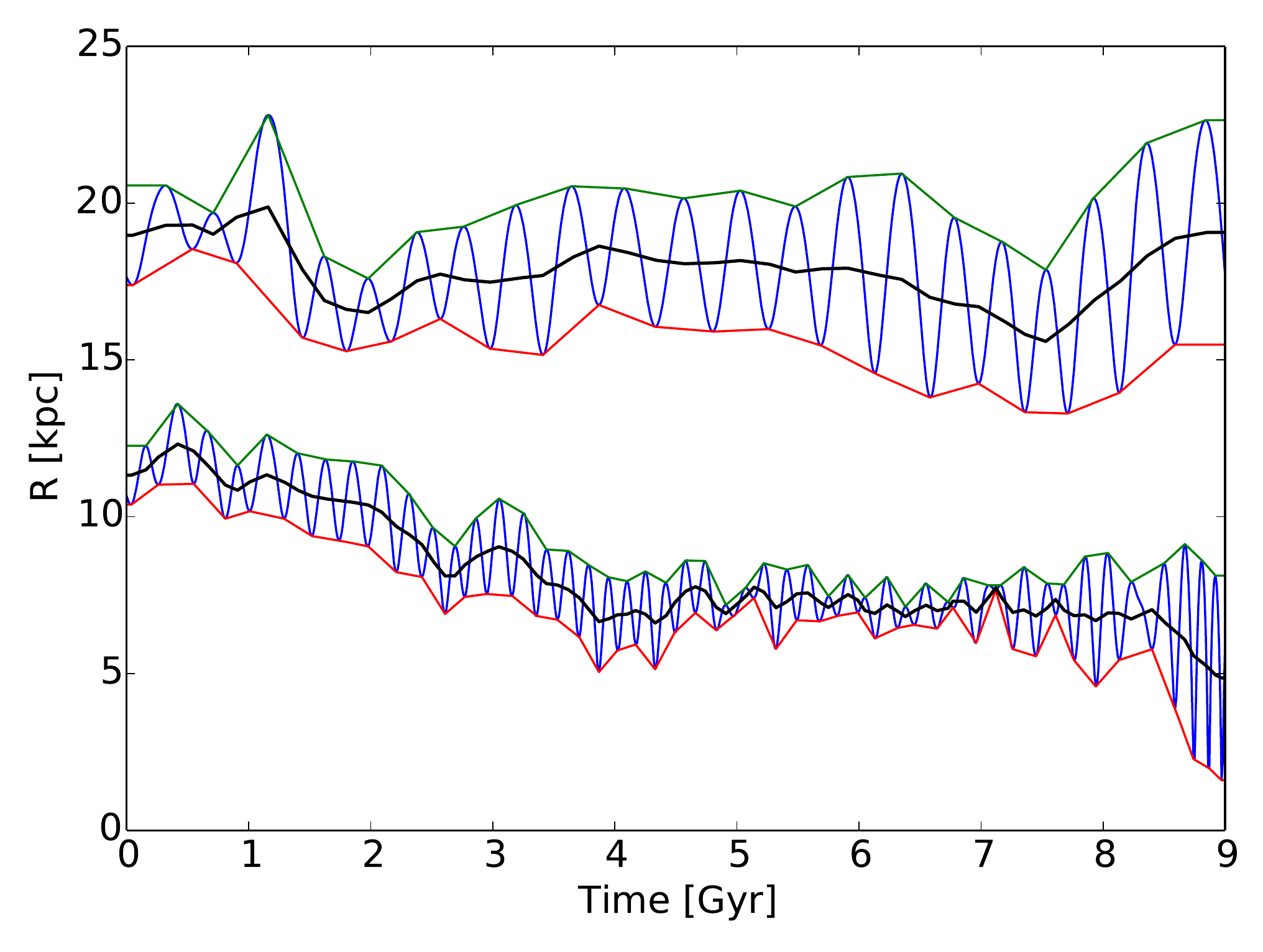} }
\caption{Determination of the guiding radius for two random stellar particles. Blue lines: galactocentric radii $R(t)$. Red lines: $R_{\rm min}(t)$. Green lines: $R_{\rm max}(t)$. Black lines: guiding radii $\langle R \rangle (t)=\dfrac{R_{\rm min}(t)+R_{\rm max}(t)}{2}$.}
\label{determRav-fig}
\end{figure}

Figure~\ref{determRav-fig} shows examples of this determination of the guiding radius for two random stellar particles. The positions of the particles are centred on the centre of mass of the whole galaxy at each time-step for this analysis. This method directly
provides the guiding radius and also the amplitude of the radial oscillations at any time $t$.

\subsection{Overall migration compared with churning alone}
\label{amp-sec}

\begin{figure*}
\centering
\includegraphics[width=15cm]{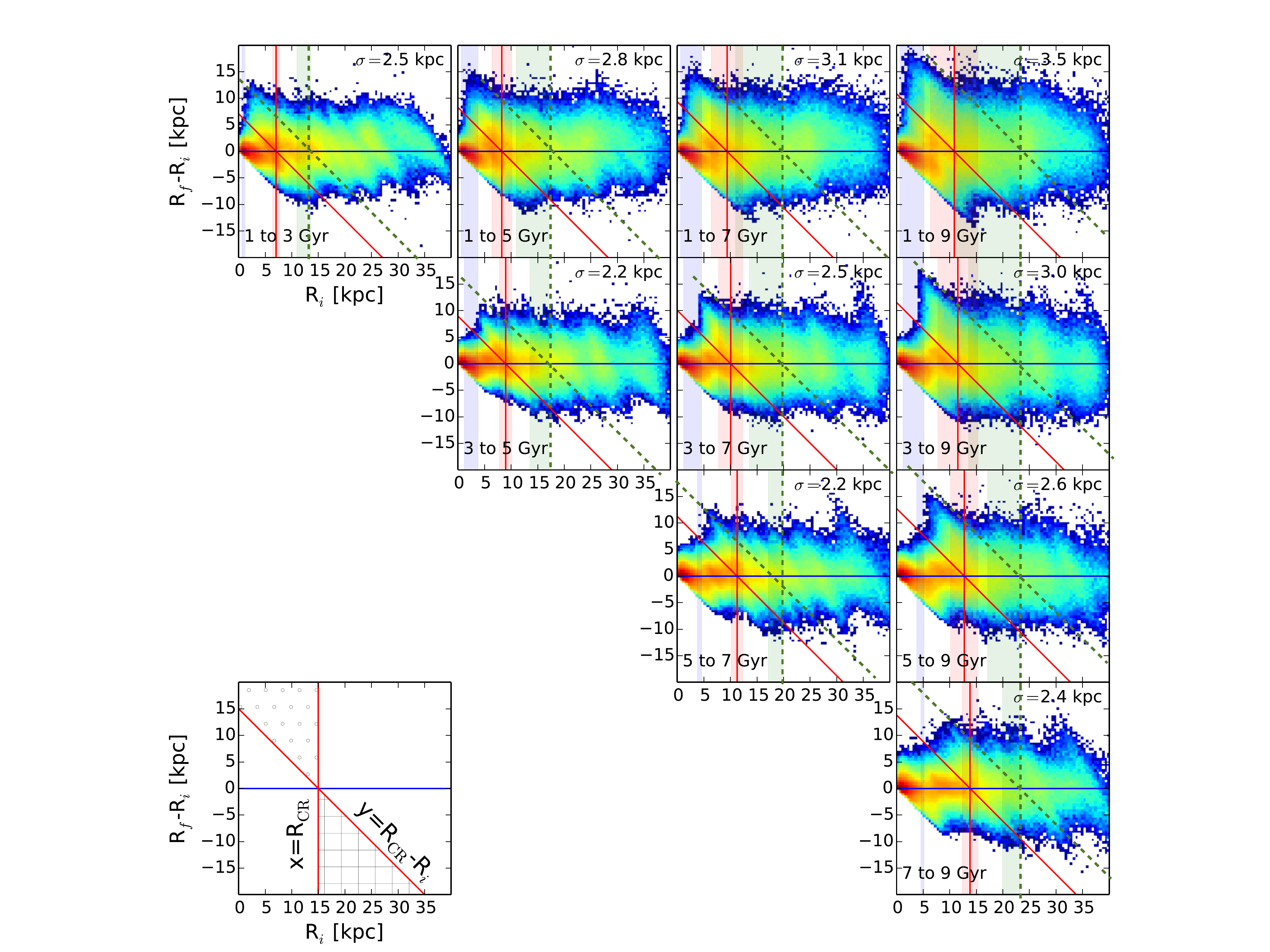}
\includegraphics[width=15cm]{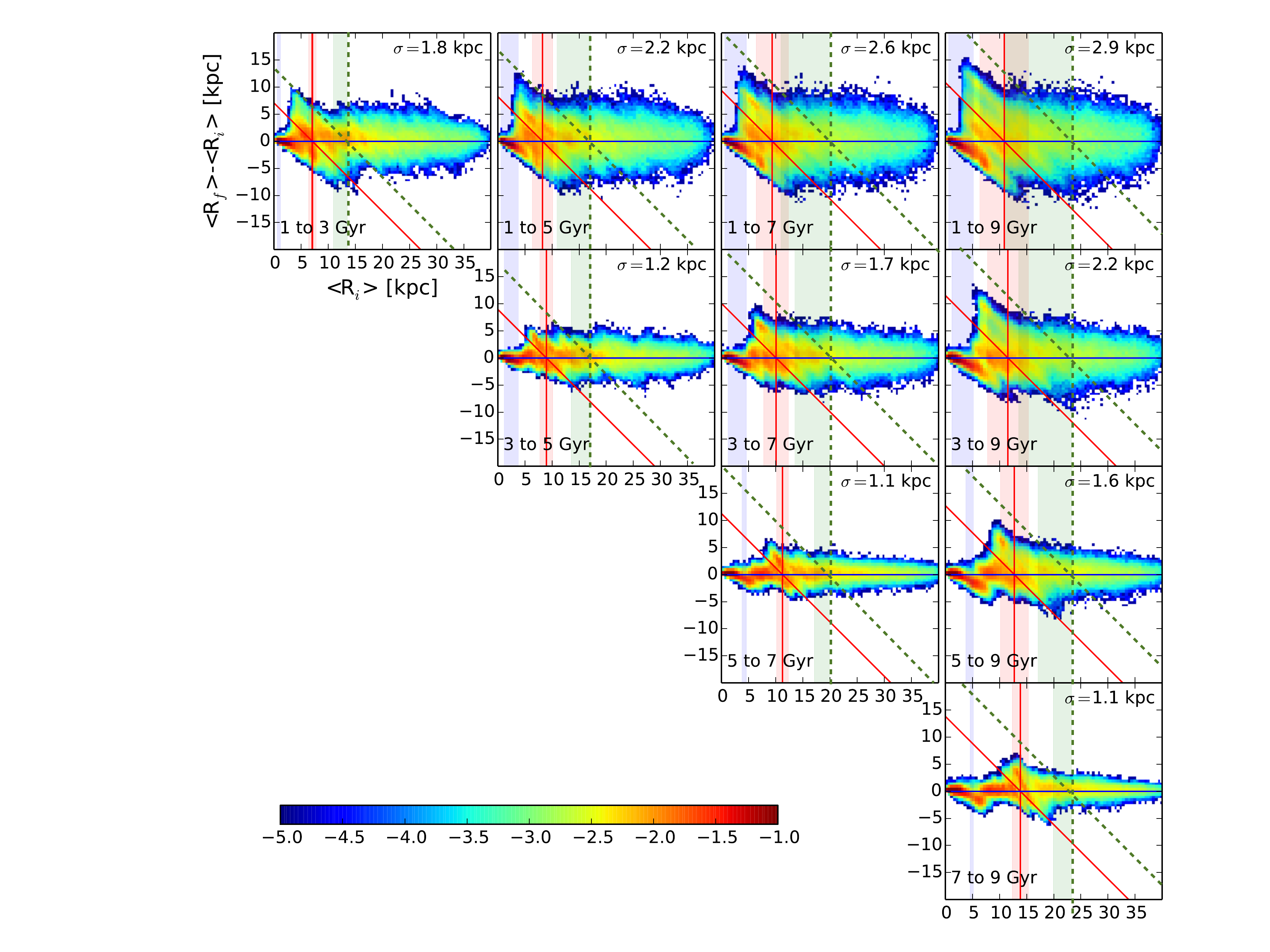}
\caption{Distributions of the variations of galactocentric radii (top half) and guiding radii (bottom half). The colour of each bin signifies the stellar mass in the bin and the colour scale shown at the bottom of the figure applies everywhere. The shaded areas represent the bar ILR (blue), CR (red), and OLR (green) radii variation in the time-span of each plot. The red vertical lines are the average of the bar CR radius during the time-span of a plot. The diagonal red lines help identify the migrators that cross the bar CR radius from lower radii (dot-shaded part of the schematic plot) or from larger radii (grid-shaded part). The diagonal green lines similarly help identify migration with respect to the OLR radius at the end of the time interval.}
\label{diffrrav-fig}
\end{figure*}

Our aim is to compare the overall population of migrating stars, by blurring and churning, with that of migrators by churning alone. For this,  we first examined the distributions of the variations of radius (/guiding radius) as a function of the initial radius (/guiding radius), for different time-intervals between 1 and 9~Gyr of evolution. The top set of plots of Fig.~\ref{diffrrav-fig} shows the difference of radii of stellar particles at final time $t_{\rm f}$ and initial time $t_{\rm i}$ as a function of the radius at $t_{\rm i}$, while the bottom set shows the difference of guiding radii as a function of the guiding radius at $t_{\rm i}$, where $t_{\rm i}$ and $t_{\rm f}$ are given a range of values. Thus, in the top set of plots the whole population of migrators is shown, whilst in the bottom plots only migrators by churning are selected. The considered stellar components are the old stellar disc and the new stars that are formed before $t_{\rm i}$. The RMS of the variations are indicated for each time interval. 
 
In each of the panels of Fig.~\ref{diffrrav-fig}, the shaded areas show the span of the values of the ILR (blue), CR (red) and OLR (green) radii of the bar determined from Fig.~\ref{fourres-fig} at $t_{\rm i}$ and $t_{\rm f}$. The vertical solid line is the average of the corotation radius and the diagonal line equation is $y=R_{\rm CR} - R_{\rm i}$ for the radii plots, $y=R_{\rm CR} - \langle R_{\rm i} \rangle$ for the guiding radii plots. The regions between the diagonal and vertical lines allow estimating the migrators that cross the corotation radius by blurring+churning or by churning alone:
\begin{itemize}
\item  In the region corresponding to the dot-filled region of the schematic diagram, the stars in the radii plots have $R_{\rm i} < R_{\rm CR}$ and $R_{\rm f} > R_{\rm CR}$, and those on the guiding radii plots have $ \langle R_{\rm i} \rangle < R_{\rm CR}$ and $ \langle R_{\rm f} \rangle > R_{\rm CR}$. 
\item  In the region corresponding to the grid-filled region of the schematic diagram, the stars in the radii plots have $R_{\rm i} > R_{\rm CR}$ and $R_{\rm f} < R_{\rm CR}$, and those on the guiding radii plots have $ \langle R_{\rm i} \rangle > R_{\rm CR}$ and $ \langle R_{\rm f} \rangle < R_{\rm CR}$.
\end{itemize}
Analogously, in each panel, the vertical dashed line shows the position of the OLR at $t=t_{\rm f}$, and the dashed diagonal line  equation is $y=R_{\rm OLR} - R_{\rm i}$ for the radii plots, $y=R_{\rm OLR} - \langle R_{\rm i} \rangle$ for the guiding radii plots. The regions between the diagonal and vertical dashed lines thus allow estimating the migrators that cross the OLR, at its final position in the time interval considered,  by blurring+churning or by churning alone.
We use these criteria in Sects.~\ref{flux-subsec} and \ref{barriers}.\\

The results presented in Fig.~\ref{diffrrav-fig} can be summarised as follows:
 \begin{enumerate}
\item As a general behaviour, the distribution for the whole population and for migrators by churning alone shows some features that appear as diagonal structures.
 These features are seeded by several resonances that are due to the presence of the bar and of the spiral arms. The most prominent diagonal features occur around the corotation of the bar, the main source of migrators in the stellar disc, as in  \citet{minchev10,minchev11,brunetti11}. 
 \item Quantifying radial migration by means of the instantaneous difference between radii at two different times leads to overestimating churning both in terms of maximum extent and RMS of the variations and in terms of fraction of migrators (see further discussion in Fig.~9). As an example, the RMS values for the 1 to 3~Gyr time-interval is 2.5~kpc for the variations of radius, while it is only 1.8~kpc for the variations of guiding radius (we note that the absolute values are specific to our model and may not be directly applicable to any specific galaxy, MW included). These differences are due to the radial excursions that can cause a stellar particle to have a radius between $\langle R \rangle (t)-A(t)$ and $\langle R \rangle (t)+A(t)$, where $A(t)=R_{\rm max}(t)-\langle R \rangle (t)$ is the semi-amplitude of the radial oscillations at time $t$. The radius $R(t_0)$ at time $t_0$ can be expressed as
\begin{equation}
R(t_0)=\langle R \rangle (t_0)+r(t_0)
,\end{equation}
where $r(t_0)$ is in the interval $[-A(t_0), A(t_0)]$. The variation in radius $R$ between time $t_{\rm i}$ and $t_{\rm f}$ is thus
\begin{equation}
R(t_{\rm f})-R(t_{\rm i})=\langle R \rangle (t_{\rm f})-\langle R \rangle (t_{\rm i})+r(t_{\rm f})-r(t_{\rm i})
,\end{equation}
and $r(t_{\rm f})-r(t_{\rm i})$ is in the interval $[-(A(t_{\rm f})+A(t_{\rm i})), A(t_{\rm f})+A(t_{\rm i})]$. 
\item 
The migration in terms of change of radius (=blurring+churning) keeps approximately the same distribution and RMS value on time-intervals of the same length, although the amplitude and RMS of the migration in terms of change of guiding radius (=churning alone) are smaller at late times. For example, for time-intervals of 2~Gyr, the RMS value of the radius variation only varies from $2.2$ to $2.5$~kpc, while the amplitude and RMS value of the guiding radius variation decreases from 1.8~kpc in a time interval of 2~Gyr from $t=1$ to 3~Gyr to 1.1~kpc between $t=5$ and 7~Gyr or $t=7$ and 9~Gyr. This difference is due to the increase, with time, of the amplitude of radial oscillations, that is, blurring, as quantified in Table~\ref{sigma_blurring}, where the RMS of the distribution\footnote{The RMS values were evaluated as the quadratic difference between the RMS value of the overall population of migrators and that migrated by churning alone.} is given for the same time intervals as shown in Fig.~\ref{diffrrav-fig}. In particular, from the comparison of  the values given in Table~\ref{sigma_blurring} with those of churning alone given in the bottom panels of Fig.~\ref{diffrrav-fig}, it can be noted that in the early phase of disc evolution, the spatial variations induced by churning overwhelm those due to blurring. This trend in reversed already after 3~Gyr of disc evolution, when the extent of the radial variations by blurring dominates the variations induced by churning alone. 
\item The disc experiences its most intense phase of migration by churning in the very early phase of its evolution, when the stellar bar forms from an axisymmetric potential and it is thin and strong. After the buckling instability and the formation of the boxy/peanut-shaped bulge at t$\sim$3Gyr, the amplitude of the variation of the guiding radii of stars migrated by churning alone diminishes. The increase in the RMS variation of the guiding radii as the time-span increases (cf., for example, the RMS variations of guiding radii in the time interval 1-3~Gyr versus 1-9~Gyr) does not necessarily imply that the number of migrators by churning increases with time, but only that migrators by churning have time to reach larger distances on longer timescales. We elaborate on this point below. 
\end{enumerate}

\begin{table}
\caption{RMS values of the variations of radius induced by blurring alone, at different times. See Fig.~\ref{diffrrav-fig} for a comparison with the variations obtained for the whole population of migrators and for those migrated by churning alone. Note that these values are obtained with an isolated galactic disc with specific size parameters (possibly more extended than the Milky Way disc) and are not applicable to any specific galaxy.}
\begin{flushleft}
\begin{tabular}{ccccc}
\hline
& $t_{\rm f}=3$Gyr & $t_{\rm f}=5$Gyr  & $t_{\rm f}=7$Gyr  & $t_{\rm f}=9$Gyr  \\
\hline
$t_{\rm i}=1$Gyr & 1.73~kpc & 1.73~kpc & 1.68~kpc & 1.95~kpc\\
$t_{\rm i}=3$Gyr  & & 1.84~kpc & 1.83~kpc & 2.04~kpc\\
$t_{\rm i}=5$Gyr  & & & 1.90~kpc & 2.05~kpc\\
$t_{\rm i}=7$Gyr   & & & & 2.13~kpc\\
    \hline
\end{tabular}
\end{flushleft}
\label{sigma_blurring}
\end{table}

\begin{figure}
\centering
\resizebox{\hsize}{!}{\includegraphics{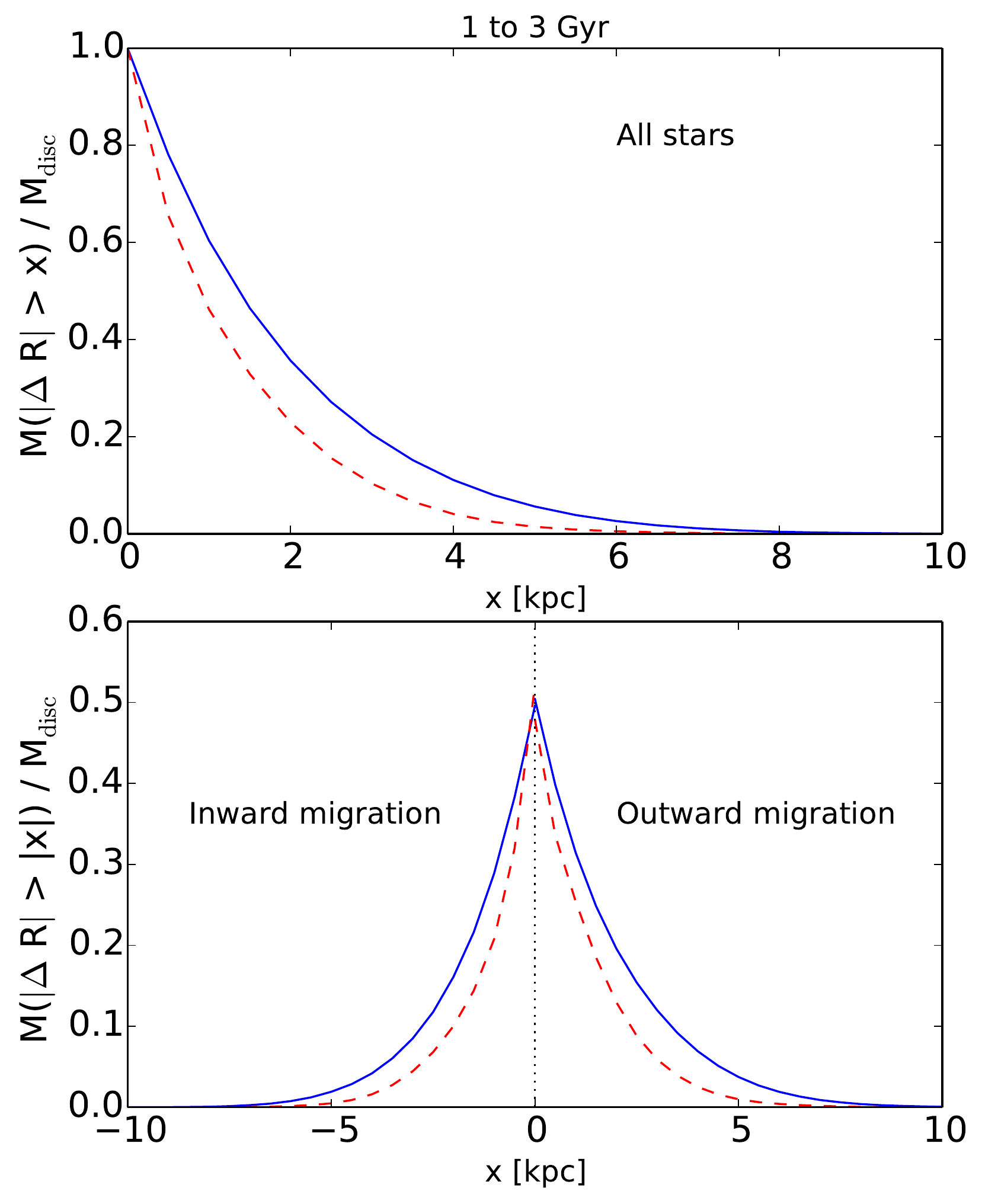} }
\resizebox{\hsize}{!}{\includegraphics{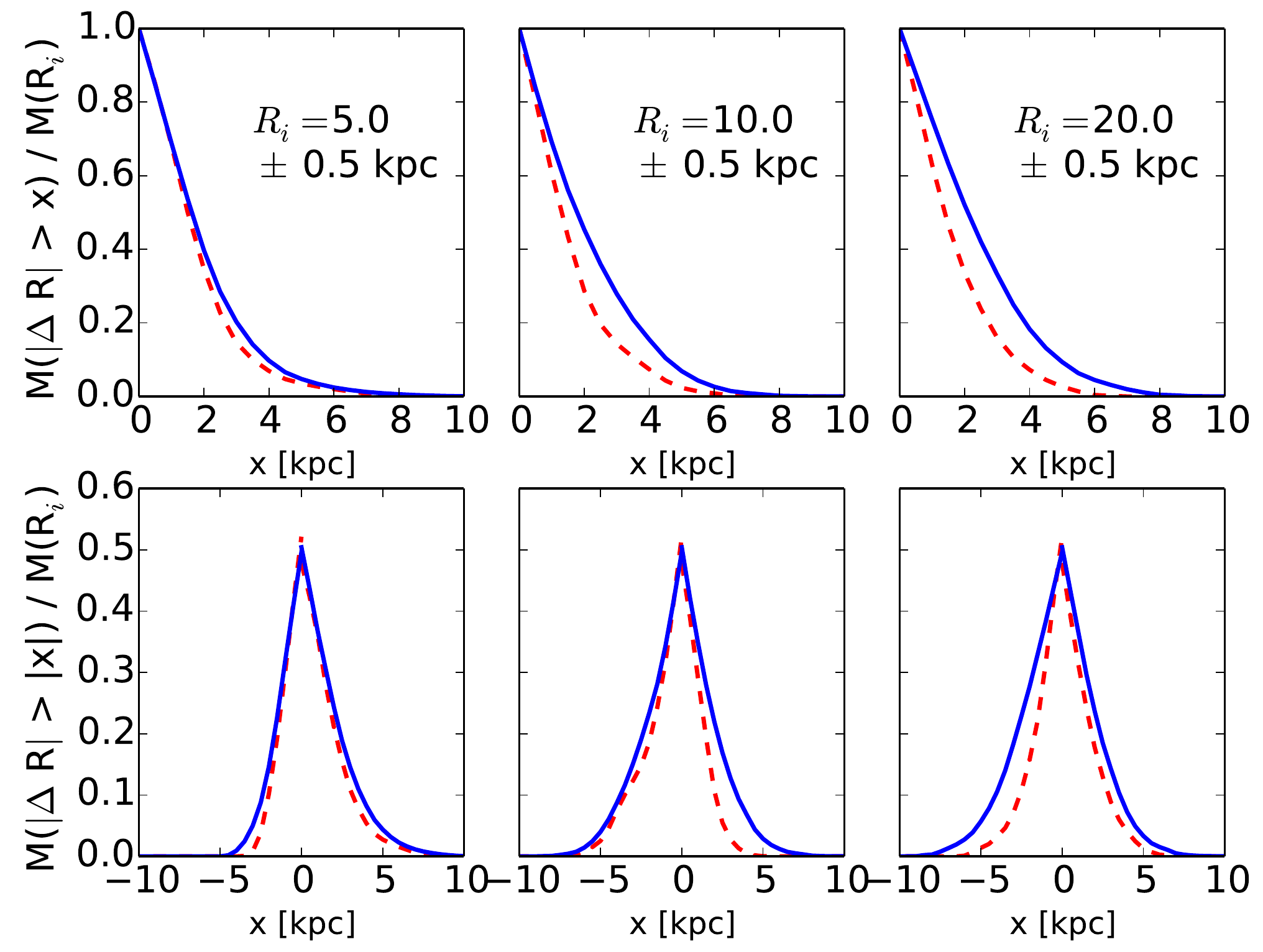} }
\caption{Migration between 1 and 3~Gyr. First line: Mass fraction of stars that migrate by more than $x$~kpc inwards or outwards in terms of radius (solid line) or guiding radius (dashed line). Second line: Mass fraction of stars that migrate outwards (right half of the plot) or inwards (left part) by more than $x$~kpc in terms of radius (solid) or guiding radius (dashed). Third and fourth lines: Same study, but only for stars with a radius (solid lines) or a guiding radius (dashed lines) around $R_{\rm i}$.}
\label{migrefrom3Ri13-fig}
\end{figure}

\begin{figure}
\centering
\resizebox{\hsize}{!}{\includegraphics{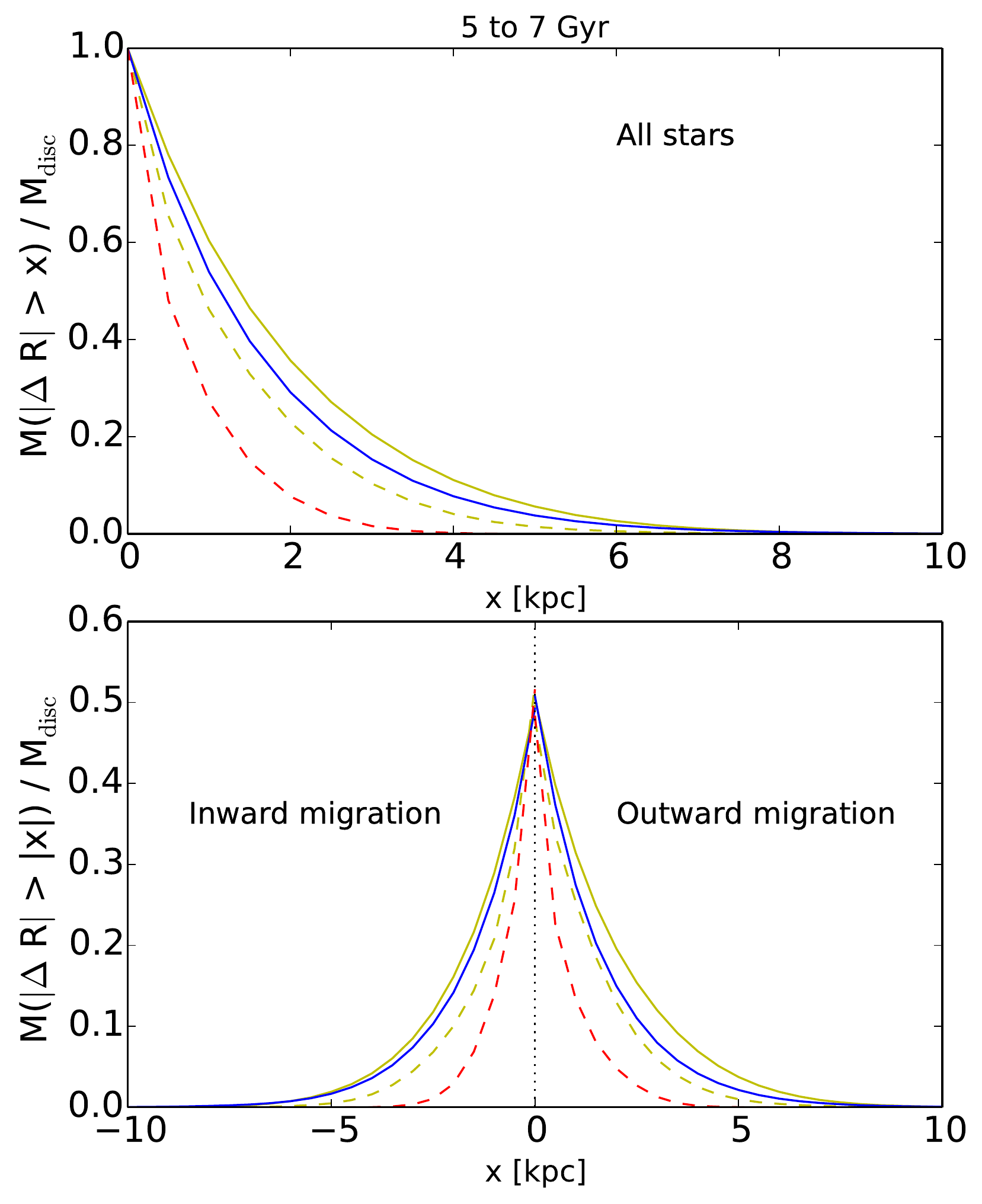} }
\resizebox{\hsize}{!}{\includegraphics{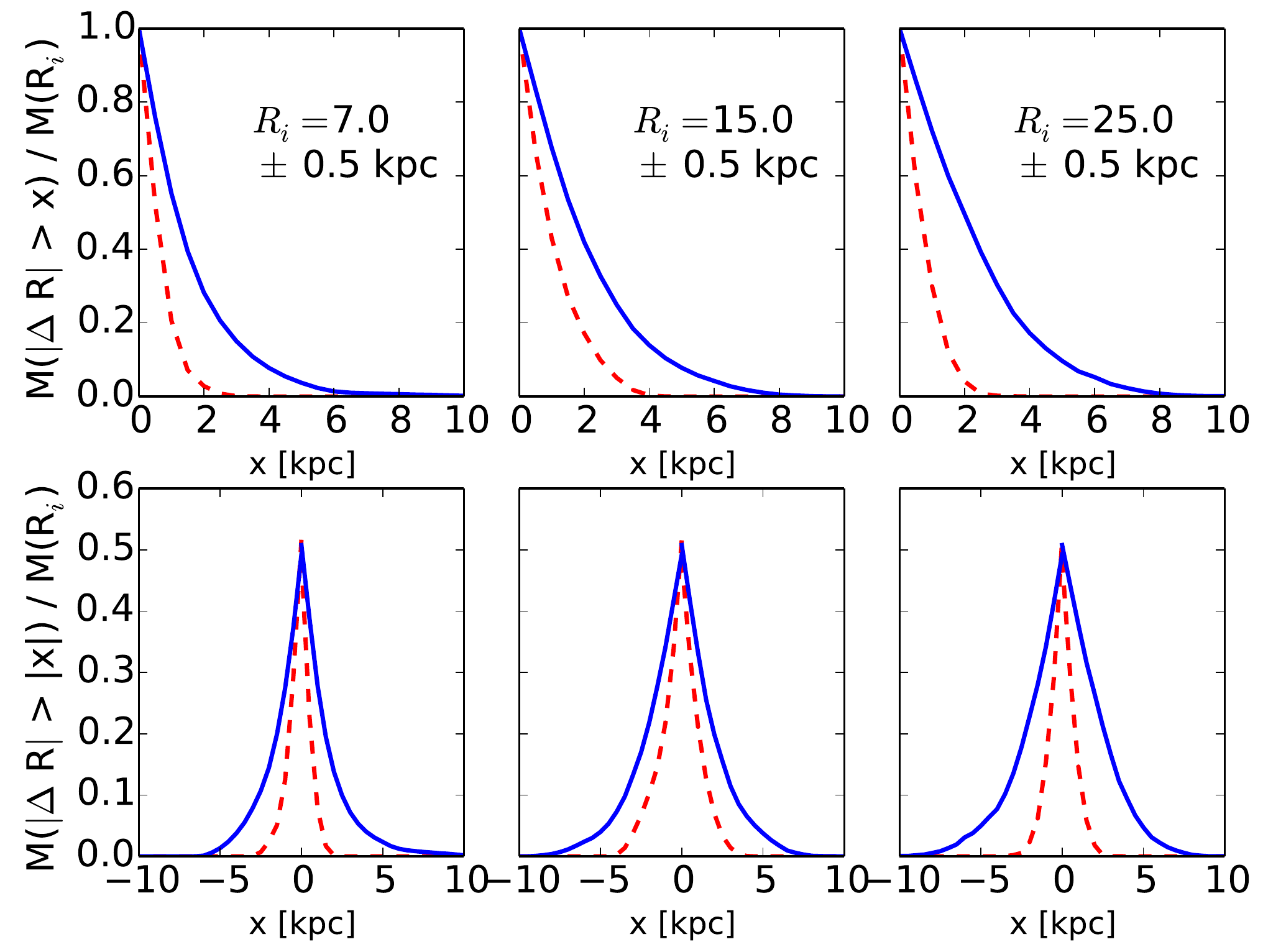} }
\caption{Migration between 5 and 7~Gyr. First line: Mass fraction of stars that migrate by more than $x$~kpc inwards or outwards in terms of radius (solid line) or guiding radius (dashed line). Second line: Mass fraction of stars that migrate outwards (right half of the plot) or inwards (left part) by more than $x$~kpc in terms of radius (solid) or guiding radius (dashed). The yellow lines in these two plots are the corresponding migration from 1 to 3~Gyr (identical to the curves of Fig.~\ref{migrefrom3Ri13-fig}). Third and fourth lines: Same study, but only for stars with a radius (solid lines) or a guiding radius (dashed lines) around $R_{\rm i}$.}
\label{migrefrom3Ri57-fig}
\end{figure}

\begin{figure}
\centering
\resizebox{\hsize}{!}{\includegraphics{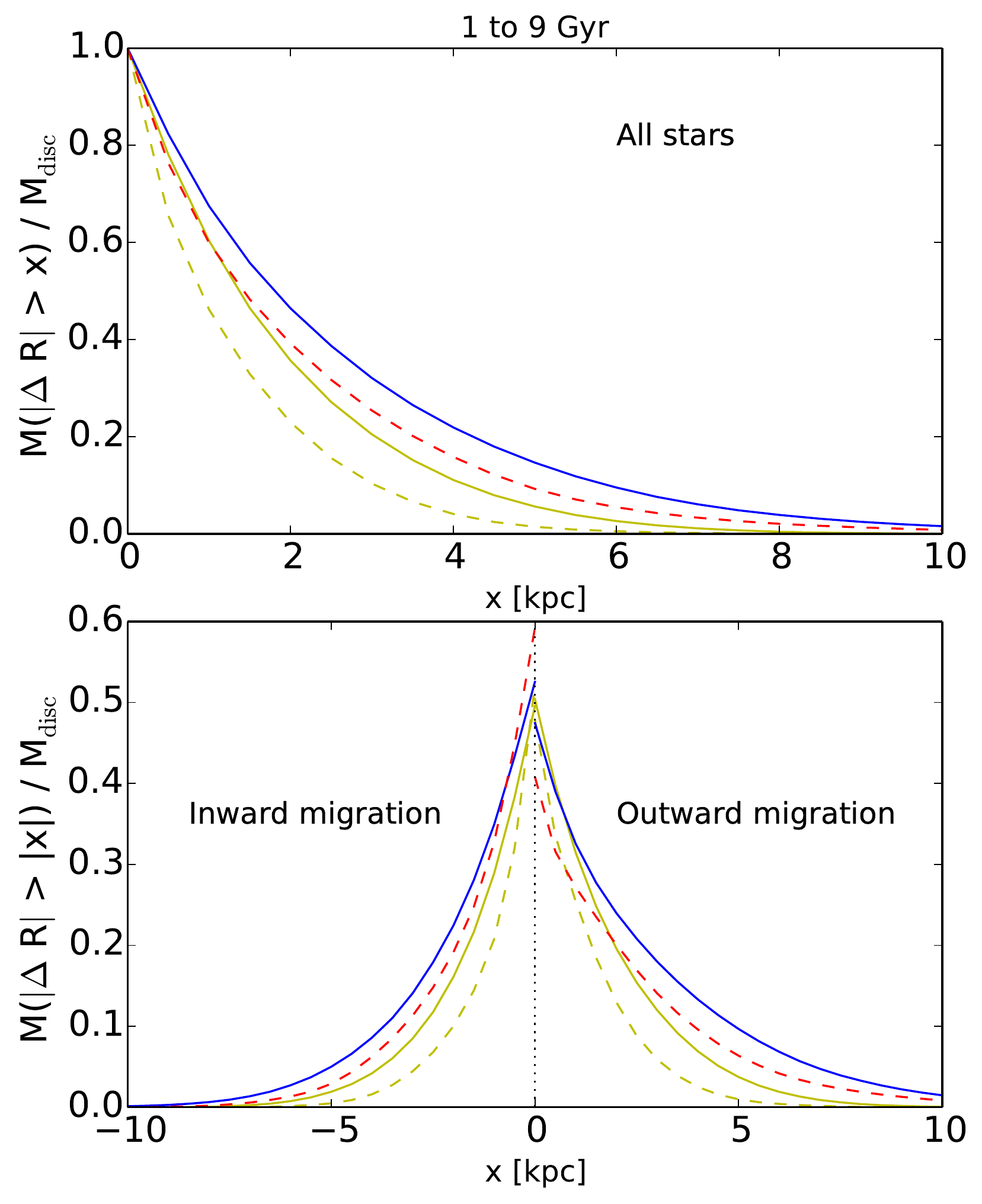}  } 
\resizebox{\hsize}{!}{\includegraphics{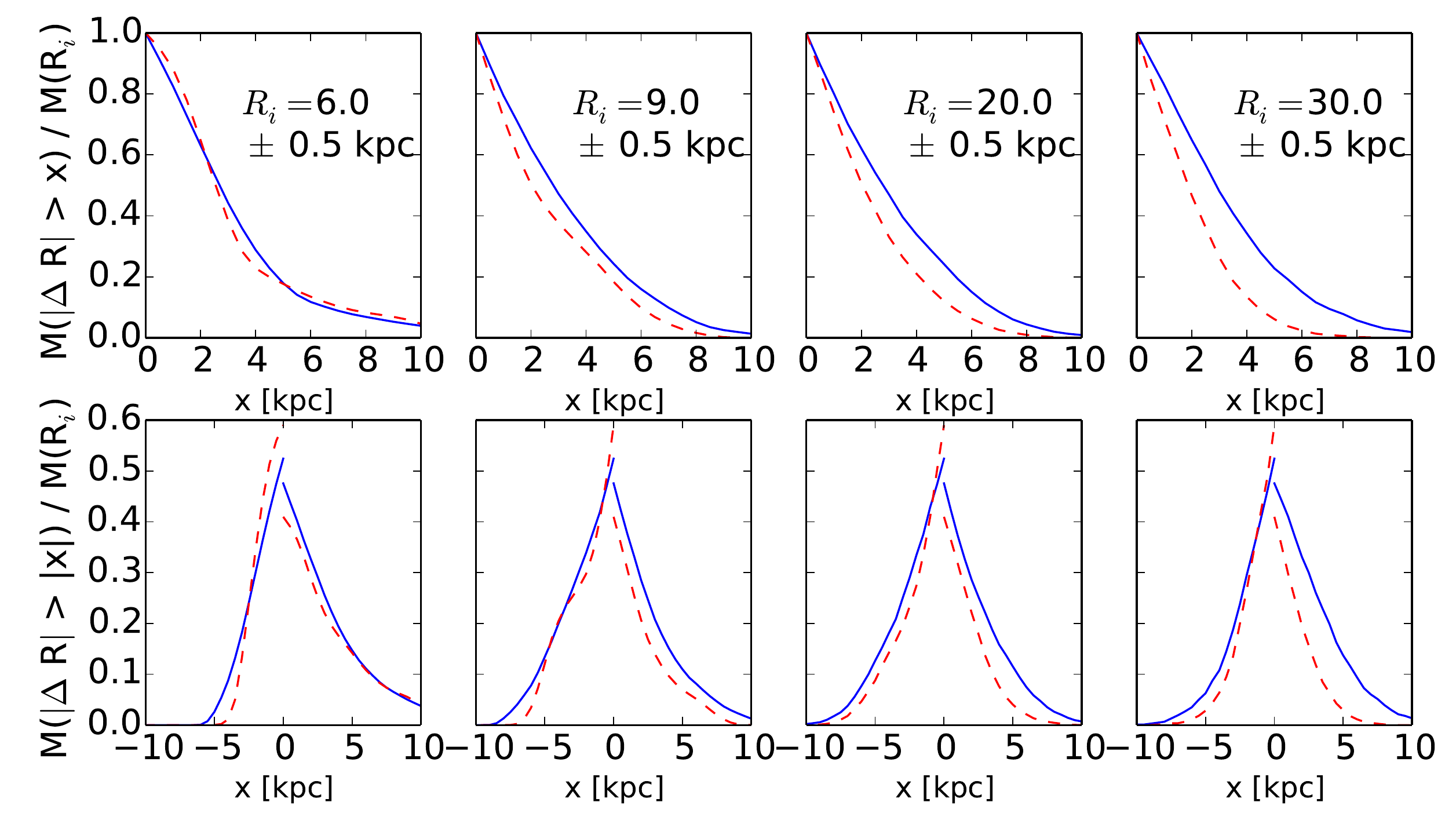} }

\caption{Migration between 1 and 9~Gyr. First line: Mass fraction of stars that migrate by more than $x$~kpc inwards or outwards in terms of radius (solid line) or guiding radius (dashed line). Second line: Mass fraction of stars that migrate outwards (right half of the plot) or inwards (left part) by more than $x$~kpc in terms of radius (solid) or guiding radius (dashed). The yellow lines in these two plots are the corresponding migration from 1 to 3~Gyr (identical to the curves of Fig.~\ref{migrefrom3Ri13-fig}). Third and fourth lines: Same study, but only for stars with a radius (solid lines) or a guiding radius (dashed lines) around $R_{\rm i}$.}
\label{migrefromRi19-fig}
\end{figure}

To emphasize the difference between migration simply defined as a change in the instantaneous radius and migration due to a change of guiding radius, the decrease of churning with time, and to quantify the fraction of stars involved in blurring and/or churning in this simulation, we represent in the top panels of Fig.~\ref{migrefrom3Ri13-fig}  the mass fraction of the stars that migrate by more than $x$~kpc in terms of radius or guiding radius from 1 to 3~Gyr. Figures~\ref{migrefrom3Ri57-fig} and~\ref{migrefromRi19-fig} then show the same fraction, but for stars migrating in the time interval between  5 and 7~Gyr and over the whole interval 1-9~Gyr, respectively. Figure~\ref{migrefrom3Ri13-fig} for example
shows that between 1 and 3~Gyr, $\simeq 40 \%$ of the ensemble of stars (with mass-weighting accounting for the different masses of stellar particles) change their instantaneous radius by more than 2~kpc, while this fraction reduces to only $\simeq 20 \%$  for a change of the guiding radius of the same amplitude. The plotted fraction decreases more rapidly with $x$ for the variation of guiding radius and it extends to a lower maximum value. The plots in the second row show the detail of the inward and outward migration. In the negative $x$ left part of the plot, the curves represent the fraction of stars migrating inwards by more than $|x|$~kpc, while in the right positive $x$ part, the curves represent the fraction of stars migrating outwards by more than $x$~kpc. The curves are almost symmetric with respect to the $x=0$ vertical axis. The migration has a lower amplitude in terms of guiding radius than in terms of radius for both inwards and outwards migration. The lower panels show the same study for stars in specific initial radius or guiding radius bins. More specifically, in these last two rows we quantify the stellar mass migrating of more than $\Delta R$ (showing both the absolute and the real value of the variation) from some initial radii $R_{\rm i}$, normalising this stellar mass to the mass initially contained at $R=R_{\rm i}$. In particular, we chose to analyse the fraction of migrating stars with respect to three different initial radii, around $R_{\rm i}=5$~kpc, $R_{\rm i}=10$~kpc and $R_{\rm i}=20$~kpc, corresponding to radii between the ILR and the CR, the CR and the OLR, and outside the OLR, respectively. From inspection of these plots, we can deduce the following:
\begin{enumerate}
\item Over the time period 1-3~Gyr, the whole population of migrators moves limitedly with respect to its initial radius, independently of the location of this initial radius with respect to the bar's resonances. As an example, the fraction of stars that has a variation of its radius in absolute values greater than $|\Delta R|=4$~kpc is  about $10\%$ at $R_{\rm i}=5$~kpc, and $20\%$ at $R_{\rm i}=10$~kpc and $R_{\rm i}=20$~kpc. These fractions decrease when considering stars that experience churning, the fraction of stars that change their guiding radii by more than 4~kpc is smaller than $10\%$ for all values of the initial radius inspected.  
\item If one considers the sign of the displacement (bottom panels in Fig.~\ref{migrefrom3Ri13-fig}) -- with positive $\Delta R$ corresponding to outward migration and negative $\Delta R$ to inward migration -- , one sees that migration is, in general, not symmetric. For radii between the ILR and the CR, there is an excess of stars migrating outwards, $\Delta R= -R_{\rm i}$ being the largest possible extent of inwards migration for migrators originating at $R=R_{\rm i}$. Between the CR and the OLR, there is an excess of inward migrators, both in terms of blurring+churning
and in terms of churning alone. We note in particular that outward migrators by churning do not cross the OLR at t=3~Gyr, the fraction of stars with $\Delta R > 4$~kpc -- this value is the smallest displacement required to cross the OLR during this time interval -- is indeed null. A limited number of stars that originated between the CR and the OLR can cross the OLR by blurring, but this number constitutes a very limited fraction of the stars born inside the OLR (fewer than $5\%$ for $R_{\rm i}=5$~kpc). We discuss this important finding in more detail in the next section. Finally, most of the stars that initially were located
at $R_{\rm i}=20$~kpc, thus outside the OLR between 1 and 3~Gyr, do not reach the OLR, with migration mostly spatially redistributing stars that are ab initio in the outer disc. Moreover,
there is a slight excess of inward migrators by blurring in this region.  
\end{enumerate}

The results found in Fig.~\ref{migrefrom3Ri13-fig} are confirmed when analysing migration over the same time duration (2~Gyr), but for a different time interval (5-7~Gyr).   
 Figure~\ref{migrefrom3Ri57-fig} shows the same study as Fig.~\ref{migrefrom3Ri13-fig}, for the time period from 5 to 7~Gyr, with the bottom panels detailing the migration for initial radii located again between the ILR and the CR ($R_{\rm i}=$7~kpc), between the CR and the OLR ($R_{\rm i}=$15~kpc), and beyond the OLR ($R_{\rm i}=$25 kpc). Because the bar slows down with respect to earlier times, the initial radii $R_{\rm i}$ are now more external than those chosen for the analysis shown in Fig.~\ref{migrefrom3Ri13-fig}. The striking difference between migration between 5 and 7~Gyr, compared to the time interval 1-3~Gyr  (Fig.~\ref{migrefrom3Ri13-fig}),  is the even smaller change in guiding radius experienced by the stars at these late times: there are no migrators by churning with $|\Delta R|>$ 4~kpc, most of the stars (80$\%$ for the region between the CR and the OLR,  almost 100$\%$ for stars originating inside the CR or outside the OLR) experience a change in their guiding radii smaller than 2~kpc. The strength of churning, in terms of spatial extent covered by the process,  thus decreases. We also confirm what we already noted in Figs.~\ref{diffrrav-fig} and \ref{migrefrom3Ri13-fig}, that is, migrators by churning originated in the inner disc, that is, inside the OLR, do not cross this resonance: the maximum displacement of outward migrators by churning whose $R_{\rm i}=15$~kpc at $t_{\rm i}=5$~Gyr is about 3.5~kpc, which is not enough to cross the OLR, whose location is at $R_{\rm OLR}=20$~kpc at time $t=7$~Gyr. 

It is not straightforward to identify the reason of the decrease of the importance of churning with time that is observed in our modelled galaxy. It may be due to a concomitance of factors that may be difficult to separate. This decrease with time can be explained by the evolution of the bar, and of its strength, the influence of other resonances in the disc, and the kinetic state of the disc. As an example, Fig.~\ref{barst-fig} shows that after an abrupt decline, the bar takes time to regain strength in the final several Gyr. It is then not as concentrated in the galactic plane (it has a peanut shape), and the tangential force it exerts on stars in the plane is thus weaker than at earlier times. In the period from 1~Gyr to about 3~Gyr, there is also a phenomenon of resonance overlap that can increase the radial migration (see \citet{minchev10}. The disc also gradually becomes hotter (as we show in Sect.~\ref{heatcool-sec}, see especially Figs.~\ref{velzdisp-fig} and~\ref{velrdisp-fig}), and stars thus become less responsive to potential perturbations. 

Finally, when analysing migration over the whole time interval 1-9~Gyr, as done in Fig.~\ref{migrefromRi19-fig}, one can see that the maximum displacement of migrators has increased. This
is because stars have had time to cross a larger portion of the disc. For example, 5\% of migrators by churning migrated by more than $|\Delta R|=6$~kpc, whilst this fraction was null between 1 and 3~Gyr. But when examining the details of migration, it is still valid that the amount of the displacement depends both on its sign (inward versus outward migrators) and on the location of the initial radius $R_{\rm i}$.  We note in particular that the largest displacement of outwards migrators occurs in the region between the ILR and the CR ($R_{\rm i}=6$~kpc and $R_{\rm i}=9$~kpc, respectively), but that even over such a long time interval, these outwards migrators do not cross the final OLR radius. Furthermore, blurring significantly contributes to shaping the outer disc -- compare for example the fraction of outwards migrators in the outer disc ($R_{\rm i}=30$~kpc) by churning+blurring versus churning alone.

\subsection{Local flux of migrators}

\label{flux-subsec}

\begin{figure*}[!htb]
\centering
\includegraphics[width=15cm]{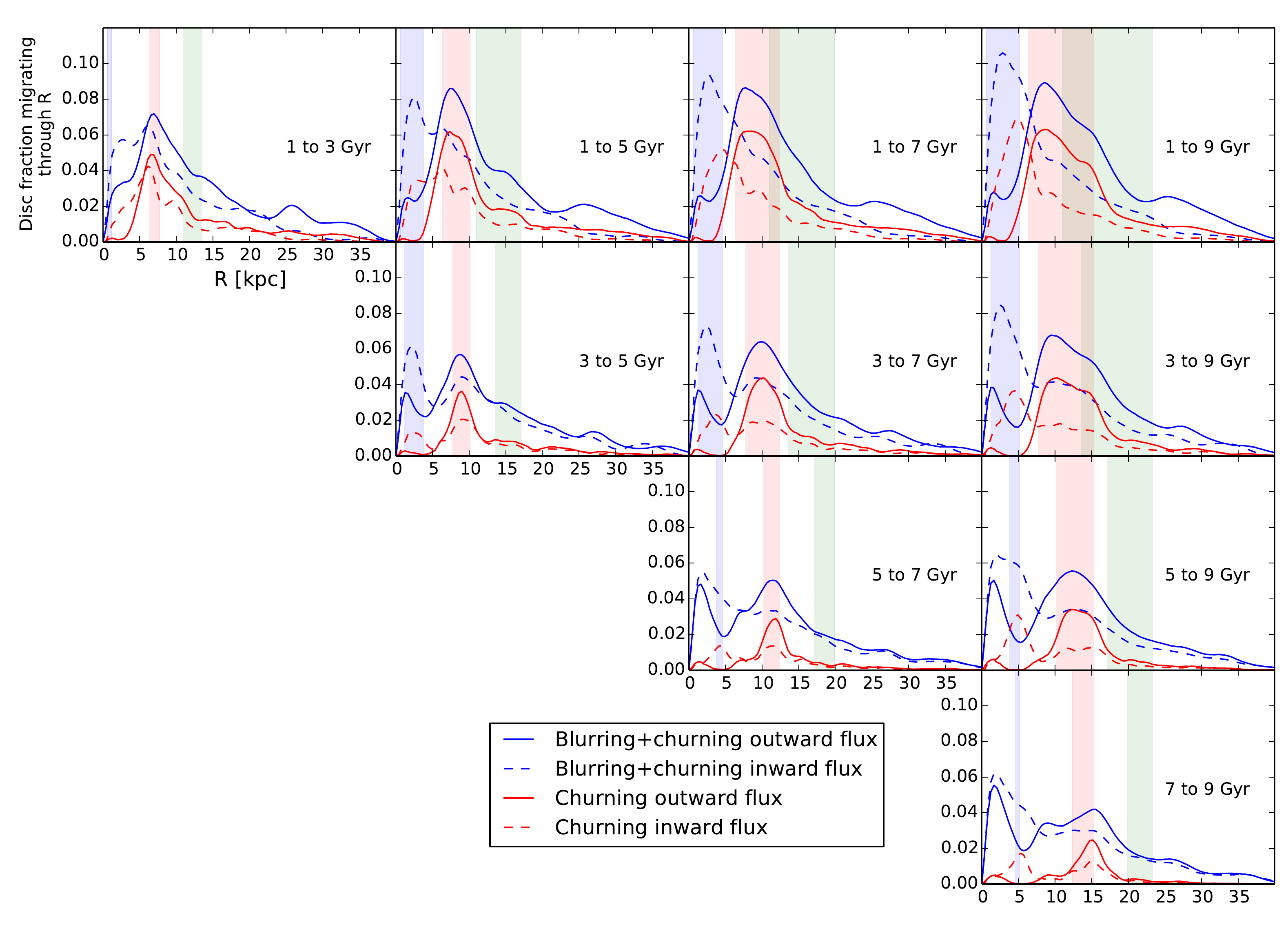} 
\includegraphics[width=15cm]{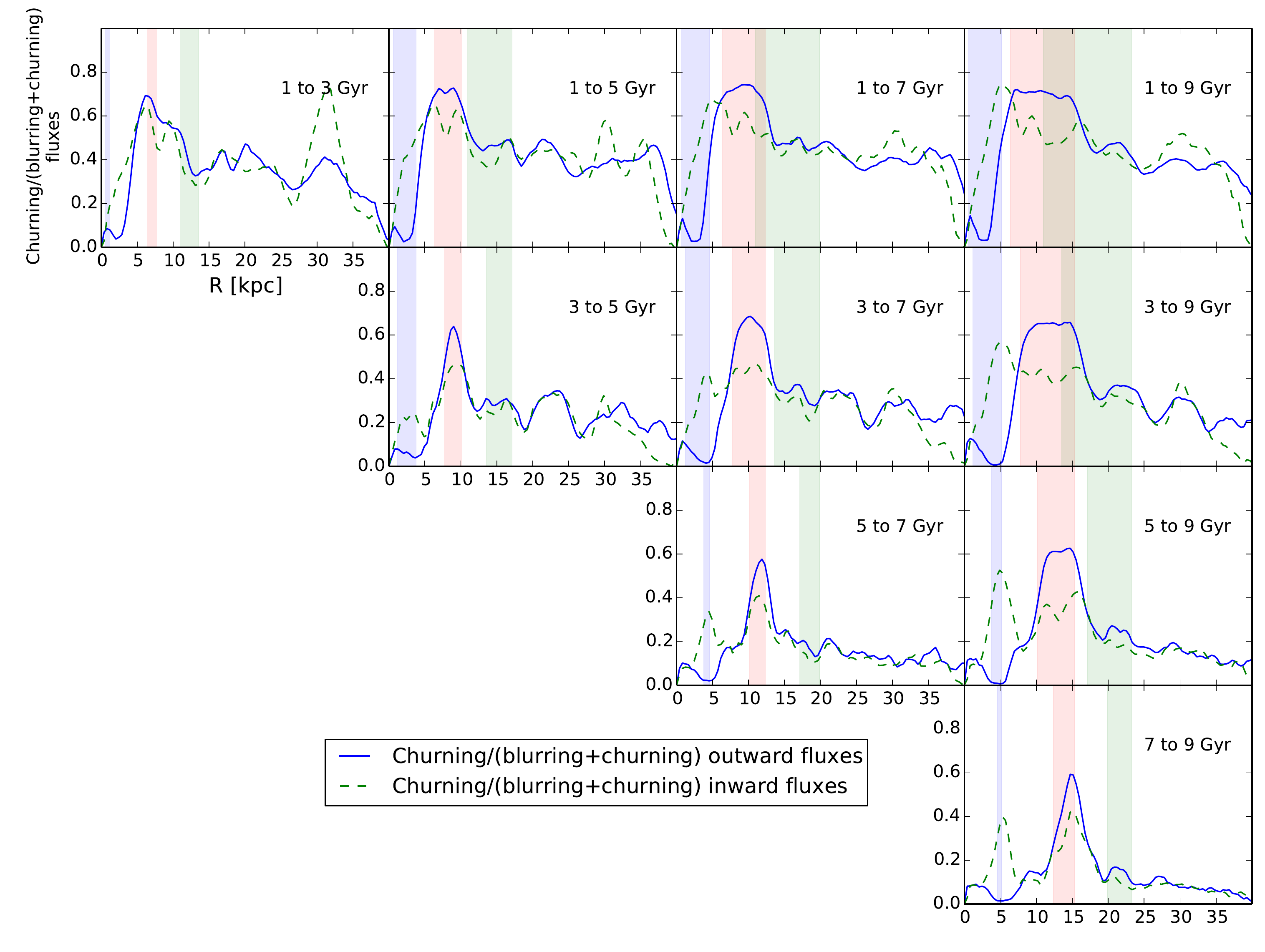}
\caption{Top half: Mass fraction of the stellar disc crossing a radius $R$ in terms of radius (apparent migration) or both radius and guiding radius (true migration). Bottom half: Ratios of the true to apparent migrators. The shaded areas represent the bar ILR (blue), CR (red), and OLR (green) radii variation in the time-span of each plot.}
\label{churnedfrac-fig}
\end{figure*}

We now attempt to determine which migration effect dominates at some radius (migration only due to epicyclic excursions away from a guiding radius or a change of guiding radius), before computing the fraction of different types of migrators at a final radius.
 
A quantification of a `migration flux' as a function of the radius can be obtained by computing the mass fraction of the disc that crosses a given radius. We represent in the top plots of Fig.~\ref{churnedfrac-fig} as a function of the galactocentric radius $R$
\begin{itemize}
\item the whole migration flux, that is, the mass fraction of the stellar disc that crosses $R$, the crossing of $R$ being quantified in terms of radius, that is: 
$R_{\rm i} < R$, $R_{\rm f}>R$ from time $t_{\rm i}$ to time $t_{\rm f}$ for outward migration (solid lines) and $R_{\rm i} > R$, $R_{\rm f}<R$ for inward migration (dashed lines)  
\item the migration flux by churning alone , that is, the mass fraction of the stellar disc that crosses $R$ in terms of radius and also of guiding radius: $R_{\rm i} < R$, $R_{\rm f}>R$ and $\langle R_{\rm i} \rangle < R,  \langle R_{\rm f} \rangle > R$ for outward migration (solid lines) and $R_{\rm i} > R$, $R_{\rm f}<R$ and $\langle R_{\rm i} \rangle > R,  \langle R_{\rm f} \rangle < R$ for inward migration (dashed lines).  
\end{itemize}
Both fluxes in the top panels of Fig.~\ref{churnedfrac-fig} are normalised to the total stellar mass in the disc. 

The whole outward migration flux at the corotation radius is thus the portion of stars that belongs to the dot-filled area of the schematic diagram of Fig.~\ref{diffrrav-fig} in the $R_{\rm f} - R_{\rm i}$ vs $R_{\rm i}$ plots, and the whole outward migration flux at a radius $R$ different from the corotation radius is simply obtained by shifting the threshold radius and the diagonal line accordingly. Similarly, the whole inward migration corresponds to the grid-filled area of the schematic diagram. The migration flux by churning alone is obtained by selecting stars that are both in the shaded zones of the radii plots and of the guiding radii plots (top and bottom panels in Fig.~\ref{diffrrav-fig}, respectively). The migration flux by churning alone is thus always inferior to the whole migration flux because it is a fraction of it. 
The inner kpc inward migrators (both in terms of change in their instantaneous radius or in their guiding radius) are stars captured by the bar. Some other much smaller peaks are observed at large radii, corresponding to spiral patterns. However, in agreement with previous results (see Introduction), the most  significant part of the migration flux occurs near the corotation of the bar because the bar is the strongest potential perturbation in our simulation and its corotation radius is located in a disc
region with a high surface density. In discussing these plots, we wish to emphasize two points:
\begin{itemize}
\item The finding that the corotation is the locus of the strongest flux of migrators by churning can be particularly appreciated considering the time intervals of different lengths, for example comparing the time interval 1-3~Gyr with those at 1-5~Gyr, 1-7~Gyr, and 1-9~Gyr. The corotation spans an increasingly larger radial extent as the duration of the time interval grows. As a result,  the  flux through the corotation region, which is mostly a thin spike around the CR for short time intervals (cf. 1-3~Gyr), transforms into a sort of  large plateau for longer time intervals, whose extent corresponds to the spatial extent spanned by the corotation during the corresponding time.  
\item The dominant role of churning near corotation can also be appreciated in the bottom panels of Fig.~\ref{churnedfrac-fig}, where we show the fraction of stars migrated by churning with respect to the whole sample of migrators crossing a given radius $R$ . While more than 60\% of migrators crossing the corotation region are migrating by churning, this fraction significantly
decreases for other regions of the disc. In particular, it can be appreciated that the fraction of migrators by churning decreases for later times outside corotation. This confirms the growing importance of radial heating (i.e. blurring) as time increases (see also discussion in the previous section).
\end{itemize}

\begin{figure*}
\centering
\includegraphics[width=15cm]{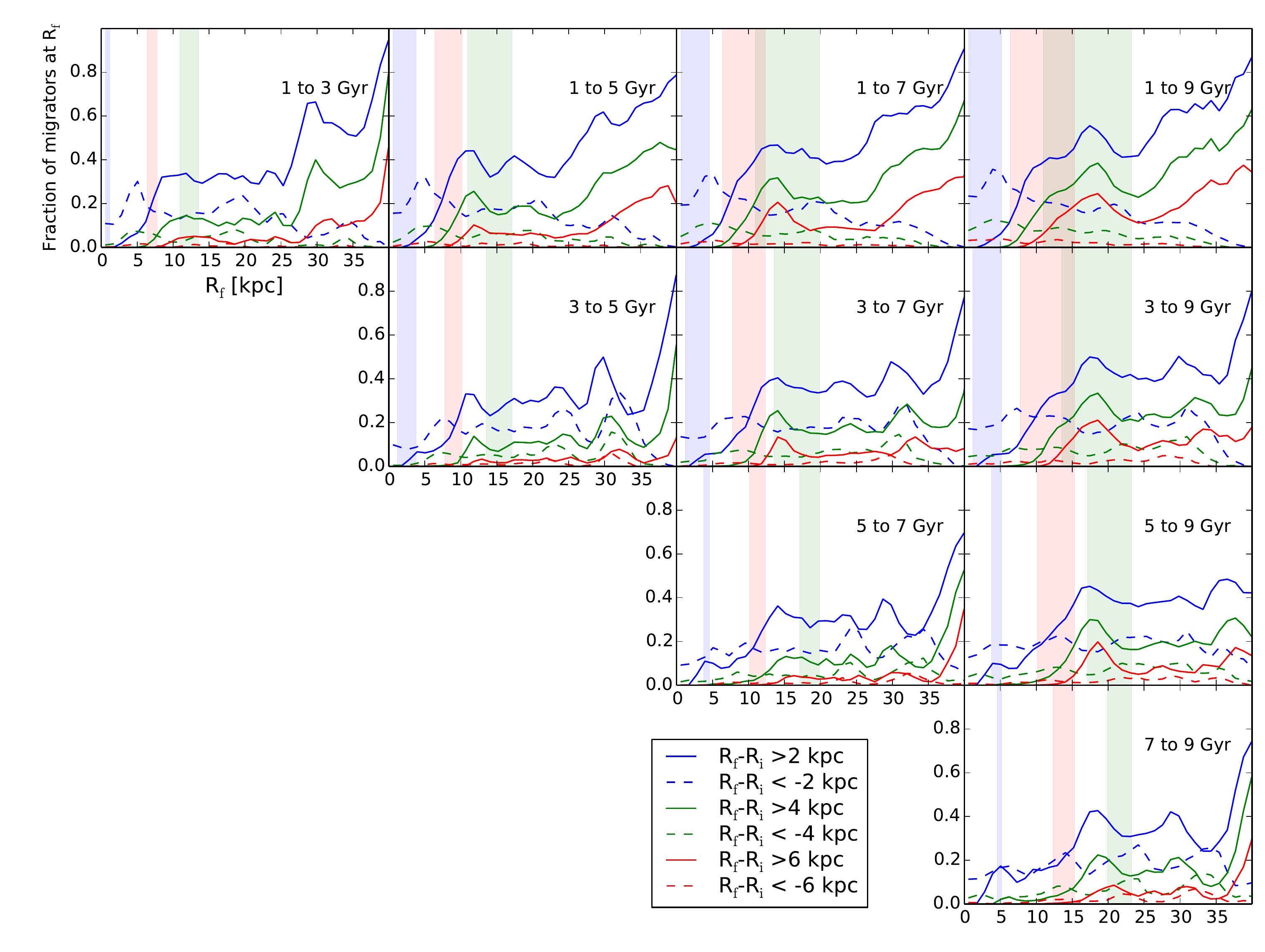}

\includegraphics[width=15cm]{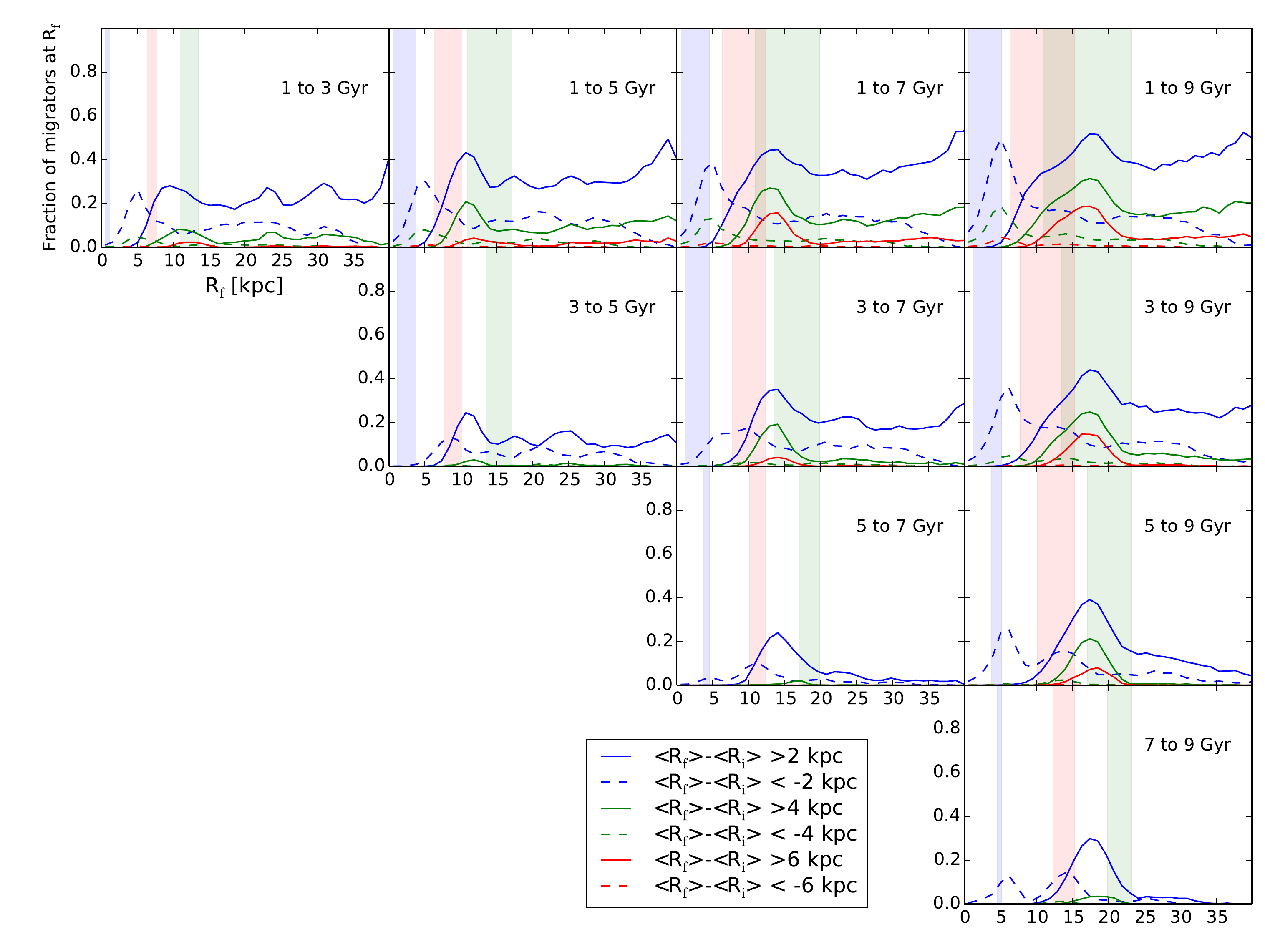}

\caption{Fraction of migrators as a function of radius (top half) or guiding radius (bottom half) at the end of the time-span specified in each plot. The shaded areas represent the bar ILR (blue), CR (red), and OLR (green) radii variation in the time-span of each plot. Note that these values are obtained with an isolated galactic disc with specific size parameters (possibly more extended than the Milky Way disc) and are not applicable to any specific galaxy.}
\label{locfracmigr-fig}
\end{figure*}

\subsection{Sources and sinks: How many migrators are there at a given radius? And how far from it did they originate?}
In the previous sections, we have quantified the main sources of migration in the disc and the flux of migrating stars as a function of distance from the galaxy centre. We now discuss where migrators are redistributed in the disc, and in particular, how many migrators can be expected at different radii throughout the disc. To this aim, we represent in Fig.~\ref{locfracmigr-fig} the fraction of stars of a radius bin around $R_{\rm f}$ at $t_{\rm f}$ that have migrated by more than $n$~kpc ($n=$2, 4, and 6) in radius since $t_{\rm i}$ (top half of the figure) or whose guiding radius has changed by more than $n$~kpc since $t_{\rm i}$ (bottom half). These fractions peak on both sides of the CR radius of the bar, consistently with the migrating fluxes of Fig.~\ref{churnedfrac-fig}, which peak at the bar CR radius: radii $R_{\rm f}>R_{\rm CR}$ receive stars migrating outwards from $R_{\rm i}<R_{\rm CR}$, while radii $R_{\rm f} < R_{\rm CR}$ receive stars migrating inwards from $R_{\rm i} > R_{\rm CR}$. However, as already discussed previously (see also next section),  inward and outward migrators cannot cross the whole disc, and in particular, outward migrators originated around the CR cannot reach the outer ($R > R_{\rm OLR}$) regions. \\
At first glance, the fraction of the whole population of migrators (churning+blurring) seems very high at certain radii (see top panels in Fig.~\ref{locfracmigr-fig}): as an example, in the time interval 1-9~Gyr, 40\% of the stars in the CR region are stars that have reached this region by migrating by more than 2~kpc from some inner radii by blurring or churning; in the OLR region between 40\% and 50\% of stars are outward 2~kpc--migrators; outside the OLR, the contribution of these stars to the local population increases nearly monotonically up to 80\%. This high fraction is due to the exponentially declining stellar surface density that causes inwards migrators to easily constitute a large part of the stars at an external final radius. The population of more extreme migrators -- those that have migrated by more than 6~kpc outwards by blurring or churning --  are still important contributors to the local (= at a given $R_{\rm f}$) stellar population: the fraction has a local maximum in the OLR region, at radii $< R_{\rm OLR}(t=t_{\rm f})$, where it peaks at  $\sim25$\%, it decreases to about 10\% at $R_{\rm f}=R_{\rm OLR}$, to finally increase again nearly monotonically in the outer disc. However, when we examine the contribution of migrators by churning alone at a given radius (bottom panels of  Fig.~\ref{locfracmigr-fig}), we can notice some different trends and absolute values. In particular, extreme ($\Delta R > 6$~kpc) outward migrators by churning never exceed 20\% of the local stellar population, with this value reached in the OLR region at radii below the final OLR radius. After this  maximum, the fraction diminishes to a value of about 5\% at the final OLR and stays constant in the outer disc. For inward migrators, the highest contribution is reached at radii just beyond the ILR region, with the contribution of long-distance migrators by churning being very limited.

\begin{figure*}
\centering
\includegraphics[width=17cm]{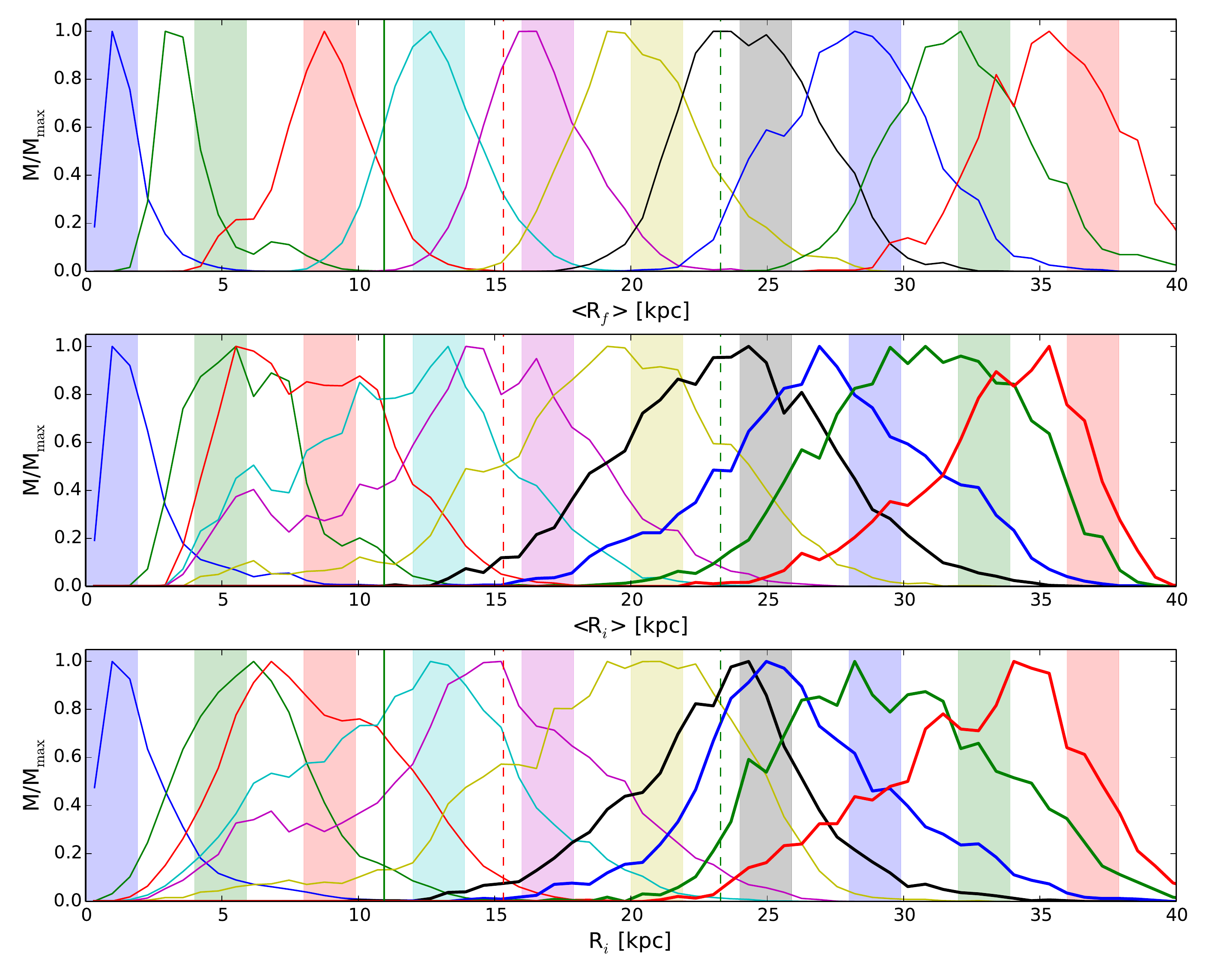}
\caption{Distributions of final guiding radii (top), initial guiding radii (middle), and initial galactocentric radii (bottom) of stars whose final galactocentric radii belong to bins of 2~kpc width. Each curve corresponds to the closest same-colour shaded bin of final galactocentric radii. The time interval is from 1~Gyr to 9~Gyr. The vertical dashed lines are the CR (red) and OLR (green) radii at 9~Gyr. The vertical solid green line is the OLR radius at 1~Gyr. In the two bottom lines, the histograms corresponding to stars in radius bins outside the OLR radius at 9~Gyr have been plotted with thicker lines (see next Section). The stellar masses $M$ in bins of each histogram are divided by the largest mass $M_{\rm max}$ contained in a bin of the histogram.}
\label{minchev-fig1}
\end{figure*}

\begin{figure}
\centering
\resizebox{\hsize}{!}{\includegraphics{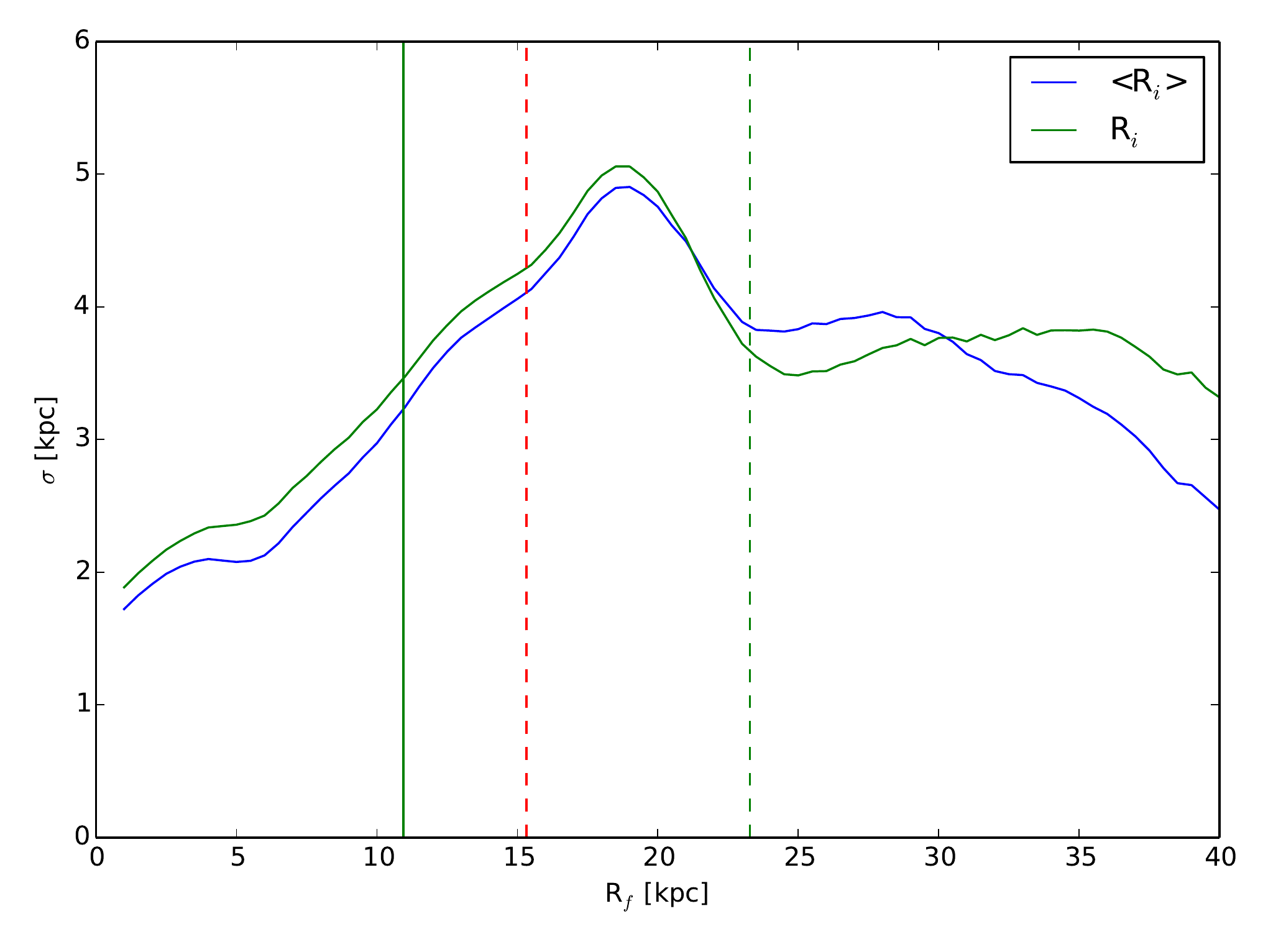} }
\caption{RMS values of galactocentric and guiding radii at 1~Gyr of stars with galactocentric radii in 2~kpc wide bins around $R$ at 9~Gyr. The vertical dashed lines, as in Fig.~\ref{minchev-fig1}, are the CR (red) and OLR (green) radii at 9~Gyr. The vertical solid green line is the OLR radius at 1~Gyr.}
\label{minchev-fig2}
\end{figure}

To quantify how migration (by churning and/or blurring) can affect a given radius, the distributions of initial and final radii for stars at different radial bins in the disc are often shown (see for example, \citep{minchev14}. To allow for a comparison of our results with \citet{minchev14}, we show in Figs.~\ref{minchev-fig1} and \ref{minchev-fig2} the results of an analysis similar to that in Figs.~3 and 4 of \citet{minchev14}. To keep some consistency with the rest of our analysis that studies migration in a fixed amount of time, we used as birth radii the radii of all stars at 1~Gyr. The top row of Fig.~\ref{minchev-fig1} shows the distribution of guiding radii at 9~Gyr of stars whose radii are in different bins (shown in shaded colours). The histograms, especially at large radii, tend to peak at smaller radii than the average radii of the shaded area, exhibiting the well-known asymmetric drift effect. The second row shows the distribution of guiding radii at 1~Gyr of the same stars (with guiding radii in the shaded bins at 9~Gyr). Comparison of the top and middle row shows that the distributions of guiding radii change between the initial and final times. The amplitude of the distributions tend to decrease at all radial bins (this is also observed by \citet{minchev14}). The initial distributions of $\langle R_i \rangle $, skewed towards low radii, translate into distributions of $\langle R_f \rangle$ that are much more symmetric. The differences between the distributions of initial and final guiding radii is evidence that migration by churning has occurred, and in particular, the increase of guiding radii with time indicates a predominance of outward migration. 
The bottom line shows the galactocentric radii of the same stars at 1~Gyr. The stars clearly have migrated significantly. The distributions are  very similar to that of the middle line, indicating that near the beginning of the simulation, at $t=$1~Gyr, most of the stars are in circular orbits. We compare the distributions of initial radii and guiding radii more quantitatively in Fig.~\ref{minchev-fig2} by representing the RMS value of the initial galactocentric radii or guiding radii of stars in bins of final radius (we use overlapping bins of 2~kpc width as in \citet{minchev14}). As in \citet{minchev14}, we see these RMS values, estimates of the widths of the histograms of Fig.~\ref{minchev-fig1}, are generally slightly higher for the initial galactocentric radii. The implications of Figs.~\ref{minchev-fig1} and \ref{minchev-fig2} on the mixing of the disc are discussed  in more detail in the next section.   

\section{OLR: a barrier for migrators}
\label{barriers}

In the previous section, we have reported that migrators by churning do not cross the OLR. In this section we develop this point further. \\
As previously described, the bottom panels of Fig.~\ref{diffrrav-fig} show the variation $\Delta R = \langle R_{\rm f} \rangle - \langle R_{\rm i} \rangle$ of guiding radii as a function of their initial guiding radius $R_{\rm i}$ in the time interval $[t_{\rm i}, t_{\rm f}]$. The dashed vertical line indicates the position of the OLR at $t_{\rm f}$, that is, at the end of the time interval under consideration, and the diagonal dashed line the equation $\Delta R=R_{\rm OLR}(t_{\rm f})-\langle R_{\rm i} \rangle$. It is evident from these figures that at all times, stars with $\langle R_{\rm i} \rangle < R_{\rm OLR}(t_{\rm i})$ do not cross the position of the OLR at the final time $t_{\rm f}$, these stars are always below the line  $\Delta R=R_{\rm OLR}(t_{\rm f})- \langle R_{\rm i} \rangle$. Migrators in the OLR region in the time interval under consideration, that is, with $R_{\rm OLR}(t_{\rm i}) \le R_{\rm i} \le R_{\rm OLR}(t_{\rm f})$,  can cross the OLR, but these are stars that at least at $t=t_{\rm i}$ had guiding radii all beyond the initial OLR position $R_{\rm OLR}(t_{\rm i})$. In other words, the only migrators by churning in a barred
galaxy at any given time that can be found beyond the OLR are stars migrating from a region between the OLR at the initial time of bar formation and the current OLR position. The larger the variations of the pattern speed of the bar over time, and consequently, the larger the variations $\Delta R_{ \rm OLR}=R_{ \rm OLR}(t_{\rm f})-R_{ \rm OLR}(t_{\rm i})$ of the OLR position, the larger will be the region of the inner disc where migrators crossing the OLR can originate from. In no case, however, can
stars born at corotation cross the final OLR. Analogously, if we were to draw a diagonal line of equation $\Delta R=R_{ \rm OLR}(t_{\rm i})- \langle R_{\rm i} \rangle$ in the bottom panels of  Fig.~\ref{diffrrav-fig}, this line separating stars that can cross the initial position of the OLR, we would find that no stars with $\langle R_{\rm i} \rangle \ge R_{ \rm OLR}(t_{\rm f})$  would migrate inward crossing the radius $R=R_{ \rm OLR}(t_{\rm i})$. This suggests that the OLR region, that is, the region between the initial (i.e. at the epoch of bar formation) and final (i.e. current) position of the OLR is a transition region for a disc galaxy, the only region where migrators by churning can be exchanged between the inner and the outer disc. But neither stars with $\langle R_{\rm i} \rangle \le R_{ \rm OLR}(t_{\rm i})$ can cross the OLR, nor, in the opposite sense, stars with $\langle R_{\rm i} \rangle \ge R_{ \rm OLR}(t_{\rm f})$ can penetrate below the radius $R_{\rm i}=R_{ \rm OLR}(t_{\rm i})$.
Note that some more contamination of the regions around the OLR can be caused by stars migrating by blurring (top panels of Fig.~\ref{diffrrav-fig}), even if, in general, these polluters originate in regions outside those spanned by the CR. The middle row of Fig.~\ref{minchev-fig1} also shows the very limited mixing between the regions inside the initial OLR (at $t=$1~Gyr) and the regions outside the final OLR (at $t=$9~Gyr). There are no stars that migrated by churning from inside the initial OLR and, vice versa, inside the initial OLR, there are no stars coming from outside the final OLR. The lack of contamination of the disc region outside the final OLR radius by the regions inside the initial OLR radius can be seen from the extent of the distributions of the initial guiding radii of stars that are beyond the final OLR (plotted as thicker lines). For example, none of the stars ending up at the final time between 24 and 26~kpc (that is, just outside the OLR at the final time $t=$9~Gyr) come from radii below $R_{\rm OLR}$($t=$1~Gyr)=11~kpc, whereas stars belonging to the bins centred between 5 and 21~kpc, inside $R_{\rm OLR}$($t=$9~Gyr), all have distributions of $\langle R_i \rangle$ that can reach values as low as 3~kpc. Figure~\ref{minchev-fig2} shows that RMS values of initial galactocentric or guiding radii of stars in a final galactocentric radius bin are monotonically increasing from the innermost regions up to a radius inside the final OLR. After this maximum, the RMS values tend to decrease in the outer disc, indicating again that there is no accumulation of migrators from the inner disc outside the final OLR. \\

To our knowledge, this result has not been pointed out before, and it recalls the similar barriers encountered by the gas in a barred galaxy \citep{schwarz81, combes88}: similarly to stars, gas in the CR-OLR region can also gain angular momentum to reach the OLR position, at most. Similarly to stars (Fig.~\ref{locfracmigr-fig}, bottom panels), gas accumulates in the OLR region, but  differently from the stellar component, gas is then able to dissipate part of its energy, shocks and forms rings of concentrated material at the OLR. Similarly, gas outside the OLR is not able to penetrate this resonance and access the inner region of the disc, until the barrier is removed (because of a significant change in the bar properties, see for example \citet{combes11}).\\

The results we presented are based on the analysis of a single simulation, and one may have concerns about the generality of the findings discussed. It is difficult to compare with other N-body  works presented in the literature, either because they usually lack information about the position of the OLR at initial times and only give its average position or position at the final time analysed or because the OLR is too close to the initial boundary of the simulated stellar discs to make robust predictions (see for example \citet{minchev12, dimatteo13}). However, an interesting comparison can be made with test particle simulations: in these experiments, the pattern speed of the main asymmetry is kept fixed, and as a consequence, the position of the OLR does not change with time; the role played by the OLR in these cases can be discussed. A comparison with the work by \citet{minchev10}, for example, shows that when only a bar pattern is present in the disc (their Fig 2, left panels), the  angular momentum exchange induced by the bar diminishes closer to the OLR and stops very clearly outside it, with a null angular momentum variation at radii outside the OLR. Independently of the bar strength, their experiments confirm the robustness of our results: the angular momentum exchange is only allowed inside the OLR, and, in particular, the maximal variation of angular momentum produced at the CR (where most of the migrators by churning are generated) is not sufficient to allow those stars to cross the outer resonance. In the right panels of the same figure, a similar analysis is made for a single spiral pattern. Unfortunately, the OLR is outside the x-range shown in these plots, and as a consequence, is not possible to discuss where the exchange of angular momentum stops. This can be seen in the work by \citet{sellwood02}, however. Their constrained model, containing a unique spiral pattern, shows that the OLR limits the exchange of angular momentum, with any variation stimulated at the OLR being extremely local and concerning a very limited number of stars. Of course, the response of a disc to asymmetric perturbations changes when many perturbations of similar amplitudes are superimposed, as is the case of several transient spiral patterns (\citet{sellwood02}, but also \citet{roskar08} and \citet{veraciro14}) or of the overlap of a bar and a spiral pattern of similar strengths (see, for example,  Fig.~3 of \citet{minchev10}). But these cases are different from the case discussed here, where a unique main pattern is present (the bar), with possibly an overlap with weak spiral arms, especially in the first 2~Gyr of evolution of the system. In this case, most of the angular momentum exchange occurs near the corotation of the bar, far from the OLR, and the contamination of the outer disc ($R > R_{\rm OLR}(t_{\rm f}$) with stars migrated from the inner disc ($R < R_{\rm OLR}(t_{\rm i}$) is very weak or null. In this case of a growing bar, the stopping of angular momentum exchange at the OLR of the bar is moreover consistent with the mechanism of bar growth: bars strengthen and grow by receiving angular momentum from stars near their ILR and by giving angular momentum to stars near their OLR \citep{lynden72}. There is thus some mixing close to the OLR, but beyond the OLR region there are no more resonant stars that may exchange angular momentum with the bar.

Why is this result important? For the Milky Way, at least for two reasons:
\begin{enumerate}
\item It may explain why the inner and the outer discs of the Galaxy have been able to maintain two different stellar populations over a time interval of $\sim$10~Gyr, which probably corresponds to the whole interval of secular evolution experienced by the Galaxy. We recall here the results of \citet{haywood13}, who extensively showed and discussed that the chemical properties
of outer-disc stars ($\langle R \rangle \ge 10$~kpc) are substantially different from those of stars of the inner disc of similar ages. These results have been confirmed by \citet{anders14} and \citet{nidever14}, who also showed that the chemical properties of the outer disc are significantly different from those of the inner disc (see also \citet{bensby12}, for a high-resolution spectroscopy study of the inner and outer disc). The current position of the OLR in the Milky Way is estimated to be slightly inside the solar radius by some observational or theoretical studies \citep{dehnen00, famaey05, minchev07}. This may therefore explain why the Sun appears to be in a transition region \citep{haywood13} and why the outer disc does not show signatures of any significant pollution by stars originating in the inner disc ($R \le ~6-7$~kpc).
\item Depending on the exact location of the Sun with respect to the OLR, the solar vicinity may have been strongly polluted by stars migrating by churning from the inner disc (see Fig.~\ref{locfracmigr-fig}, bottom panels, for stars inside the final OLR position), or not at all. If the Sun is outside the OLR position, as has been deduced by \citet{dehnen00, famaey05, minchev07} and the bar is the main source of asymmetric perturbations in the Milky Way, then the impact of churning at the solar vicinity may have been significantly overestimated (see Fig.~\ref{locfracmigr-fig}, bottom panels, for stars beyond the OLR region). In particular, stars that migrated by churning at the solar vicinity may well originate in regions much closer to the solar radius than previously thought, and
in this case, invoking substantial migration from the corotation and the inner disc is not possible.
\end{enumerate}
This result may also lead to doubts about the interpretation of U-turn in age profiles or inversion in colour-profiles found in the outer disc of external galaxies. When these inversions occur outside the OLR position for bar-dominated
galaxies, it is difficult to explain them in terms of strong migration from the inner disc, since, according to our results, no strong migration from the corotation is expected to contaminate the outer disc, unless another pattern with lower speed existed before the present one, with an OLR extending to outer radii.

\section{Migration and cooling/heating}

\label{heatcool-sec}

\subsection{Vertical heating/cooling}

The effect of radial migration on the vertical structure of discs has been studied in a number of works \citep[e.g.]{minchev12z, solway12, roskar13, veraciro14}. The assumption that outward migration would help create a thick disc because of the higher vertical velocity dispersion of the outward migrators originating from the inner hotter disc has been debated because of the vertical cooling outward migrators should undergo if the vertical action of a stellar orbit is conserved \citep{minchev12z, solway12, roskar13}. 

\begin{figure}[h!]
\centering
\resizebox{\hsize}{!}{\includegraphics{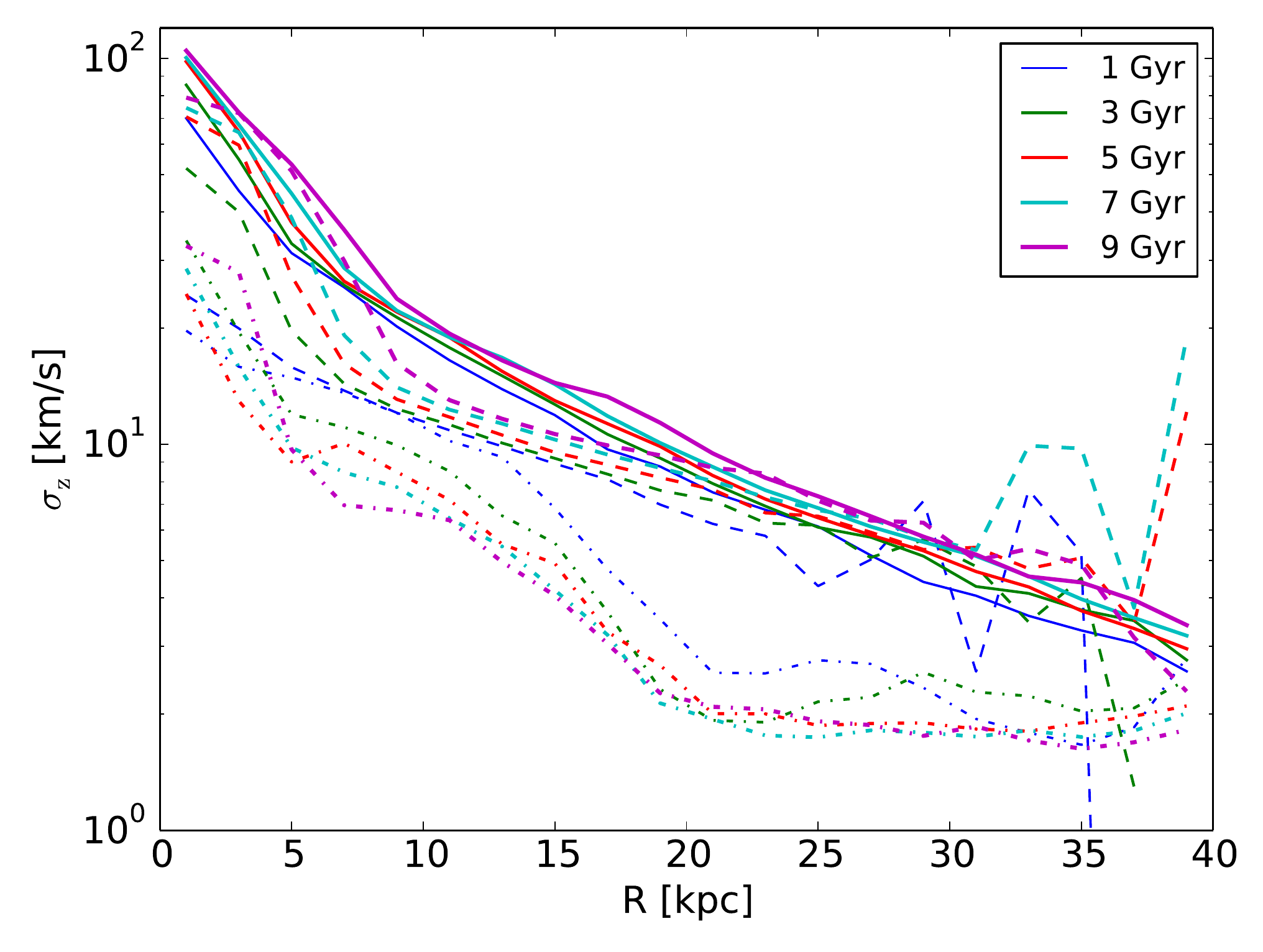}}
\caption{Time evolution of the vertical velocity dispersion of the old stellar disc without the bulge (solid), the young stellar disc (dashed), and the gas disc (dotted-dashed).}
\label{velzdisp-fig}
\end{figure}

We observe that the vertical velocity dispersion of the stellar disc slightly increases with time at all radii. In Fig.~\ref{velzdisp-fig} we plot the vertical velocity distribution of the old stellar disc, the stellar disc of stars formed during the simulation, and the gas disc. The velocity dispersion of the new stars, formed from the colder gas disc, is generally lower than the velocity dispersion of the old stellar disc. The mass of new stars is only a small fraction of the total stellar mass (see Fig.~\ref{surfdens-fig}), however, so star formation has no net cooling effect on the total stellar disc.  

\begin{figure*}[!htb]
\centering
\includegraphics[width=13cm]{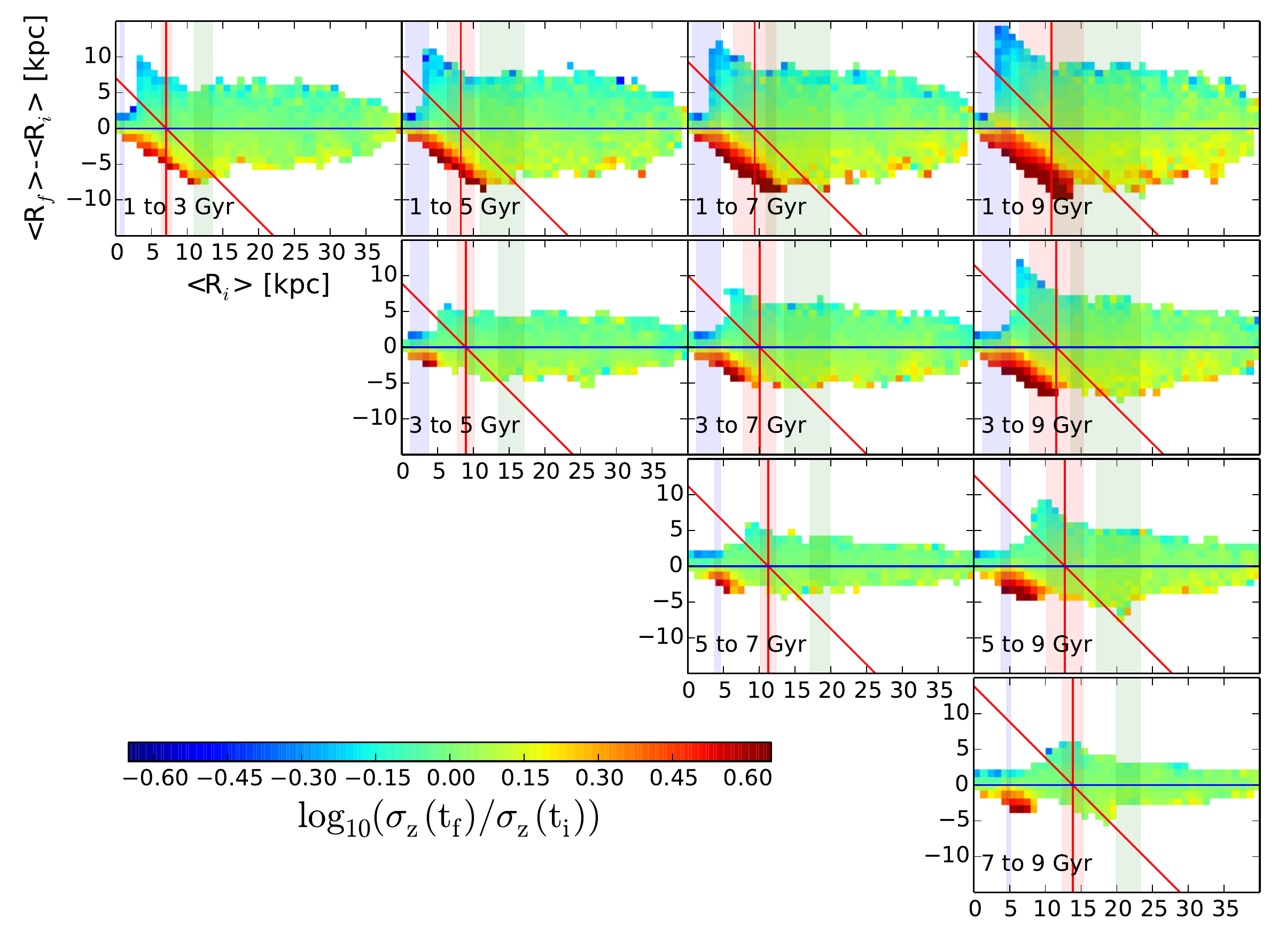} 
\caption{Ratio of vertical velocity dispersion of stars in bins in $\langle R_{\rm i} \rangle$ and $\langle R_{\rm f} \rangle  - \langle R_{\rm i} \rangle $ to the vertical velocity dispersion of all stars in radial bin centred on $\langle R_{\rm i} \rangle$. The shaded areas represent the bar ILR (blue), CR (red), and OLR (green) radii variation in the time-span of each plot. The red vertical lines are the average of the bar CR radius during the time-span of a plot.}
\label{sigvssigri-fig}
\end{figure*}

\begin{figure*}[!htb]
\centering
\includegraphics[width=13cm]{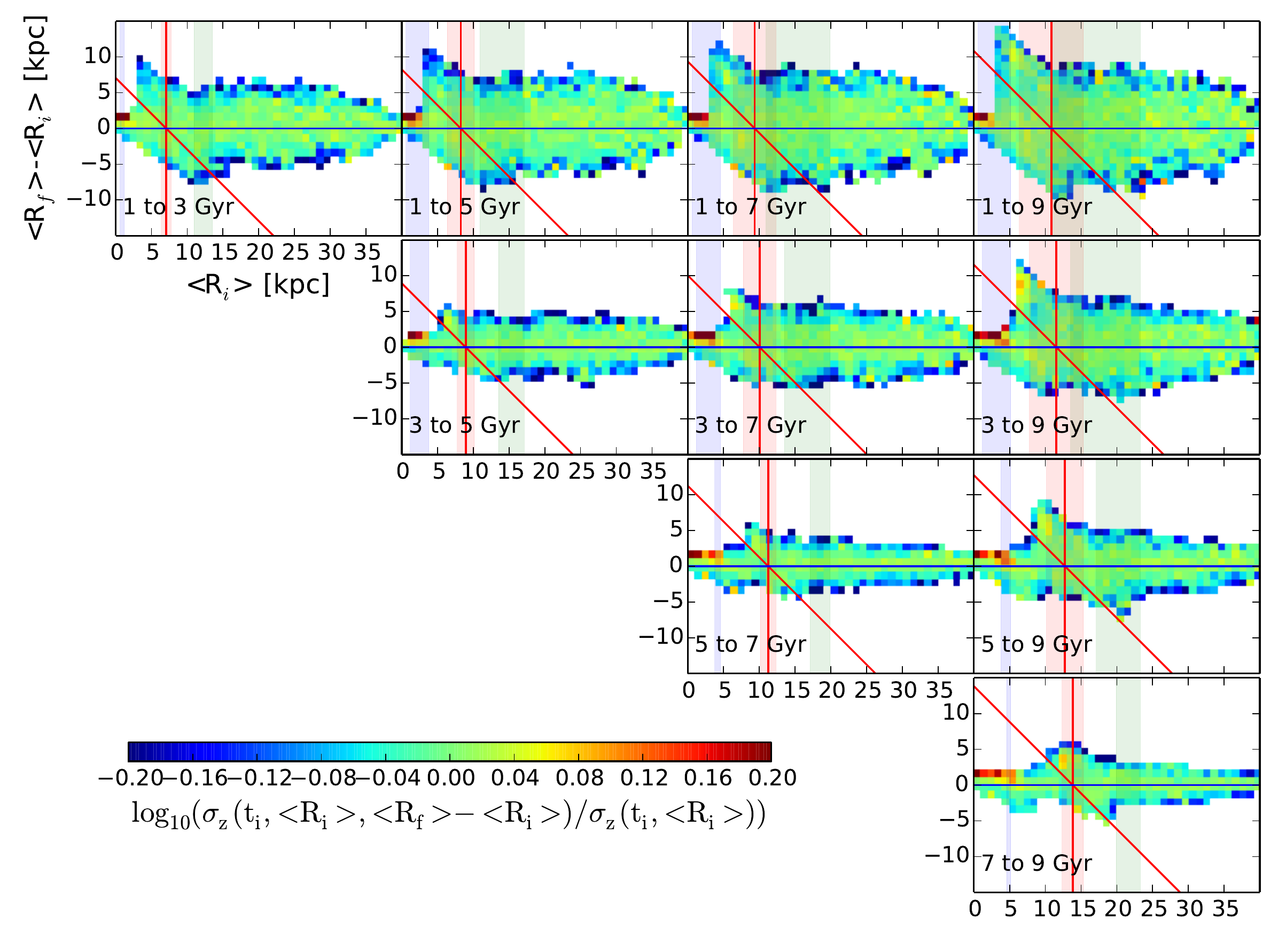} 
\caption{Distribution of the ratio of final to initial vertical velocity dispersion of stars migrated by $\Delta R=\langle R_{\rm f} \rangle  - \langle R_{\rm i} \rangle$ from their initial guiding radius $\langle R_{\rm i} \rangle$ in the time interval $[t_{\rm i},t_{\rm f}]$.  The shaded areas represent the bar ILR (blue), CR (red), and OLR (green) radii variation in the time-span of each plot. The red vertical lines are the average of the bar CR radius during the time-span of a plot.}
\label{sigevol-fig}
\end{figure*}

\begin{figure*}[!htb]
\centering
\includegraphics[width=13cm]{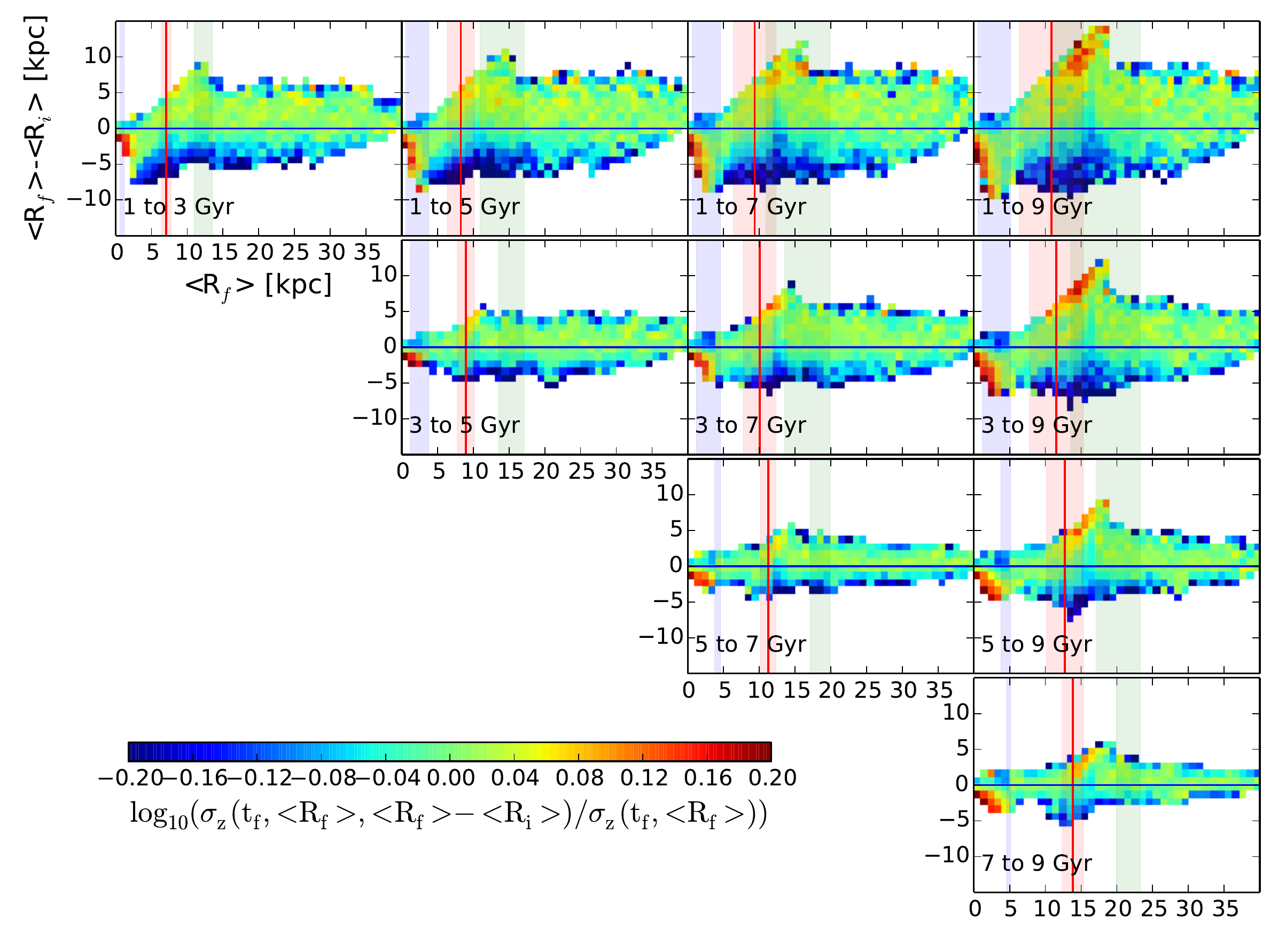} 
\caption{Distribution of the ratio of vertical velocity dispersion of stars in bins in $\langle R_{\rm f} \rangle$ and $\langle R_{\rm f} \rangle  - \langle R_{\rm i} \rangle $ to the vertical velocity dispersion of all stars with guiding radii centred on the final value $\langle R_{\rm f} \rangle$. The shaded areas represent the bar ILR (blue), CR (red), and OLR (green) radii variation in the time-span of each plot. The red vertical lines are the average of the bar CR radius during the time-span of a plot.}
\label{sigvssigrf-fig}
\end{figure*}

We would like to know if the stars that are going to migrate are a special sub-population in terms of kinetics of stars with the same initial guiding radius, whether the migration affects their kinetic state, and if the migrators affect the population of stars at their final guiding radius. 

\citet{veraciro14} found a provenance bias of migrators in terms of kinetic state. They found that radial migration driven by spiral arms (seeded in their simulations by massive perturbers) preferentially affect stars with a lower velocity dispersion than the average velocity dispersion at the initial radius. We find a similar trend, as shown in Fig.~\ref{sigvssigri-fig}, where the ratio of the vertical velocity dispersion of stars in bins in $\langle R_{\rm i} \rangle$ and $\langle R_{\rm f} \rangle  - \langle R_{\rm i} \rangle $ (we only used bins that
contained at least the mass of ten old-disc stellar particles) to the  vertical velocity dispersion of all stars in a bin at $\langle R_{\rm i} \rangle$ (a column in the plots) is represented. The stars migrating the most from an initial guiding radius tend to be colder in the z-direction than the average for all the stars at this initial guiding radius. 

We next investigated whether the migrators can be subject to heating (or cooling) when they reach hotter (respectively colder) regions. In Fig.~\ref{sigevol-fig} we represent the distribution of the ratio of the final vertical velocity dispersion to the initial one for stars that have migrated by $\Delta R=\langle R_{\rm f} \rangle- \langle R_{\rm i}\rangle $ from their initial radius $R_{\rm i}$ in a time interval $[t_{\rm i}, t_{\rm f}]$.  We observe that, especially in the inner regions of the disc, that is, inside the OLR, outward migrators tend to lower their velocity dispersion, while inward migrators tend to increase it. The effect increases with the amplitude of the variation (it is more visible for the extreme cases of migration). 

\begin{figure}[h!]
\centering
\resizebox{\hsize}{!}{\includegraphics{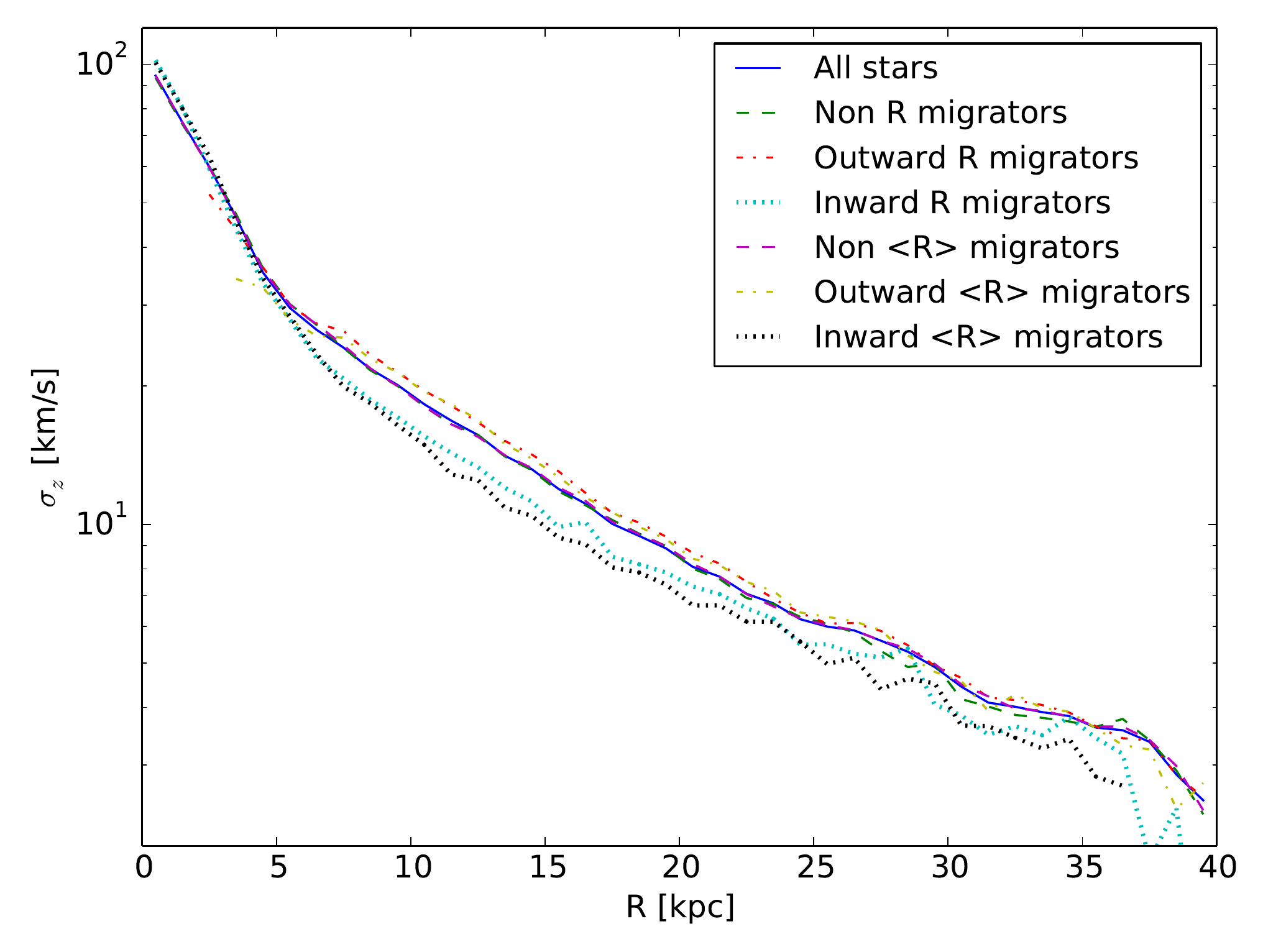}}
\caption{Vertical velocity dispersion at $t=3$~Gyr at $R$ of all stars, non-migrators, or migrators, that migrated by more than 2~kpc in terms of radius or guiding radius from 1 to 3~Gyr.}
\label{migreff-fig}
\end{figure}

We now discuss how the final vertical velocity dispersion of migrators compares to that of stars with the same final guiding radii $\langle R_{\rm f} \rangle$.
In Fig.~\ref{sigvssigrf-fig} we represent the distribution of the ratio of the velocity dispersion of migrators in a bin in $\langle R_{\rm f} \rangle$ and $\langle R_{\rm f} \rangle  - \langle R_{\rm i} \rangle $ to the velocity dispersion of all stars of guiding radius $\langle R_{\rm f} \rangle$. We observe different trends, depending, among others, on the location of the radius $R_{\rm f}$ under consideration. We start by discussing what occurs inside the OLR at the final time. We observe that in the very inner regions, stars migrating inwards to reach a guiding radius $\langle R_{\rm f} \rangle$ have a higher velocity dispersion than the whole population of stars at $\langle R_{\rm f} \rangle$. These stars must be taking part in the peanut-shaped bar with a high vertical velocity dispersion. This agrees with the results reported by \citet{dimatteo14}, who found that outside-in migrators participating in a boxy/peanut shaped structure tend to be kinematically hotter than in situ stars. At higher radii, however, the inward migrators tend to have lower velocity dispersions than the average at their final guiding radius. Outward migrators tend to have a slightly higher ($\sim$20\%) vertical velocity dispersion than all stars at their final guiding radii, except for the most extreme migrators --those with the highest $\Delta R$ variations, found  between the CR and the OLR region. These extreme  migrators can be significantly hotter (up to 50\% more) than the whole population at the same final guiding radius.  We note that this effect is only evident when a selection on the amplitude of the migration $\Delta R$ is made. When the overall population of (inward or outward) migrators is analysed as a whole, the final vertical velocity dispersions of migrators and non-migrators are much more similar, which is because apart from the extreme cases of migration, many stars that have migrated by small $\Delta R$ tend to have a final kinematic similar to that of the overall population. We can see this in Fig.~\ref{migreff-fig}. In this figure, we have plotted the vertical velocity dispersion of all stars as a function of radius at $t=3$~Gyr and the velocity dispersion of inward and outward migrators that are in a radius bin and that have migrated by more than 2~kpc in terms of radius or guiding radius since $t=1$~Gyr, and the non-migrators defined as the stars of the radius bin that have not changed their radius or guiding radius by more than 2~kpc. The inward migrators have a slightly lower vertical velocity dispersion, the outward migrators have a slightly higher velocity dispersion, but the total velocity dispersion is very close to the one of the non-migrators, indicating a weak effect of the overall migration on the local velocity dispersion. We insist, however, on the fact that the most extreme migrators, in particular the most extreme outward migrators, can have dispersions significantly different (50\%) from the average. While we agree with the results by \citet{minchev12z},
therefore, that  migration overall contributes little to disc thickening, extreme migrators can depart from the average small increase of the heating found in Fig.~\ref{migreff-fig} and found also by \citet{minchev12z}, contributing with significantly higher vertical velocity dispersions to the whole sample of migrating stars that end up at the same final guiding radius. 

\subsection{Radial heating/cooling}

\begin{figure}[!htb]
\centering
\resizebox{\hsize}{!}{\includegraphics{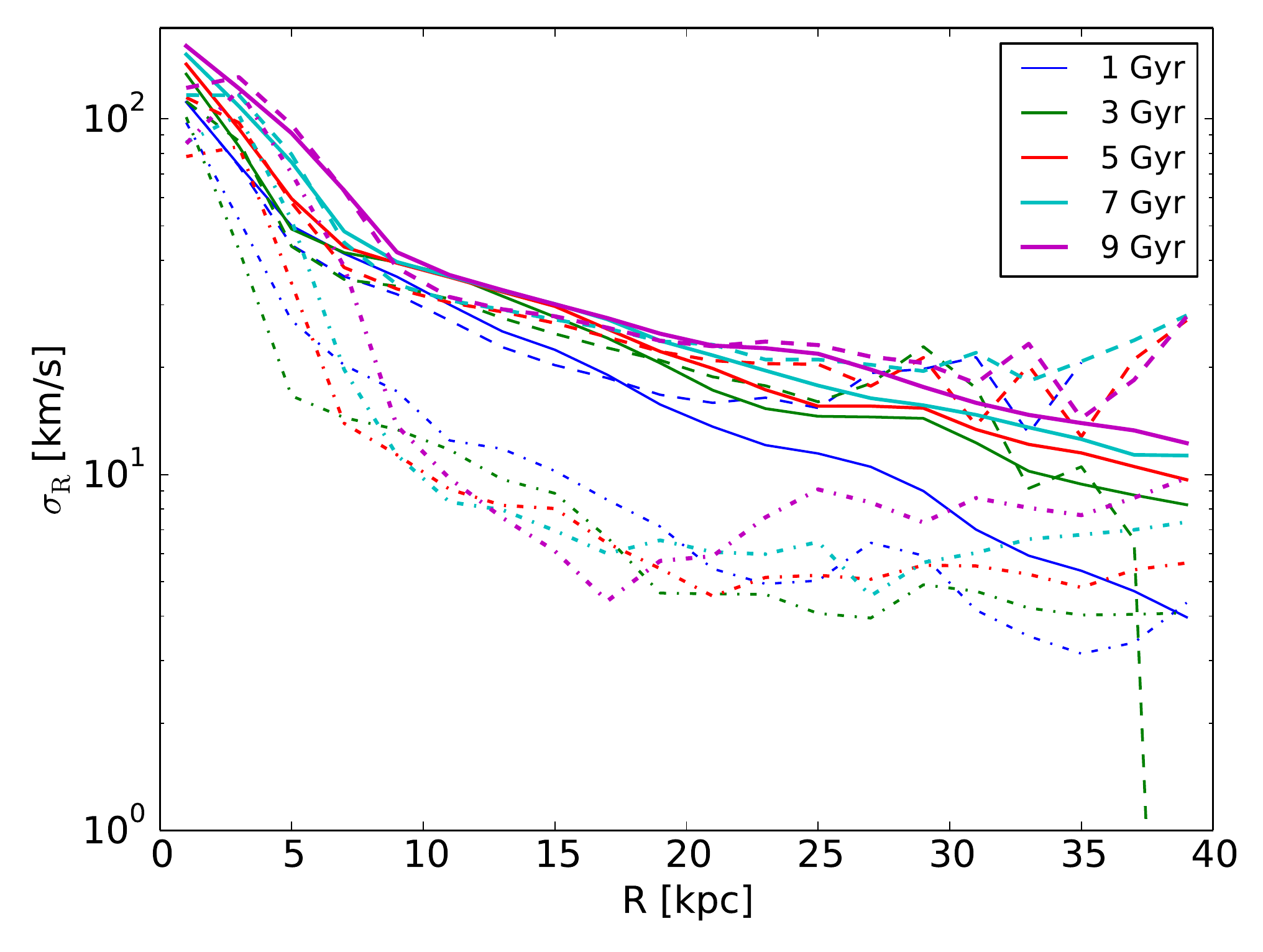} }
\caption{Time evolution of the radial velocity dispersion of the old stellar disc without the bulge (solid), the young stellar disc (dashed), and the gas disc (dotted-dashed).}
\label{velrdisp-fig}
\end{figure}

\begin{figure}[!htb]
\centering
\resizebox{\hsize}{!}{\includegraphics{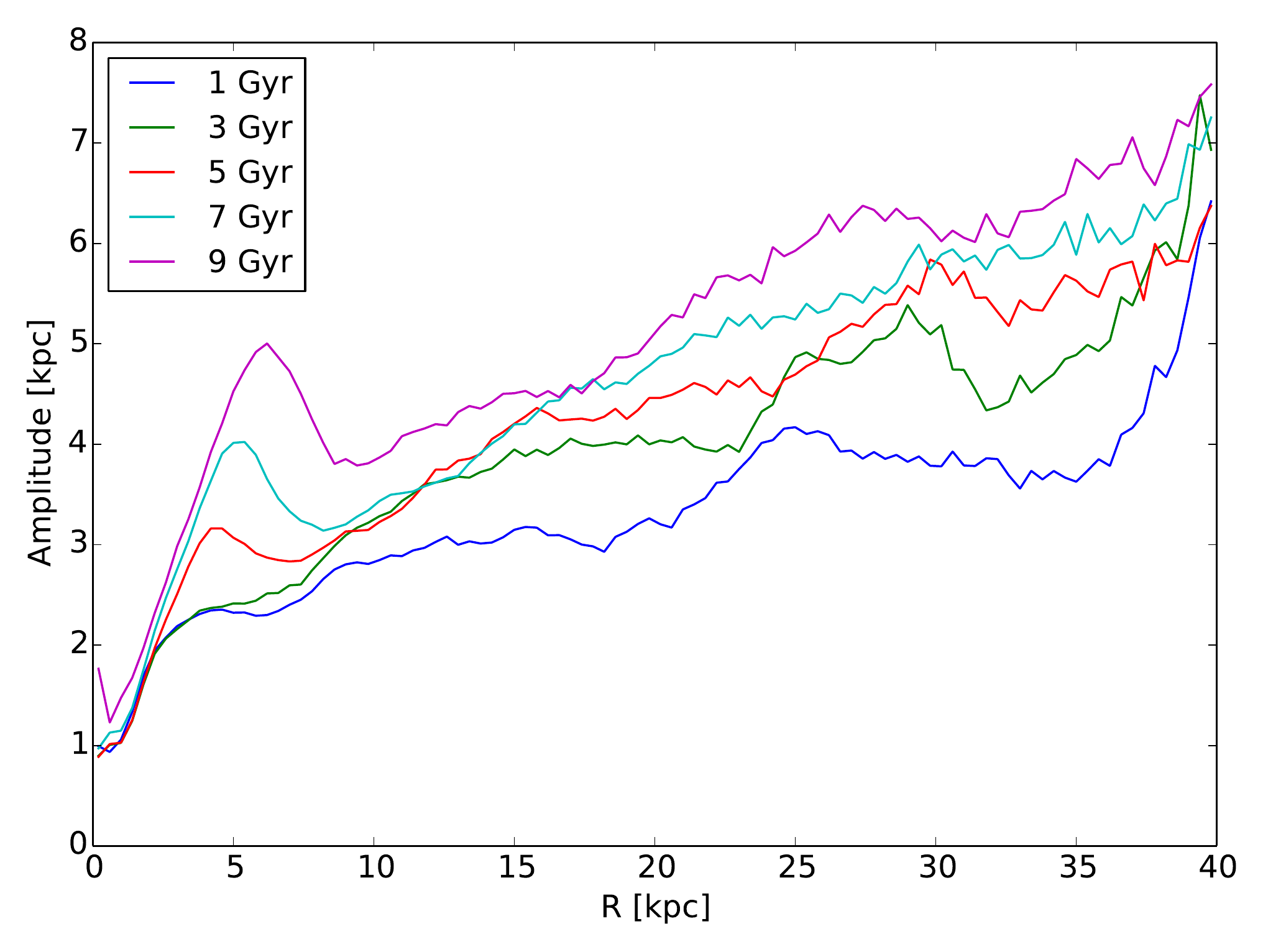}}
\resizebox{\hsize}{!}{\includegraphics{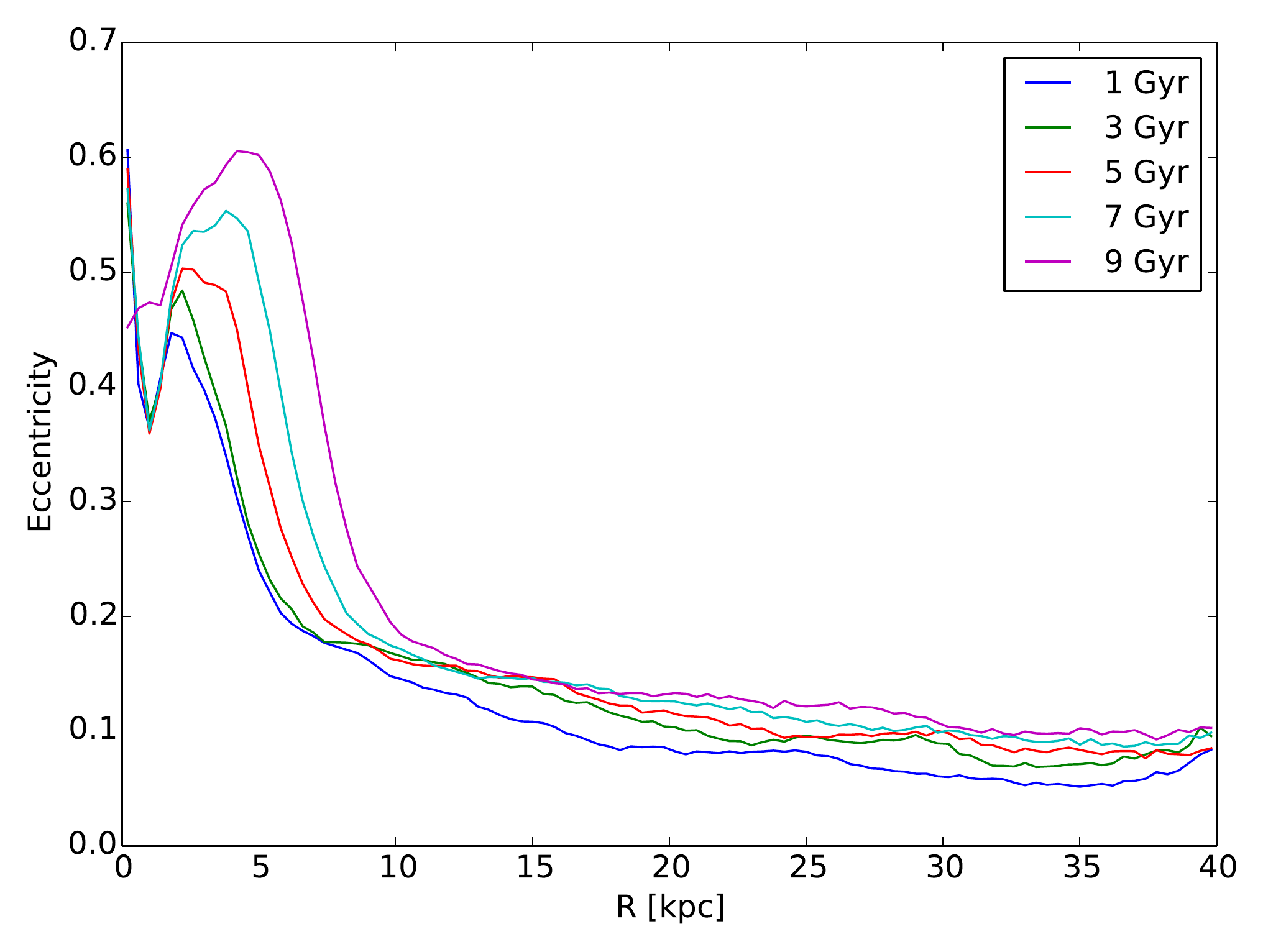}}
\caption{Time evolution of the amplitude of radial oscillations (top) and eccentricity (bottom) with time.}
\label{ampecc-fig}
\end{figure}

\begin{figure}[!htb]
\centering
\resizebox{\hsize}{!}{\includegraphics{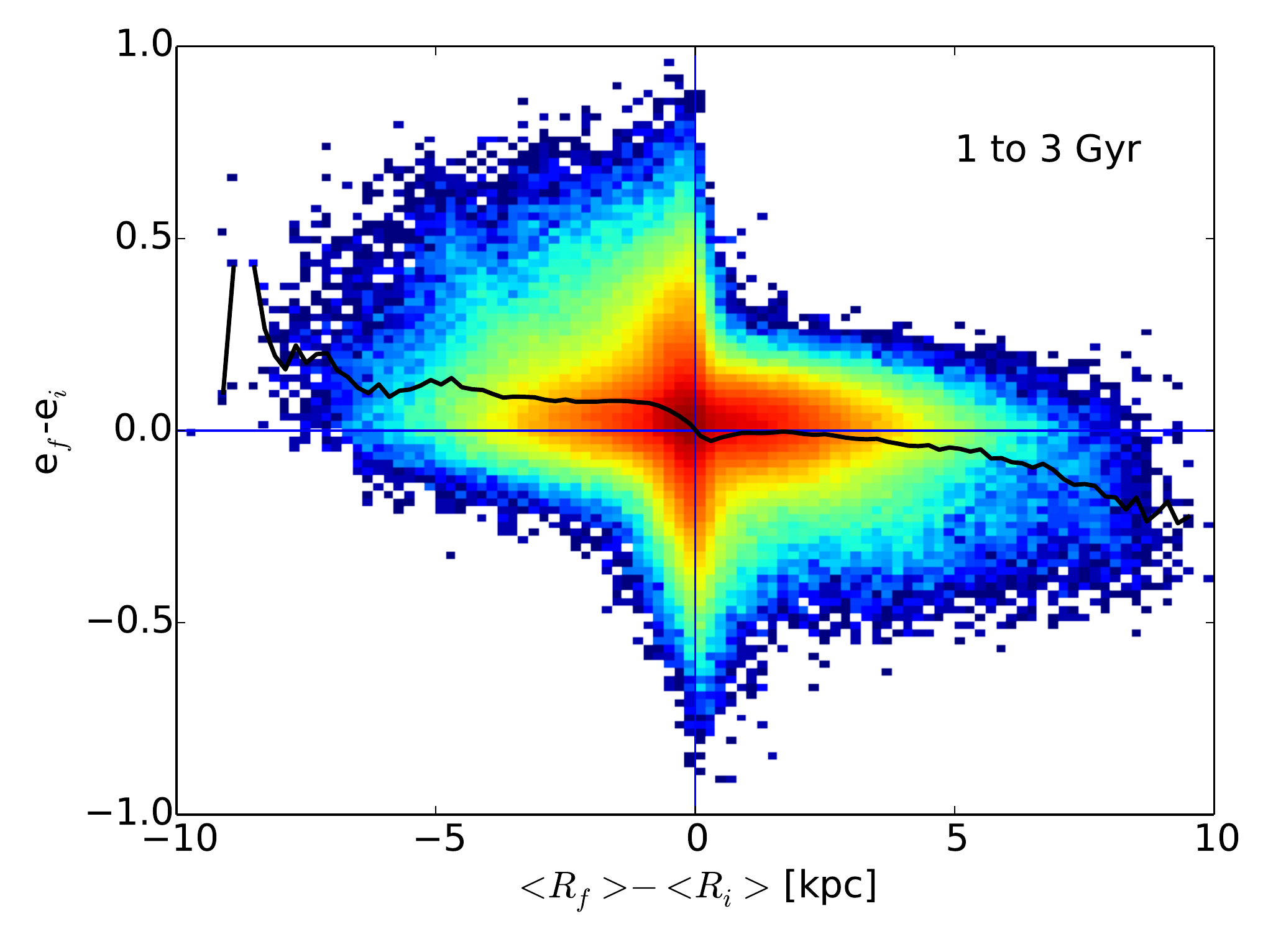}}
\resizebox{\hsize}{!}{\includegraphics{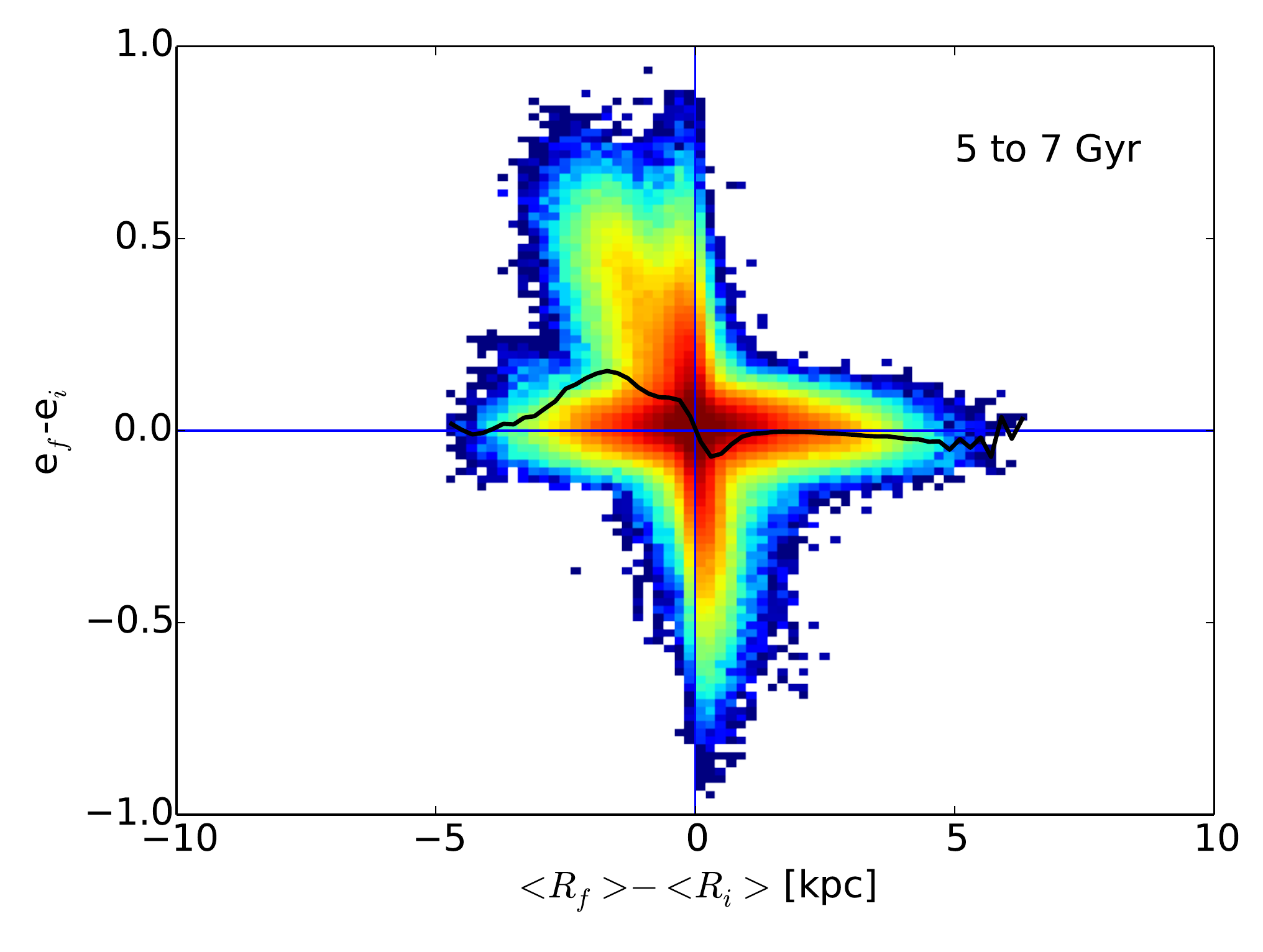}}
\caption{Change of eccentricity as a function of change of guiding radius from 1 to 3~Gyr (top) and 5 to 7~Gyr (bottom). The colour scale represents the mass in each bin and is logarithmic. The solid black line is the average of the change in eccentricity by bin of change of guiding radius (i.e. the average of the corresponding column).}
\label{eccvar-fig}
\end{figure}

The radial velocity dispersion also increases on average with time, as can be seen in Fig.~\ref{velrdisp-fig}. We can follow the amplitude of the radial oscillations as well as the eccentricity of the trajectories of the stellar particles defined as
\begin{equation}
e=\dfrac{R_{\rm max} - R_{\rm min}}{R_{\rm max} + R_{\rm min}}
.\end{equation}

The eccentricity varies between 0 for circular orbits and 1 for radial orbits. The mean eccentricity of the disc stars increases with time. This increase is mainly caused by the growth of the bar: More and more stars are gradually trapped by the bar and gain eccentricity as their orbits become elongated. The evolution of the amplitude of radial oscillations ($2 A(t)$ where $A(t)$ is the semi-amplitude introduced in Sect.~\ref{amp-sec}) and eccentricity $e$ as a function of radius can be seen in Fig.~\ref{ampecc-fig}. The bar growth is visible from the increase of eccentricity and amplitude with time in the inner kpc. The amplitude of radial oscillations increases on average with time at all radii, which is consistent with the increase of radial velocity dispersion shown in Fig.~\ref{velrdisp-fig} and with the growing role of blurring with time, as discussed in Sect.~\ref{amp-sec}. The eccentricity also slightly increases at all radii.

\begin{figure*}
\centering
\resizebox{\hsize}{!}{\includegraphics{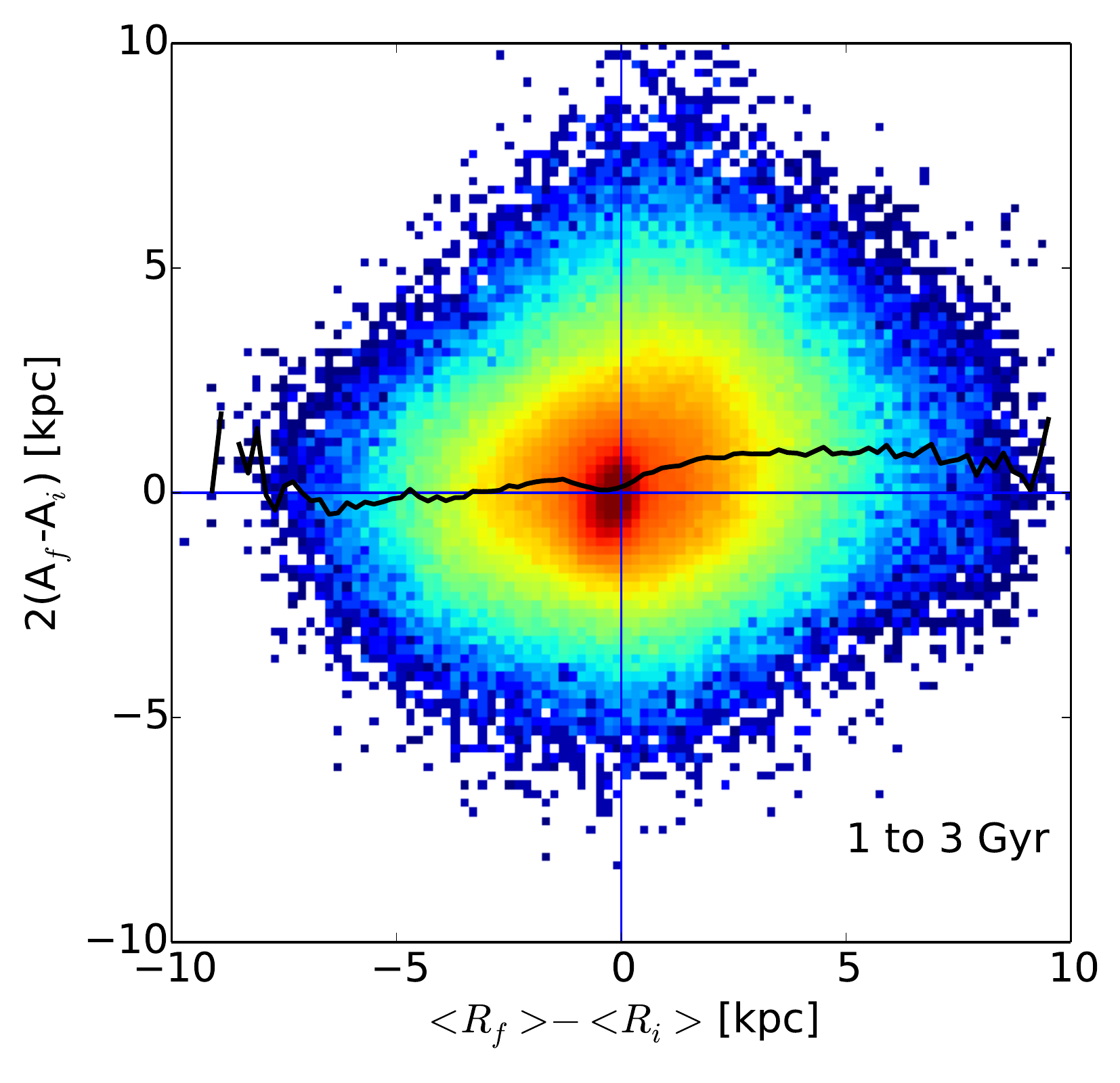}
\includegraphics{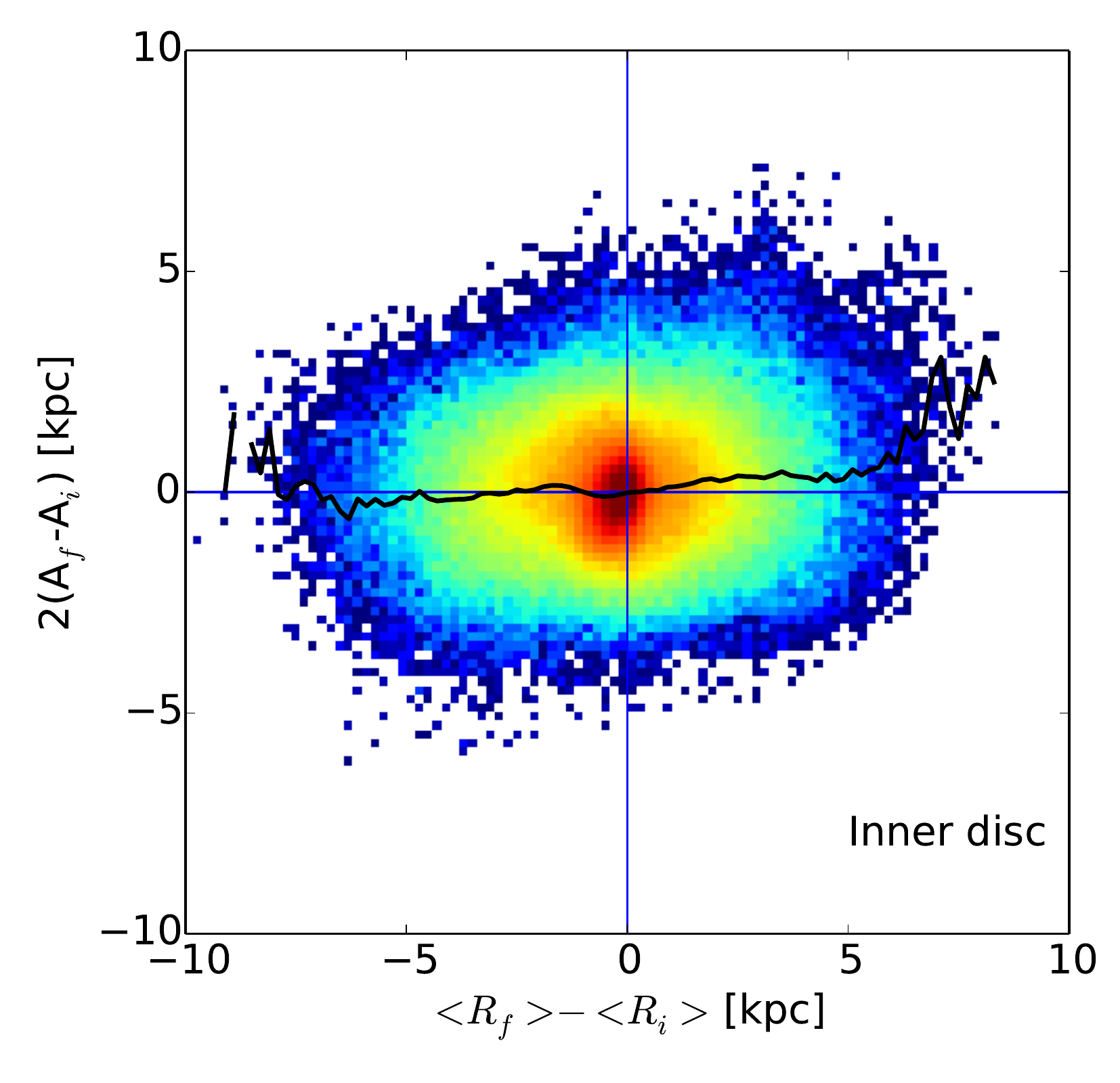}
\includegraphics{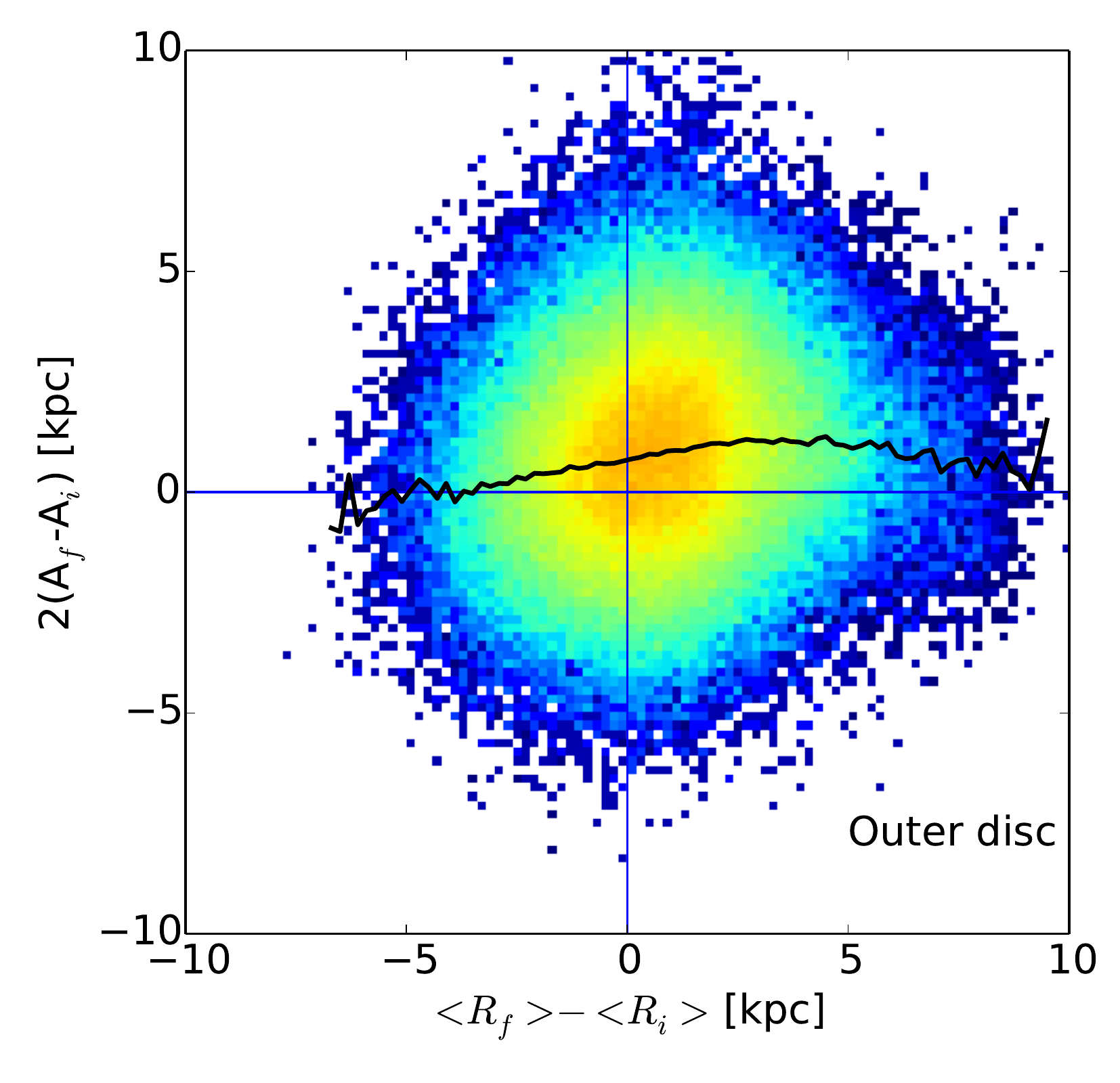}}
\resizebox{\hsize}{!}{
\includegraphics{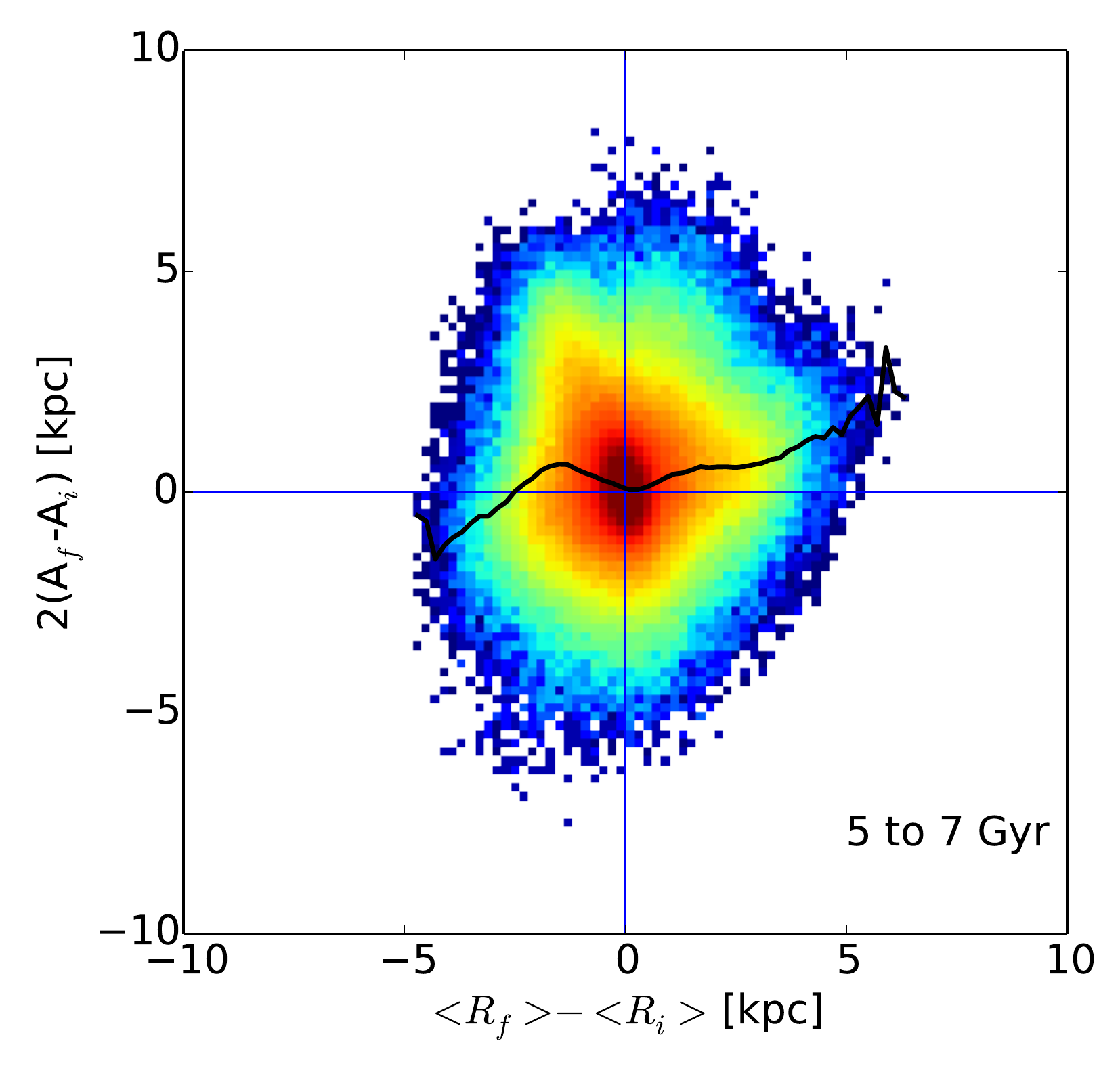}
\includegraphics{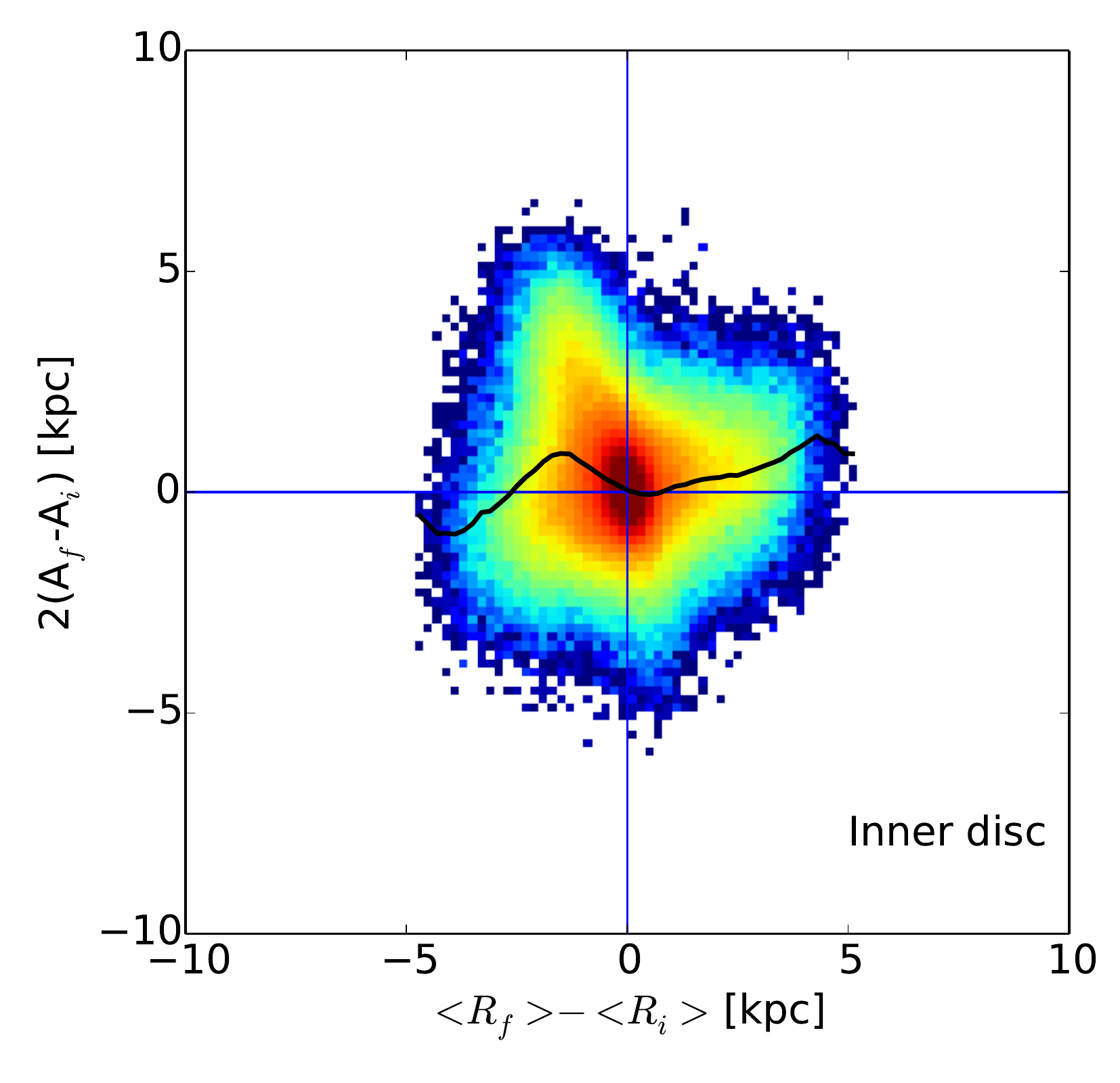}
\includegraphics{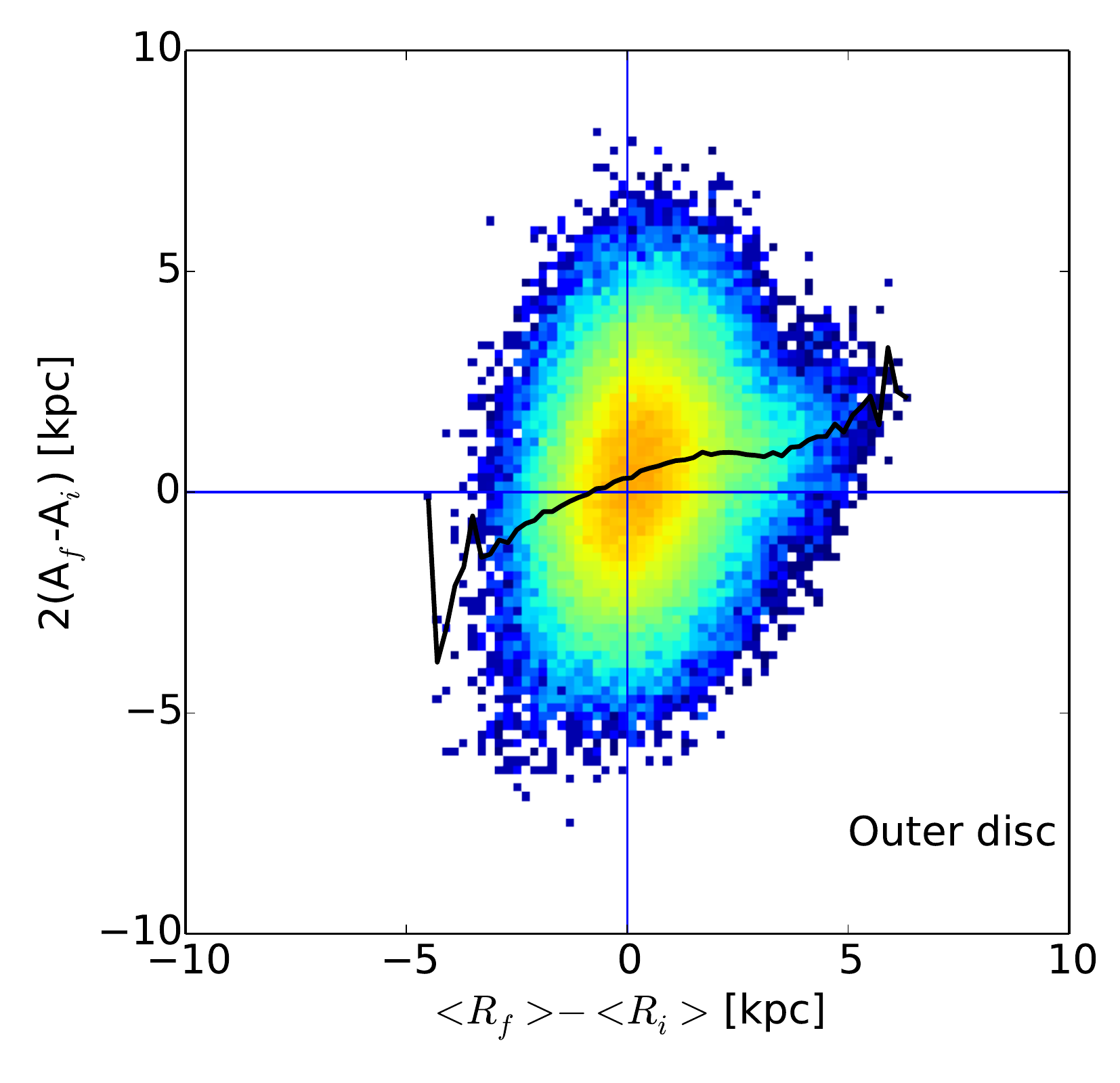}}
\caption{Change of radial oscillation amplitude as a function of change of guiding radius from 1 to 3~Gyr (top) and 5 to 7~Gyr (bottom). The colour scale represents the mass in each bin and is logarithmic. The solid black line is the average of the change in amplitude by bin of change of guiding radius (i.e. the average of the corresponding column).}
\label{ampvar-fig}
\end{figure*}

It is interesting to investigate whether radial migration changes the radial amplitude and the eccentricity of the stars trajectories. In Fig.~\ref{eccvar-fig}, we represent the change in eccentricity as a function of the change in guiding radius between $t=1$ and 3~Gyr in the top panel and between $t=5$ and 7~Gyr in the bottom panel. The distribution is broad. The majority of stars only
slightly change their eccentricity, but on average (black line), stars migrating outwards (with positive $\langle R_{\rm f} \rangle  - \langle R_{\rm i} \rangle $) tend to decrease their eccentricity while the opposite stands for stars migrating inwards. The inward effect increase of the eccentricity (especially visible in the bottom panel of Fig.~\ref{eccvar-fig} from 5 to 7~Gyr) is again dominated by the capture of stars by the central bar. The decrease of eccentricity does not necessarily mean that the orbits are circularised in terms of decrease in amplitude of radial oscillations because of the denominator in the expression of the eccentricity, which can increase in the case of outwards migration. In Fig.~\ref{ampvar-fig} we therefore represent the variation of the amplitude of radial oscillations and separate in the middle and right columns the disc into an `inner disc' and an `outer disc' by a cut between the CR and OLR bar radii at 10~kpc for the first time-interval and at 15~kpc for the second one, so that the effect of the bar growth on the stellar orbits is not present in the `outer disc' plots. In the inner-disc plot of the time-span from 5 to 7~Gyr, a clear trend of an increase of the amplitude of the radial oscillations is visible for inward migration, corresponding to stars migrating inwards and contributing to the bar. All the distributions are again broad, with the largest variations of amplitude occurring for stars that do not migrate significantly. Stars that migrate outwards the most increase their radial oscillations amplitude on average, but the possible increase is limited compared with the rest of the distribution.

\section{Conclusions} 

We have studied stellar radial migration in a simulation of an Sb-type extended galactic disc. We confirmed the main role of corotations of the density resonances as seeds for radial migration in terms of both blurring and churning and observed a strong radial migration when there is a bar-spiral resonance overlap;
this is consistent with the results of \citet{minchev10, minchev11}.    

We have quantified the effects of blurring and churning. Migration defined as the simple difference in galactocentric radii between two  times can lead to significantly overestimating  the fraction of migrators by churning -- and also the spatial extension of this migration.  The intrinsic nature of the epicyclic orbits and the increase in the amplitude of radial oscillations with time -- both effects being included when migration is quantified by a change of instantaneous radius -- can indeed give the impression of a much stronger migration than that really experienced by stars in the galaxy. This is particularly true, for example, for the fraction of migrators in  the outer disc of our simulated galaxy. When blurring is excluded, the fraction of extreme migrators contributing to the outer disc population decreases by a factor between 2 and 8 (compare Fig.~13, top and bottom panels, for the time interval 1 to 9~Gyr).    

Whilst the spatial extent covered by migrating stars increases with time, our simulation suggests that migrators encounter barriers. In particular, stars migrating by churning from corotation cannot cross the OLR, and vice versa, stars born beyond the OLR cannot reach the inner disc. The OLR region -- defined as the region  between the position of the OLR at the epoch of bar formation, and at the final epoch -- is a transition region, the only region where some pollution between the inner and the outer disc is allowed.

Even though our model is not intended to reproduce the Milky Way -- the pattern speed of our simulated bar is lower than the pattern speed measured for the Galaxy \citep{gerhard11}, and in consequence, the main resonances are located at much larger distances from the centre than those measured for the Milky Way, and the study assumes an initially already formed thin disc -- we think that the previous result might help understand the puzzling nature of the outer Galactic disc and its significantly different stellar populations. It has recently been shown that the Milky Way outer disc followed a different chemical evolution history than did the inner disc \citep{haywood13,snaith14}. Regardless
of the formation mechanism of the outer regions of the Galaxy, stars there have been able to evolve independently of the inner disc. Our model suggests that the Galactic OLR and its location (estimated to be located slightly inside the solar radius, see \citet{dehnen00}) have a major role in explaining this finding. The OLR indeed acts as a barrier for gas \citep{combes88}, but - as we showed here - for stars as well, inhibiting migration from the inner to the outer disc and vice versa. This may explain why stars with inner thin disc chemistry are not observed in the outer disc \citep{haywood13}. This finding, if confirmed with future studies, may also lead to doubts about the interpretation of the U-shape in age profiles, or inversion in colour-profiles found in the outer disc of external galaxies. We do not yet know whether our Galaxy has the U-profile in stellar ages that is sometimes observed in external galaxies like M33. When these inversions are observed outside the OLR position in barred galaxies, it is difficult to explain them in terms of strong migration from the inner disc, with the observed bar/spiral pattern speed, since they occur beyond their OLR. However, it is possible that previous bar/spiral waves have developed with lower pattern speeds, implying OLR and migration at larger radii. Another possible explanation is that the settlement of the Galactic disc in the outer regions permitted the formation of a significant number of old stars in situ, as proposed for the Milky Way in \citet{haywood13} and \citet{snaith14}.

Finally we have analysed the kinematics of migrating stars. \\
We confirmed the results found by \citet{veraciro14} for spiral galaxies. Similarly to that case, also when migration is mainly induced by a stellar bar, there is a provenance bias of migrators in terms of their kinetic state. The stars migrating the most from an initial guiding radius tend to be colder in the z-direction than the average of all the stars at the same initial guiding radius. \\
We also confirmed the results by \citet{minchev12z}: migration contributes little to disc thickening, but we point out, consistently with \citet{minchev12}, that there is a trend of increasing vertical velocity dispersion with the extent of migration: the most extreme outward migrators that end up at a given final radius tend to also have the highest velocity dispersions when compared to the velocity dispersion of all the stars found at that final radius. Thus, while the overall effect of heating at a given radius is weak, we suggest that at a given radius, extreme outward migrators from the inner disc  are identifiable as stars that have the highest velocity dispersions among those measured for stars of the same age, at the same radius. We recall that this signature is different from that induced by mergers, which could heat the outer disc enough for extreme migrators from the central parts of the galaxy to have a cooling effect on the outer disc \citep{minchev14}.

Overall, our findings challenge the current view of the effect that radial migration from the inner disc may have in the outer regions of disc galaxies when only a main asymmetric pattern is present. This may fundamentally affect the understanding of stellar populations in bar-dominated galaxies, which we will investigate in future studies.

\label{conclu}

\begin{acknowledgements}
The authors wish to thank A. Gomez for her support, encouragements, and suggestions. This work also benefited of several enriching discussions with D.~Katz, M.~D.~Lehnert, and O.~N.~Snaith. AH thanks the Observatoire de Paris, which funded her work through an ATER grant. 
\end{acknowledgements}

\bibliographystyle{aa} 

\bibliography{papermigr}

\end{document}